\documentclass[11pt]{article}
\usepackage[english]{babel}
\usepackage{amsmath,amsfonts,amssymb,amstext,amsmath,amsthm,amscd}
\usepackage{mathtools}
\usepackage[a4paper,top=1.5cm,bottom=2cm,left=2cm,right=2cm]{geometry}
\usepackage{breakcites}

\usepackage{booktabs}
\usepackage{mathtools}
\usepackage{mathrsfs}
\usepackage{dsfont}

\usepackage{graphicx}
\usepackage{subfigure}
\usepackage{wrapfig}
\usepackage{indentfirst}
\usepackage{latexsym}
\usepackage{sidecap}
\usepackage{booktabs}
\usepackage{hyperref}

\usepackage{lipsum}
\usepackage{cleveref}
\usepackage[dvipsnames]{xcolor}

\newcommand\myshade{85}
\colorlet{mylinkcolor}{violet}
\colorlet{mycitecolor}{YellowOrange}
\colorlet{myurlcolor}{RoyalBlue}

\hypersetup{
    colorlinks=true,
    linkcolor=myurlcolor,
    citecolor=mycitecolor!\myshade!black,
    filecolor=magenta,      
    urlcolor=magenta,
}
\usepackage{setspace}
\usepackage{breakcites}
\usepackage{mathrsfs}
\usepackage{textcomp}
\usepackage{braket}
\usepackage{colortbl}
\usepackage{bbm}
\usepackage{comment}
\usepackage[colorinlistoftodos]{todonotes}

\usepackage{enumitem}
\usepackage{graphicx}
\usepackage{adjustbox}
\usepackage{framed}
\usepackage{color}
\usepackage{sidecap}
\usepackage{mathtools}
\usepackage{amsmath}
\usepackage{hyperref}
\usepackage{xcolor}
\usepackage[english]{babel}

\newcommand{\etal}{\textit{et al.}}
\newcommand{\ie}{\textit{i.e.}}

\begin{document}

\title{Linking microdosimetric measurements to biological effectiveness: a review of theoretical aspects of MKM and other models}
\author{V.E. Bellinzona\,$^{1,3}$, F. Cordoni\,$^{2,3} $, M. Missiaggia\,$^{1,3}$,\\ F. Tommasino$^{1,3} $,  E. Scifoni\,$^3 $, C. La Tessa\,$^{1,3}$ and  A. Attili\,$^4$}
\date{}
\maketitle

\renewcommand{\thefootnote}{\fnsymbol{footnote}}
\footnotetext{{\scriptsize $^1 $University of Trento, Physics, Trento, Italy.}}
\footnotetext{{\scriptsize $^2 $University of Verona, computer science, Verona, Italy.}}
\footnotetext{{\scriptsize $^3 $TIFPA-INFN, physics, Trento, Italy.}}
\footnotetext{{\scriptsize $^4 $INFN, physics, Roma, Italy.}}

\begin{abstract}
Different qualities of radiation are known to cause different biological effects at the same absorbed dose. Enhancements of the biological effectiveness are a direct consequence of the energy deposition clustering at the scales of DNA molecule and cell nucleus whilst absorbed dose is a macroscopic averaged quantity which does not take into account heterogeneities at the nanometer and micrometer scales.
Microdosimetry aims to measure radiation quality at cellular or sub-cellular levels trying to increase the understanding of radiation damage mechanisms and effects.
Existing microdosimeters rely on the well-established gas-based detectors or the more recent solid-state devices. They provide specific energy $\varepsilon$ spectra and other derived quantities as lineal energy (\textit{y}) spectra assessed at the micrometer level. The interpretation of the experimental data in the framework of different models has raised interest and various investigations have been performed to link in-vitro and in-vivo radiobiological outcomes with the observed microdosimetric data. 
A review of the major models based on experimental microdosimetry, with an emphasis on the Microdosimetric Kinetic Model (MKM) will be presented in this work, enlightening  the advantages of each one in terms of accuracy, initial assumptions and agreement with experimental data. The MKM has been used to predict different kinds of radiobiological quantities such as the Relative Biological Effects for cell inactivation or the Oxygen Enhancement Ratio (OER). Recent developments of the MKM will be also presented, including new non-Poissonian correction approaches for high linear energy transfer (LET) radiation, the inclusion of partial repair effects for fractionation studies and the extension of the model to account for non-targeted effects.
We will also explore developments for improving the models by including track structure and the spatial damage correlation information  by using the full fluence spectrum and, briefly, nanodosimetric quantities to better account for the energy-deposition fluctuations at the intra- and inter-cellular level.
\end{abstract}

\textbf{Keywords or phrases: } Microdosimetry, MKM, Microdosimetric Kinetic Model, RBE, OER, biophysical modeling, particle beam radiation.

\maketitle

\section{Introduction}
Particle therapy is becoming a well established clinical option for tumour
treatment, particularly advantageous for the highly localized dose deposition and for the radiobiological properties \cite{durante2017,mcnair1981icru}.
\\
 { While the first feature is obvious, for the macroscopic energy deposition profile, characterized by the Bragg peak in depth, and also often by a sharper lateral penumbra, due to the small multiple Coulomb scattering of fast and heavy particles,   the second one is related to microscopic features of the peculiar ionization pattern induced by particle radiation, for different charge and energy, down to the molecular scale of the biological target (DNA)}

\begin{figure}
    \centering
    \includegraphics[width= 0.85\textwidth]{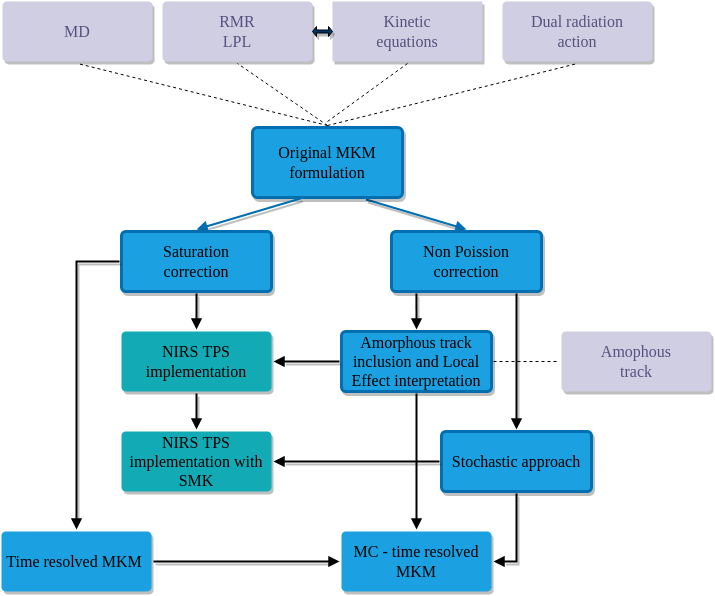}
    \caption{Conceptual map of the evolution of some of the microdosimetric kinetic models (blue) considered in this review. Some of these models are currently used for RBE and RBE-weighted dose evaluations in TPS applications (dark cyan). Solid lines refer to the consequentially of the corrections and extensions while the dotted lines mark the theoretical bases of the considered formulation \textit{(light gray color)}.}
    \label{fig:map}
\end{figure}

 {The accurate prediction of relative biological effectiveness (RBE) in different positions of an irradiating field is a fundamental requirement, in order to correctly estimate treatment responses \cite{durante2010}. Moreover, particle RBE  depends on several factors, of different nature,  biological, patient and treatment specific, because of  the complexity of the mechanisms
of action underlying tumour and normal tissue responses in radiation therapy.
A numbers of models have been presented, historically,  to predict RBE, attempting to account for such effects.}
 { Among these models, four main categories can be identified}:
\begin{enumerate}
\item purely phenomenological  {models}: NIRS\footnote{National Institute of Radiological Sciences (NIRS,  Chiba,  Japan)} mixed beam approach \cite{Kanai1997, kanai1999, gueulette2007,tsujii2004};
\item linear LETd-based models (developed  {mainly} for protons) \cite{Carabe2012a,Wedenberg2014,Jones2015,McNamara2015};
\item Local Effect Model (LEM)-based models \cite{LEMscholz1997,LEMscholz1992,LEMscholz1996,LEMelsasser2007,LEMelsasser2008,LEMelsasser2010,friedrich2012};
\item models based on microdosimetry concepts:
\begin{enumerate}
    \item models
 {based on the} Microdosimetric Kinetic Model (MKM), proposed initially by Hawkings in 1994 \cite{hawkins1994} and  {then} explored and extended till nowadays \cite{hawkins2003, kase2006, kase2007,sato2012, manganaro2017};
\item other models, such as the repair-misrepair-fixation RMF model \cite{Carlson2008,Frese2012,Stewart2018}, and phenomenological models based on RBE-weighting functions \cite{pihet1990other,menzel1990,wambersie1994other,wambersie1990other}.
\end{enumerate}
\end{enumerate}
 {All the different models present different advantages and limitations. While RBE is not measurable with physical methods, the 4th category allows a strong link with physics measurements through different types of microdosimeters.} The present paper is focused on reviewing the biological effect modellization based on the microdosimetry concepts with particular  {emphasis} to the MKM,  {a widely used model to predict the cell survival and the RBE by using microdosimetric data}.
 This topical review is organized as follows:
 the fundamental microdosimetric quantities \cite{zaider1980,icru1983report}, required for  {addressing} the problem, are defined in   { Sec.\ref{SEC:MM} together with} a focus on relevant experimental quantities.
Then, the original formulation of MKM is presented in Sec.\ \ref{SEC:MKM} with its theoretical bases (Sec.\ \ref{SSEC:basis}) and followed by the main extensions such  as non-Poission and saturation corrections \cite{hawkins2003,kase2006} (Sec.\ \ref{SSEC:satcor}), the incorporation of a track model \cite{kase2007}, a variable $\beta$ parameter deriving from the effects of the lesion yield fluctuations in the cell nucleus and domains, \cite{sato2012,manganaro2017} in Sec.\  \ref{sec:stochastic}, and the generalization of the model in case of a time structured irradiation introduced in Sec.\ \ref{SSEC:timestructure}.  The available experimental in-vitro and in-vivo validation are also reported for each extensions.
Fig.\ \ref{fig:map} represents a conceptual scheme of the main MKM formulations and extensions presented in this paper.
Further, an example of treatment planning systems (TPS) implementation of the MKM \cite{inaniwa2010} will be given in Sec. \ref{SSEC:amtrack}. Other applications of the MKM, such as the oxygen enhancement ratio (OER) modeling \cite{Bopp2016,strigari2018} and the incorporation of non-targeted effects  \cite{matsuya2018} will be described in Sec. \ref{sec:extensions}.
Finally, other models based on microdosimetry as well, \ie\  {the distribution function by} Loncol \etal\ \cite{loncol1994} and the RMF model \cite{Carlson2008,Frese2012}, will be presented in Sec.\ \ref{SEC:NonMKM}.


\section{Microdosimetric distributions and their moments}\label{SEC:MM}

The main microdosimetric quantities of interest are the \textit{specific energy} $z$ and \textit{lineal energy} $y$ \cite{rossi1991,zaider1996,icru1983report,lindborg2017}. The \textit{specific energy} $z$ is the ratio between energy imparted by ionizing radiation $\varepsilon$ and the mass $m$ of the matter that has received the radiation, that is
\begin{equation}
    z = \frac{\varepsilon}{m}\,.
\label{def:z}
\end{equation}
The energy imparted by ionizing radiation $\varepsilon$ can be represented by a sum of a finite number of energy depositions events. The \textit{lineal energy} $y$ is the ratio between energy imparted to the matter in a volume of interest by a single energy-deposition event, $\varepsilon_1$, and the mean chord length in that volume, $\bar{l}$, that is
\begin{equation}
    y = \frac{\varepsilon_1}{\bar{l}}\,.
\label{def:y}
\end{equation}

The stochastic nature of $\varepsilon_i$, implies that also $z$ and $y$ are stochastic quantities. In the following, given a probability distribution $f(z)$, we will assume that the probability that a specific energy $z$ is produced in the interval $[z_a,z_b]$ is given by
\begin{equation}
\int_{z_a}^{z_b} f(z)dz\,.
\end{equation}
When dealing with \textit{specific energy spectra}, it is important to distinguish between the \textit{single-event distribution} and the \textit{multi-event distribution}.  It is worth stressing that, although experimental microdosimetry determines single event quantities such as the $\varepsilon_1$ or the \textit{lineal energy} $y$, the starting point for models are multi-event quantities such as the \textit{specific energy} $z$ and its distribution.
\\
The \textit{single--event distribution}, denoted by $f_1(z)$, is the probability distribution of $z$ conditioned to the fact that precisely a single-event happened. The single--event distribution is the building block to define the more-general $n-$event distribution $f_n(z)$ and the \textit{multi--event distribution} $f(z)$.
\\
The $n-$event distribution $f_n(z)$, that is the probability distribution conditioned to the fact that precisely $n$ events occurred, can be computed as the $n-$fold convolution of the single-event distribution $f_1(d)$, as follows
\begin{equation}\label{eqn:conv}
\begin{split}
f_{2}(z) &:= \int_0^\infty f_{1}(\bar{z})f_{1}(z-\bar{z})d\bar{z}\,,\\
&\dots\,,\\
f_{n}(z) &:= \int_0^\infty f_{1}(\bar{z})f_{n-1}(z-\bar{z})d\bar{z}\,,\\
\end{split}
\end{equation}
see \cite{zaider1996} for details.
\\
Using the $n-$event distributions defined above, we can define the general \textit{multi--event distribution} as
\begin{equation}\label{EQN:MutiEv}
f(z;\lambda_n) := \sum_{n=0}^\infty p(n;\lambda_n) f_n(z)\,,
\end{equation}
with $p(n;\lambda_n)$ an integer valued probability distribution with average $\lambda_n$, meaning that
\[
\lambda_n := \sum_{n=0}^\infty np(n;\lambda_n)\,.
\] 
\\
The \textit{multi--event distribution} $f(z;\lambda_n)$ plays a crucial role in the development on microdosimetric-based radiobiological models. It is worth noticing that $f(z;\lambda_n)$ depends on the number of events $n$ only through $p(n;\lambda_n)$, which is independent of specific energy $z$. Also, given $p(n;\lambda_n)$, the single--event distribution $f_1$ completely determines the multi-event distribution $f(z;\lambda_n)$.
\\
Typically, since events are statistically independent, $p(n;\lambda_n)$ is assumed to be a Poisson distribution with mean value $\lambda_n$, so that equation \eqref{EQN:MutiEv} becomes
\begin{equation}\label{EQN:MutiEvP}
f(z;\lambda_n) := \sum_{n = 0}^\infty e^{-\lambda_n} \frac{\lambda_n^n}{n!} f_n(z)\,.
\end{equation}
\\
Denoting by $\langle z\rangle$ the first moment of the distribution $f(z;\lambda_n)$, formally
\begin{equation}\label{EQN:Mom1}
\langle z\rangle := \int_0^\infty z f(z;\lambda_n)dz\,,
\end{equation}
it follows that the following relation holds true,
\begin{equation}\label{EQN:MeanMultiEvent}
\langle z\rangle = \lambda_n z_F\,,
\end{equation}
being $z_F$ the \textit{frequency average} of the single-event specific energy defined as
\begin{equation}\label{EQN:Mom1zF}
z_F := \int_0^\infty z f_1(z) dz\,,
\end{equation}
see, \cite[Chapter II]{zaider1996}. In microdosimetry, $\bar{z}$ is often identified with the absorbed dose $D$; we shall use this identification in the following of the paper.
\\
Above argument, with particular reference to equation \eqref{EQN:MeanMultiEvent}, yields the form for the average value $\lambda_n$ of the \textit{multi--event distribution} to be
\begin{equation}\label{EQN:MeamLn}
\lambda_n = \frac{D}{z_F}\,,
\end{equation}
see, \cite{zaider1996,icru1983report}. Again, in the following, if not differently specified, we will consider $\lambda_n$ to be defined as in equation \eqref{EQN:MeamLn}.
\\
Further computations, see, \cite{zaider1996,icru1983report}, shows that regarding the second moment it holds
\begin{equation}\label{EQN:Mom2}
\int_0^\infty z^2 f(z;\lambda_n)dz = D^2 + z_D D\,,
\end{equation}
with $z_D$ the \textit{dose average} of the single-event specific energy
\begin{equation}\label{EQN:Mom2zD}
z_D := \frac{1}{z_F} \int_0^\infty z^2 f_1(z) dz = \frac{\int_0^\infty z^2 f_1(z) dz}{\int_0^\infty z f_1(z) dz} \,.
\end{equation}
Notation and computations performed in the current section will be extensively used thorough the work to formally  {derive} analytical solution for some relevant biological endpoints, typically the cell--survival probability, starting from a mathematical model for DNA damage. 
\\
In the following,  {we} assume that a cell nucleus is divided into $N_d$ domains, so that above microdosimetric distributions will be used both on single domain and on the whole cell nucleus. In particular, the superscript $(c,d)$ will denote that the corresponding quantity, such as as a microdosimetric distribution or a corresponding average value, is considered on the domain $d$ of the cell $c$. Further, the subscript $n$ denotes that microdosimetric distributions are on the cell-nucleus whereas if no subindex is specified it is assumed that the corresponding distribution is on the domain.
\\
In the following treatment, in order to make computations less heavy as possible, whenever we will say that we average a function $g(z)$ over all domains of a cell nucleus, denoted for short by $\langle g \rangle_d^{(c)}$, it formally means
\begin{equation}
\label{eqn:avg_d}
\langle g \rangle_d^{(c)} := \frac{1}{N_d}\sum_{d=1}^{N_d}\int_0^\infty g(z) f^{(c,d)}(z;z_n)\,dz\,.
\end{equation}
where $f^{(c,d)}(z;z_n)$ denotes the probability density of z in a domain for cell with nucleus specific energy $z_n$.
\\
Similarly, by averaging over all cell population a function $g_n(z)$ defined over a nucleus, denoted by $\langle g_n \rangle_c$, we mean
\begin{equation}
\label{eqn:avg_c}
\langle g_n \rangle_c := \frac{1}{N_c}\sum_{c=1}^{N_c}\int_0^\infty g_n(z) f^{(c)}_n(z;D))\,dz\,.
\end{equation} 
where $N_c$ is the total number of the considered cells and $f^{(c)}(z;D)$ denotes the probability density of $z$ in a nucleus for a population of cells irradiated with macroscopic dose $D$. Notice that in practical computations of an irradiated population of cells, such as those described in the immediate next sections, the probability densities are reasonably considered equals among different cells and domains. In this case we will drop the indexes $c$ and $d$ and the sums in equations \ref{eqn:avg_d}  and \ref{eqn:avg_c} can be carried out implicitly:
\begin{align}
    \label{eqn:avg_simple}
    \langle g \rangle_d &= \int_0^\infty g(z) f(z;z_n)\,dz\\
    \langle g_n \rangle_c &= \int_0^\infty g_n(z) f_n(z;D))\,dz\,.
\end{align}

\subsection{Experimental quantities}
\label{SSec:Experimenthal_quantities}

In order to account for the different densities and sizes of the sites of radiobiological interests (e.g. the cell nucleus and the domain), the specific energy $z$ used in the models as described in the following sections can be obtained experimentally through the \textit{lineal energy} $y$ defined in equation \eqref{def:y}.
\\
The lineal energy can be measured through a microdosimeter detector, where the most frequently used are the tissue-equivalent proportional counters (TEPC) \cite{conte2013,lindborg1999,wilson1970,nikjoo2002}; analogous information can be achieved also by solid-state detectors and gas electron multiplier (GEM) detectors \cite{byun2009GEM, orchard2011GEM}, recently investigated for their use in microdosimetric measurements \cite{braby2015EXP, schuhmacher2002EXP}. 
The relationship between $\bar{l}$ of the tissue-equivalent volume of the microdosimeter, from which the lineal energy is calculated, is given approximately by
\begin{equation}
\bar{l} = \bar{l}_{\mathrm{det}}\frac{\rho_{\mathrm{det}}}{\rho}
\end{equation}
where $\bar{l}_{\mathrm{det}}$ is physical mean cord of the detector and $\rho$, $\rho_{\mathrm{det}}$ are the densities of the tissue and the detector material respectively. For more general conversion methods of microdosimetric spectra between different materials and shapes see for example \cite{Magrin2018}.
\\
The theoretical single-particle imparted energy, $z_1$, can be estimated from the lineal energy $y$ as:
\begin{equation}
\label{eqn:link_zy}
    z_1 = y\bar{l}_t/m_t
\end{equation}
where $\bar{l_t}, \, m_t $ are the mean chord length and the mass of the biological site of interest respectively. The mean dose-averaged specific energy $z_D$ can be obtained from the mean-dose lineal energy $y_D$ as:
\begin{equation}\label{eqn:zD}
    z_D = \frac{\bar{l}_t}{m_t y_F} \int_0^\infty y^2f(y)\ \text{d}y = \frac{\bar{l}_t}{m_t} y_D
\end{equation}
where $y_F$ is the frequency-mean lineal energy.
\\
In the case of a spherical volume $ \rho_{\mathrm{t}} = 1\ $g/cm$^3$, the specific energy $z_1$ is linked to the lineal energy $y$ \cite{kase2012dosimetry} as
\begin{equation}
\label{eqn:sphere_zy}
z_1\ (\mathrm{Gy}) = 0.204 \times \frac{y\ (\mathrm{keV} / \mu \mathrm{m})}{[2r_\mathrm{det}\ (\mu \mathrm{m})]^{2}}
\end{equation}
the constant factor is due to the Gy-keV conversion (1 Gy= 1.6$ \times10^{-16}$keV) and the consideration that the mean chord length in the case of a sphere is $\bar{l}=4/3r_{det} $

\section{Microdosimetric Kinetic Model}\label{SEC:MKM}

The Microdosimetric Kinetic (KM) model has been developed by  {Roland} B. Hawkins \cite{hawkins1994} by
taking inspiration from the theory of dual radiation action (TDRA) \cite{Kellerer1972,kellerer1978}, the repair-misrepair model \cite{tobias1980,tobias1985} and the lethal-potentially lethal (LPL) model \cite{curtis1986,curtis1988}. In the following sections, after a brief description of the historical bases of the model and the details of its original formulation, we compare and contrast the more recent developments of the model.

\subsection{Historical bases}\label{SSEC:basis}

This section presents a brief explanation of the theoretical formulations on which the considered models are based on.

The theory of dual radiation action (TDRA) \cite{Kellerer1972,kellerer1978} assumes that, after the cell irradiation, the number of lethal lesions in a small volume of the cell nucleus, defined \textit{site}, is proportional to the square of the
specific energy $z$ deposited in that site:
\begin{equation}
\varepsilon(z) = K z^{2}
\end{equation}
where the $K$ factor expresses the rate of combination of sub-lesions and formation of lesions. In the development of the MKM \cite{hawkins1994}  this postulate is generalized, adding the quadratic proportionality to the lethal damages and therefore assuming a linear-quadratic dependence on \textit{z}.
\\
MKM inherits the concept of damage time evolution for the repair or conversion into a lethal unreparable lesion (chromosome aberration) \cite{van2018,schurmann2018} of the primary potentially lethal radiation induced lesions in DNA from the \textit{repair-misrepair (RMR) model},  developed by Tobias \etal\ to interpret radiobiological experiments with heavy ions \cite{tobias1980,tobias1985}.  The RMR model considers that the amount of DSBs in the DNA, $U(t)$, is linearly proportional to the radiation dose rate; a number of DSBs  evolve in lethal lesions, $L(t)$, while most breaks are successfully repaired with a first-order process. The model includes also the possibility of a  misrepair as a second-order process since it involves two broken DNA strands to form a chromosomal aberration. The idea of misrepair was initially applied by Lea and Catcheside \cite{Lea1942} to describe the formation of chromosome aberrations in tradescantia.
\\
These assumptions yield the following kinetic equations:
\begin{equation} \label{eqn:rmr}
\begin{aligned}
\frac{d U}{d t} &=\underbrace{\delta \dot{D}}_{\text {damage }}-\underbrace{\lambda U}_{\text {repair }}-\underbrace{\kappa U^{2}}_{\text {misrepair }} \\
\frac{d L}{d t}&=\underbrace{(1-\phi) \lambda U}_{\text {unsuccessful repair }}+\underbrace{\sigma \kappa U^{2}}_{\text {lethal misrepair }}
\end{aligned}
\end{equation}
where  $\delta$ is the number of DSBs induced per Gy of radiation, $\lambda$ is the rate at which DSBs are repaired  $\kappa$ is the rate constant for second-order DSB interaction and $\phi$ is the fraction of simple repairs that are successful. The fraction of misrepairs that
result in a lethal lesion is $\sigma$.

Like the RMR model, the \textit{lethal-potentially lethal (LPL)} model \cite{curtis1986,curtis1988} accounts that the damage caused by ionizing radiation at the molecular level to cell death can be separated into two broad classes: that which has the potential of being lethal, $P(t)$ (by fixing or binary misrepair) but also can be repaired correctly and that which is lethal \textit{ab initio} and cannot be repaired correctly, $L(t)$. Both lesions are linearly proportional to the radiation dose-rate \cite{kuang2016} and after a prescribed time, the remaining potentially lethal lesions become lethal as described in the following equations:
\begin{equation}
    \begin{aligned}
\frac{d P}{d t} &=\underbrace{\delta \eta \dot{D}}_{\text {reparable damage }}-\underbrace{\lambda P}_{\text {repair }}-\underbrace{\kappa P^{2}}_{\text {misrepair }} \\
\frac{d L}{d t} &=\underbrace{\delta(1-\eta) \dot{D}}_{\text {irreparable damage }}+\underbrace{\kappa P^{2}}_{\text {lethal misrepair }}
\end{aligned}
\end{equation}
where $ \eta$ is the amount of  radiation induced DSBs that are repairable, while all the other parameters corresponds in meaning to the ones in Eq.\ \ref{eqn:rmr}.
\\
The solution of the model equations are similar in form to those for the RMR model and the LPL. However, in contrast to the RMR, the LPL predicts that the probability of the interaction between potentially lethal lesions is strongly dependent to the dose rate and becomes negligible for low dose rates, where only the channel of the direct creation of lethal events through $\lambda$ dominate.

\subsection{Original formulation and general considerations}\label{mkm}
The MK model computes the cell survival in a way that emphasizes subcellular microdosimetry while abstracting the specific description and modeling of the radiation-induced damage to the cell by using the general categories of lethal and potentially lethal lesions as defined in \cite{curtis1986}. More specifically, the MK model is based on the following funding assumptions \cite{hawkins1994,hawkins1996,hawkins1998}:
\begin{enumerate}
\item the cell nucleus is the sensitive target and it is divided into $N_d$ sub-units, called domains, similar to the sites of TDRA. In general domains have a variety of shapes that fit together to fill the nucleus. In the case of mammalian cells, the domain diameter is usually considered to be in the range $ 0.5 \leq d_d \leq 1.0  \, \mu \mathrm{m} $ and the number of domains per nucleus is in the order of few hundred;	
\item radiation can create two different types of DNA damages, called of type I and II;
\item type I lesions represents damage that cannot be repaired, for this reason will be also called lethal lesion. On the contrary type II lesions, also called sublethal or potentially-lethal lesions, can repair or convert into a lethal lesion either by spontaneous conversion or by binary combination with another sublethal lesion;
\item type I and II lesions are confined to the domain in which they are created. This assumption defines a sub-nuclear correlation length among lesions in a way that the interaction of two lesions can happen only if they are in close spatial proximity. Specifically, a pair of type II lesions can combine to form a type I lesion only if they are created in the same domain;
\smallskip \\ 
A remark on this assumption is needed. The idea behind the division of a cell into subvolumes arises because couples of type II lesions are all likely to happen in a short time period, even for lesions that are far away in the cell-nucleus. In order to overcome such a problem, a possible approach is to divide the nucleus into smaller subdomains so that interactions might happen solely inside a single volume, as it is assumed in the MKM. It is important to stress the key role that the choice of such domains plays. In fact, if too big domains imply that far away lesions can interact, on the contrary, too small domains yield that the overall number of lesions inside a single domain is so small that couple interactions is less likely to happen. Therefore, the choice of the best possible division of the cell nucleus into smaller domain is a key aspect of the model and different choices of domains can in principle lead to different results. A possible solution to reduce the sensibility of the model from the arbitrary choice of the domains, it is to assume that interactions are possible also within different domains, allowing therefore lesions to move from one domain to another or pairs of lesions to interact if in adjacent domains.
\item the initial number of type I and II lesions in a single domain $d$ is proportional to the specific energy $z$ deposited in the domain.
\end{enumerate}

If above assumptions hold then the following further assumptions are made regarding the reproductive survival of the cell:
\begin{enumerate}[resume]
\item if at least one domain suffers a lethal lesion it is considered ``dead'';
\item if at least one domain is dead, then the whole cell is ``dead''.
\end{enumerate}

\begin{figure}[h!]
\includegraphics[width=\textwidth]{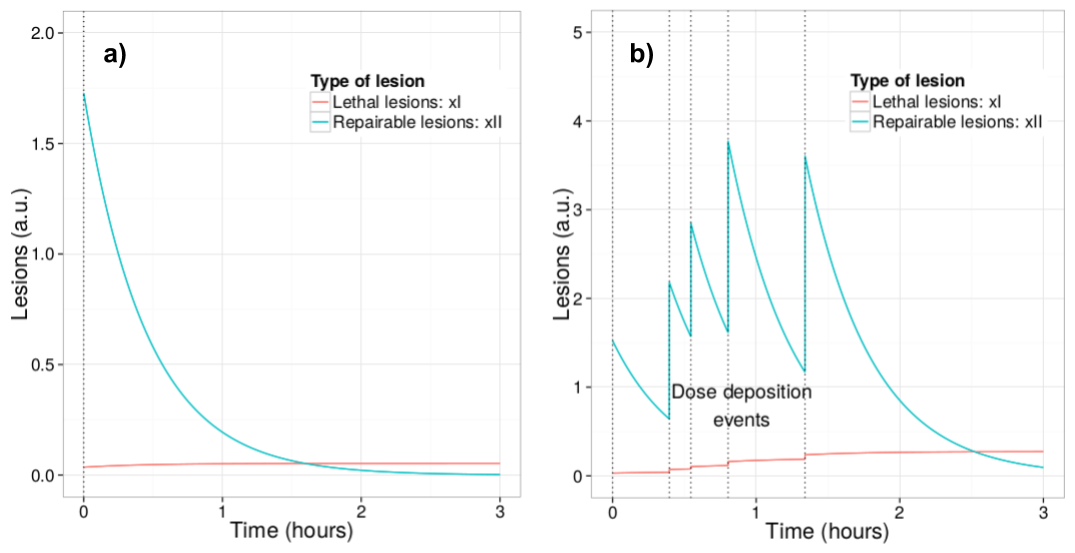}
\caption{Time evolution of $x_I $ and $x_{II}$ damages for  a single instantaneous irradiation as described by Eq.\ \ref{EQN:SolXI} and Eq.\ \ref{EQN:SolXII} respectively (a). Generalization of the temporal evolution  for any time structured irradiation as describe in Sec.\ \ref{SSEC:timestructure} and \ref{Sec:Manganaro}. The dotted vertical lines represent the energy deposition events in the cell nucleus due to the passage of ionizing particles  (b). Figure from \cite{manganaroPHD}}
\label{fig:lesions}
\end{figure}

It has to be noted that, while the MK assumptions reported in this section are general, in many studies \cite{Matsuya2014,chen2017} the lethal lesions are intended to represent a specific complex DNA damage (e.g. lethal chromosome aberrations) that cannot be repaired, whereas the creation of sublethal lesions are explicitly associated to the induction of double-strand breaks (DSB) that can be repaired.
\\
Following the MK notation, we denote by $x_{I}^{(c,d,z)}(t)$ and $x_{I}^{(c,d,z)}(t)$ the time-dependent average number of type I and type II lesions for a cell-domain $(c,d)$ caused by an acute dose $z^{(c,d)}$ at $t=0$ deposited in the cell $c$ and domain $d$. Starting from the concept, introduced in the TDRA, that a cell experiments a randomly varying dose in a microscopic volume \cite{icru1983report, rossi1991}, the microscopic specific energy $z^{(c,d)}$ is considered as a random variable with $\langle\langle z^{(c,d)} \rangle_d\rangle_c = D$, the macroscopic dose experienced by the cell population.
\\
Type II lesions are assumed that can be repaired with a constant repairing rate $r$ or can be converted to irreparable lesions through a first order process with constant rate $a$, or at the second order, representing pairwise combinations, with constant rate $b$. The average number of type I and II lesions at time 0 is proportional to the amount of specific energy $z^{(c,d)}$ with factors $\lambda$ and $\kappa$. These assumptions formally define the following set of coupled ODE similar in concept to equations \eqref{eqn:rmr}
\begin{eqnarray}\label{MKM_kinEQ}
\left\{\begin{array}{l}
\dot{x}_{I}^{(c,d,z)}=a x_{I I}^{(c,d,z)}+b\left(x_{I I}^{(c,d,z)}\right)^{2} \,,\\
\dot{x}_{I I}^{(c,d,z)} = -(a+r) x_{I I}^{(c,d,z)}-2 b\left(x_{I I}^{(c,d,z)}\right)^{2}\,,
\end{array}\right.
\end{eqnarray}
subject to the initial average number of lesions 
\begin{equation}\label{eqn:MKM_kinEQ_initial}
x_{I}^{(c,d,z)} (0) = \lambda z^{(c,d)}\,,\quad x_{II}^{(c,d,z)} (0) = \kappa z^{(c,d)}\,.
\end{equation}
In the case of ion radiation, typically the rate of pairwise combination between type II lesions is negligible with respect to the first order evaluation of $x_{II}$ for low does \cite{hawkins1996}, that is,
\begin{equation}
 2 b\left(x_{I I}^{(c,d,z)}\right)^{2} \ll (a+r) x_{I I}^{(c,d,z)}  
\end{equation}
so that the time-evolution of the average number of type II lesion can be rewritten as
\begin{equation}
\dot{x}_{I I}^{(c,d,z)} =-(a+r) x_{I I}^{(c,d,z)}\,.
\label{typeII}
\end{equation}
The solution to equation \eqref{typeII} can be seen to be
\begin{equation}\label{EQN:SolXII}
x_{I I}^{(c,d,z)} (t) = \kappa z^{(c,d)} e^{-(a+r) t}\,.
\end{equation}
\\
Substituting equation \eqref{EQN:SolXII} into the kinetic equations \eqref{MKM_kinEQ} and integrating $x_{I I}^{(c,d,z)}$ with respect to time, it follows that
\begin{equation}\label{EQN:SolXI}
x_{I}^{(c,d,z)} (t) = \lambda z^{(c,d)} + a\kappa z^{(c,d)}\left (\frac{1 - e^{-(a+r)t}}{a+r}\right ) + b \kappa^2 \left ( z^{(c,d)}\right )^2 \left (\frac{1 - e^{-2(a+r)t}}{a+r}\right ) \,.
\end{equation}
An example of the temporal evolution of lesions in a cell is depicted in Fig.\ \ref{fig:lesions}. 

It is important to remark that the exponential decay in Eq.\ \ref{EQN:SolXI} derives from the assumption of first order repair kinetics and that it could likely represents an approximation of more complex repair kinetics present in the real cell \cite{Dikomey1986,Fowler1999,Dale1999b,Carabe-Fernandez2011}. Postulating that the total number of lesions $x_{I}(t) + x_{II}(t) \sim N_\mathrm{DSB}(t)$ describes specifically the number of double strand breaks (DSBs) in the DNA, the repair kinetics represented in Eq.\ \ref{EQN:SolXI} can be verified through H2AX phosphorylation mapping experiments ($\gamma$-H2AX) \cite{schettino2011,Mariotti2013}. In the case of high-LET particle irradiation, such as carbon ions, the presence of a plateau (offset) in the observed $N_\mathrm{DSB}(t)$  \cite{Asaithamby2008,Carabe-Fernandez2011,Asaithamby2011}, suggests the presence of irreparable complex clustered damage that can be related directly to the parameter $\lambda$ of the kinetic equations, and hence to the linear parameter $\alpha_0$ of the macroscopic cell survival LQ formulation that will be introduced in the following (see Eqs.\ \ref{EQN:SurvDomain} and \ref{EQN:Survn1}).
\\
In order to connect above explicit solution of Eq.\ \ref{EQN:SolXI}, \textit{i.e.} the average number of type I and II lesions given a certain energy deposition $z^{(c,d)}$, to the survival probability one more fundamental assumption must be made.

\begin{enumerate}[resume]
\item the lethal lesion distribution given a specific energy $z$ follows a Poisson distribution.
\end{enumerate}
Under the Poisson distribution assumptions stated above, the probability that the domain $d$ survives at time $t \to \infty$ when exposed to the specific energy $z^{(c,d)}$, denoted by $s^{(c,d)}(z^{(c,d)})$, can be computed as the probability that the random outcome of a Poisson random variable is null. Therefore $s^{(c,d)}$ is given by
\begin{equation}\label{EQN:SurvMKMD}
s^{(c,d)}(z^{(c,d)}) = e^{-\lim_{t \to \infty} x_{I}^{(c,d,z)} (t)}\,.
\end{equation}
Using equation \eqref{EQN:SolXI}, it can be seen that that the average number of lethal lesion given $z^{(c,d)}$ as $t \to \infty$ can be computed as
\begin{equation}
\lim_{t \to \infty} x_{I}^{(c,d,z)} (t) = \left (\lambda + \frac{a \kappa}{a+r}\right )z^{(c,d)} + \frac{b \kappa^2}{2(a+r)} \left (z^{(c,d)} \right )^2\,,
\end{equation}
so that the log-survival for the domain $d$ is given by
\begin{equation}\label{EQN:SurvDomain}
\log s^{(c,d)}(z^{(c,d)}) = -A z^{(c,d)} - B \left (z^{(c,d)} \right )^2\,,
\end{equation}
with $A$ and $B$  defined as
\begin{equation}
    A = \left (\lambda + \frac{a \kappa}{a+r}\right )\,,\quad B = \frac{b \kappa^2}{2(a+r)}\,.
    \label{EQN:mkmAB}
\end{equation}
We remark that these constants are independent of the domain $d$ and specific energy $z^{(c,d)}$ of deposited in the domain $d$.

Indicating with $S^{(c)}_n(z_n^{(c)})$ the probability of the reproductive survival of the cell $c$ that has received exactly a specific energy $z_n^{(c)}$ in the nucleus, the log-survival of this quantity, $- \log S_n^{(c)}(z_n) = x^{(c)}_{I,n}(z_n)$, represents the total lesions in the whole cell nucleus, and can be therefore evaluated by summing of the single--domain log--survival $-\log s^{(c,d)}(z) = x^{(c,d)}_I(z)$ over all the domains of the cell, or, equivalently, by formally using the average of this quantity over the domains. Assuming that the probability density function are the same over all domain and cell, we can drop the index $c$ and $d$ and use equations \ref{eqn:avg_simple} to write
\begin{equation}\label{EQN:SurvMKMC}
\begin{split}
\log S_n (z_n) &:= - x_{I,n}(z_n) \\
&= - N_d  \langle x_{I}(z) \rangle_d =  -N_d  \langle \log  s(z) \rangle_d\\
&= - N_d \left(A\langle z \rangle_d + B\langle z^2 \rangle_d\right) \\
&= - N_d A \int_0^\infty z f (z;z_n)\,dz -N_d B \int_0^\infty z^2 f (z;z_n)\,dz\,,
\end{split}
\end{equation}
where $f (z;z_n)$ denotes the probability density of $z$ in a domain for a cell with a mean specific energy in the nucleus $z_n$.
In particular, as shown in Section \ref{SEC:MM}, the following holds
\begin{equation}
z_n = \langle z \rangle_d =\int_0^\infty z f (z;z_n) dz \,.
\end{equation}
\\
Using equations \eqref{EQN:Mom2} and \eqref{EQN:Mom2zD} derived in Section \ref{SEC:MM}, the log survival in equation \eqref{EQN:SurvMKMC} can be written as
\begin{equation}\label{EQN:Survn1}
\log S_n (z_n) = -(\alpha_0 + z_{D}\beta_0) z_n - \beta_0 \left (z_n \right )^2\,, 
\end{equation}
with $\alpha_0 := N_d A$ and $\beta_0 := N_d B$. Also, $z_{D}$ is the dose average $z$ per event in a domain, obtained from equation \eqref{EQN:Mom2zD} applied to the domain.
\\
Notice that in equation \eqref{EQN:SurvMKMC} we have used the notation $f (z;z_n)$ to denote the \textit{multi--event distribution}, rather than $f (z;\lambda_n)$ as done in Section \ref{SEC:MM}. This is due to the fact that, since the following relation holds true
\begin{equation}
\lambda_n = \frac{z_n}{z_{F}}\,,
\end{equation}
we have preferred to specify the dependence upon the \textit{multi--event distribution} average.
\\
In order to obtain the cell survival $S(D)$ for a population of cells irradiated with macroscopic dose $D$, the quantity $S_n(z_n)$ defined in equation \ref{EQN:SurvMKMC} should be averaged accounting for the distribution of the specific energy $z_n$ over the cell population. In terms of the logarithm of the cell population survival, $\log S$, under again the assumption that all the cells have the same probability distribution of specific energy $z_n$, this can be written as:
\begin{equation}\label{EQN:SurvMKMP}
\begin{split}
\log S(D) &:= \log \langle S_n (z_n) \rangle_c\\
&=\log\left (\int_0^\infty S_n (z_n)  f_n(z_n;D)\,dz_n\right )\,,
\end{split}
\end{equation}
where similar to above we have denoted by $f_n(z_n;D)$ the probability density of $z_n$ for a macroscopic absorbed dose $D$ over the cell population, \textit{i.e.}
\begin{equation}
D = \langle z_n \rangle_c = \int_0^\infty z_n f_n(z_n;D) dz_n\,.
\end{equation}

We remark that equation \eqref{EQN:SurvMKMP} is fundamentally different from equation \eqref{EQN:SurvMKMC} since the it  considers the average of the argument of the logarithm whereas in equation \eqref{EQN:SurvMKMC} the average of the logarithm has been taken. This basically indicates that, due to the stochastic nature of $z_n$, the distribution of lethal lesions $\log S_n(z_n)$ over the cell population is in general non-Poisson and hence that the log of the survival cannot be directly related to the average number of lethal lesions per cell, $\log S(D) \neq - \langle x_{I,n}(z_n) \rangle_c$. However, provided that the variance of $z_n$ is small, a Poission approximation is assumed and the same procedure used to obtain equation \eqref{EQN:SurvMKMC} can be used. In this approximation equation \eqref{EQN:SurvMKMP} can be written as follows,
\begin{equation}\label{EQN:SurvMKMPFin}
\begin{split}
\log S(D) &= \log \langle S_n (z_n) \rangle_c \\
&\approx - \langle x_{I,n}(z_n) \rangle_c =  \langle \log S_n (z_n) \rangle_c\\
&= \int_0^\infty \log\left (S_n (z_n) \right ) f_n(z_n;D)\,dz_n \\
&= -(\alpha_0 + (z_{D}-z_{n,D}) \beta_0)D - \beta_0 D^2\,,
\end{split}
\end{equation}
with $z_{n,D}$ the dose average $z_n$ in the nucleus per event. All the quantities $z_{n,D}$; $z_{D}$ and $z_n \approx \langle z_n \rangle_c = D$ are assumed to be the same for each cell or domain. All other notations are used as previously introduced. Since the size of the domain is usually much smaller than the size of the nucleus, it holds that $z_{n,D} \ll z_{D}$, (see \cite{hawkins1996}), so that we eventually obtain
\begin{equation}\label{EQN:SurvMKMPFin2}
\begin{split}
\log S &= -\alpha_\mathrm{P} D - \beta D^2\,,
\end{split}
\end{equation}
with
\begin{equation}
\label{EQN:SurvMKMPFin2_ab}
\alpha_\mathrm{P} := \alpha_0 + z_{D} \beta_0\,,\quad \beta := \beta_0\,.
\end{equation}
where the subscript P indicates that the relationships hold when the assumption of Poisson distribution of lethal lesions among the irradiated cell population is reasonable, \textit{i.e.} for low-LET irradiation, as it is discussed in the following section.
\\
A further refinement of the MKM kinetic equations involves a fourth type of possible interaction, that happens at time $t_r$. The following is assumed:
\begin{enumerate}[resume]
\item after a time $t_r > 0$, all sub--lethal lesions that are not either dead or repaired, automatically transform into lethal lesions.
\end{enumerate}
The mathematical formulation of the main kinetic equations remain the same as in equations \eqref{MKM_kinEQ}--\eqref{EQN:SolXII}--\eqref{EQN:SolXI} in the time interval $t \in [0,t_r)$. As soon as $t_r$ passes, all type II lesions that have not been either repaired or died, the will immediately be converted into type I lesions, meaning
\begin{equation}
x_{II}(t) = 0\,,\quad t > t_r\,.
\end{equation}
The solution for the average number of type I lesion can be now explicitly found for $t > t_r$ adding all type II lesions that persisted after $t_r$ passes, that is 
\begin{equation}
x_{I}^{(c,d,z)} (t) =  x_{I}^{(c,d,z)} (t) + x_{II}^{(c,d,z)} (t_r)\,,
\end{equation}
so that we obtain
\begin{equation}
\begin{split}
\lim_{t \to \infty} x_{I}^{(c,d,z)} (t) &=  \lim_{t \to \infty} x_{I}^{(c,d,z)} (t) + x_{II}^{(c,d,z)} (t_r) =\\
&= \left (\lambda + \frac{a \kappa}{(a+r)} +\frac{\kappa r}{(a+r)}e^{-(a+r)t_t}\right )z^{(c,d)} +\\
&+\frac{b \kappa^2}{2 (a+r)}\left (1-e^{-2(a+r)t_r}\right ) \left (z^{(c,d)} \right )^2\,.
\end{split}
\end{equation}
Proceeding as above, taking therefore the average over all cell domains and cell population, we obtain the generalization of equation \eqref{EQN:SurvMKMPFin2} to be
\begin{equation}
\log S = -\alpha D - \beta D^2\,,
\end{equation}
with
\begin{equation}
\begin{cases}
\alpha &:= \bar{\alpha}_0 + z_{D}^{(c,d)} \bar{\beta}_0\,,\quad \beta := \bar{\beta}_0\,\\
\bar{\alpha}_0 &:= N_d \left (\lambda + \frac{a \kappa}{(a+r)} +\frac{\kappa r}{(a+r)}e^{-(a+r)t_t}\right )\,,\\
\bar{\beta}_0 &:= N_d \frac{b \kappa^2}{2 (a+r)}\left (1-e^{-2(a+r)t_r}\right )\,.
\end{cases}
\end{equation}

\subsection{Link to the radiobiological observables}
\begin{figure}[h!]
\centering
\includegraphics[width= 0.8\textwidth] {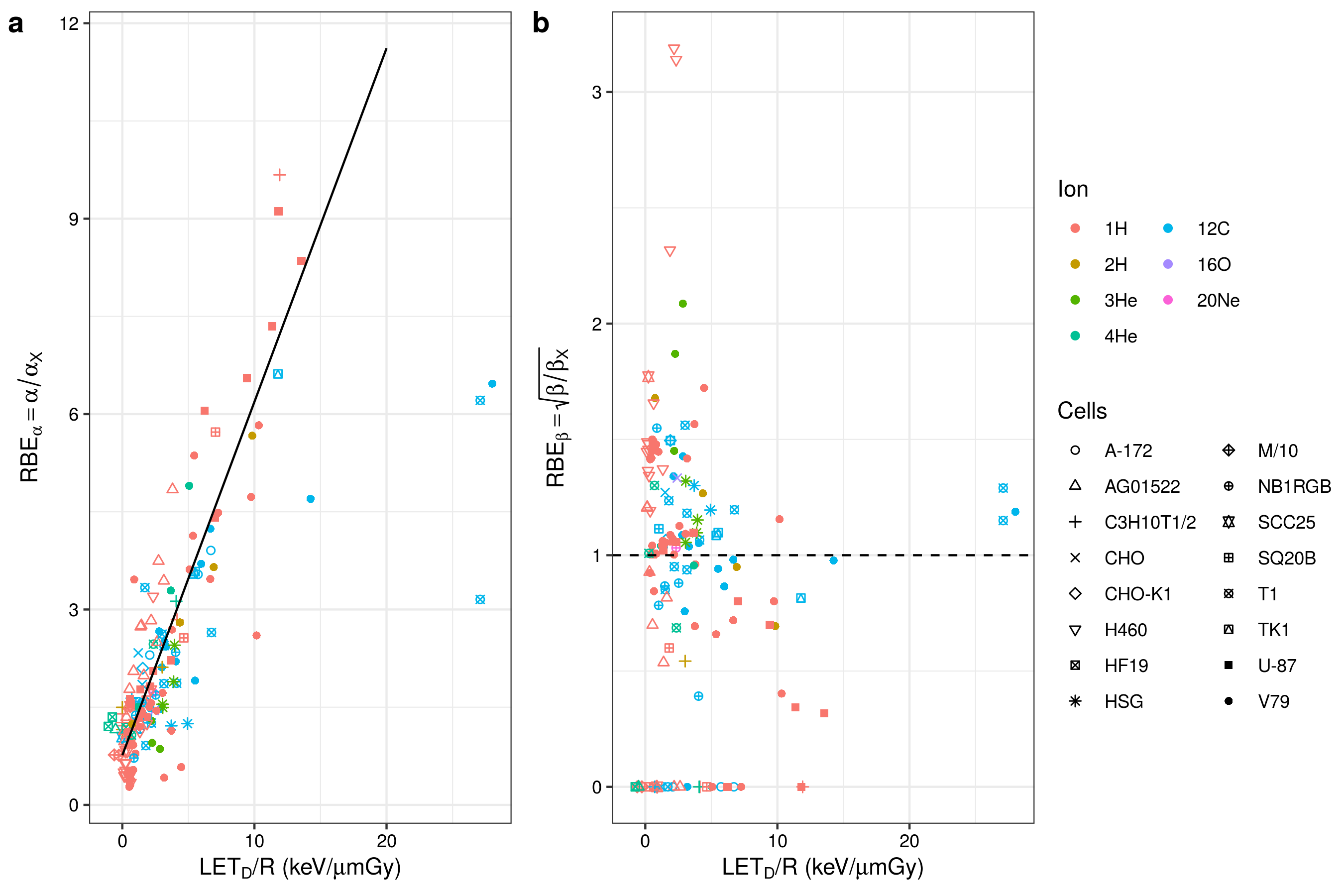}
\caption{a) Experimental \textit{in vitro} $\mathrm{RBE}_\alpha$  (panel a) and $\mathrm{RBE}_{\beta}$ (panel b) versus $\mathrm{LET}_D / R$. The data have been taken from the Particle Irradiation Data Ensemble (PIDE v3.2) database \cite{Friedrich2012b}. The continuous line in (a) is the fit of the linear Poisson solution of the MK model  (Eq.\ \ref{eqn:rbe_alpha_gen}) carried out in the low-LET region ($\mathrm{LET}_D < 20$ keV/$\mu$m) for ion irradiation with charge $Z \le 11$. The dashed line in (b) corresponds to the constant $\mathrm{RBE}_\beta = 1$}
\label{rbea_poisson}
\end{figure}

From equations \eqref{EQN:SurvMKMPFin2} and \eqref{EQN:SurvMKMPFin2_ab} it is possible to obtain the  direct link of the model to the phenomenological LQ formulation of the cell survival.
The $\alpha $ coefficient is therefore explicitly dependent on the radiation quality through a single term, the dose average specific energy per single event $z_{D}$, that can be related to microdosimetric measurements (equation \ref{eqn:zD}). It has to be noted that, in this formulation of the MK model, there is no explicit dependence to the radiation quality in the quadratic coefficient $\beta_0$, that is considered constant, analogously to the result of the TDRA \cite{Kellerer1972}. The latter is an approximation of the model that is in contrast with experimental observations \cite{prise1998review,anderson2010,Friedrich2012b} although in many cases, considering in particular the experimental uncertainties associated to the $\beta_0$ determination (see for example Fig. \ref{rbea_poisson}), it is assumed to be reasonable. In an evolution of the model which accounts for the stochastic aspects of the irradiation, as described in sections \ref{SEC:sato2012} and \ref{Sec:Manganaro}, this approximation will be relaxed and the $\beta$ coefficient will be considered dependent on the quality of the radiation.

From the knowledge of the LQ parameters is possible to derive the dose ($D$) and radiation quality ($z_{D}$) dependent RBE \cite{Dale1999,Carabe-Fernandez2007}:
\begin{equation}
\label{eqn:rbe}
\text{RBE}(D,z_{D}) = \frac{1}{2D}
\left(
-1 + \sqrt{1 + \frac{4}{R} \left( \text{RBE}_\alpha(z_{D}) D + \frac{(\text{RBE}_\beta D)^2}{R} \right)}
\right)
\end{equation}
where $R=\alpha_X/\beta_X$, $\text{RBE}_\alpha(z_{D}) = \alpha(z_{D})/\alpha_X$, $\text{RBE}_\beta = \sqrt{\beta/\beta_X}$, and $\alpha_X$ and $\beta_X$ are the phenomenological LQ coefficient for the photon reference radiation.

Since the parameters $\alpha_0$ and $\beta_0$ are assumed to be independent on the radiation quality, and $\beta = \beta_0$, it is possible to identify $\alpha_0 = \alpha (\text{LET} \to 0)$ and $\beta_0 \simeq \beta_X$  ($\text{RBE}_\beta \simeq 1$). In the case of high-LET evaluations, since the single-particle dose averaged specific energy $z_{D}$ for photons can be considered negligibly small relative to that of high-LET radiation, $\alpha_0 \simeq \alpha_X$ and hence is possible to write
\begin{equation}
\label{eqn:rbe_alpha}
\text{RBE}_\alpha = \frac{\alpha_0}{\alpha_X} + \frac{\beta_0}{\alpha_X} \times z_{D}^{(c,d)} \approx 1 + \frac{1}{R} \times z_{D}
\end{equation}
where ratio $R$ can be derived from a nonlinear regression analysis of measured cell survival data for a low-LET reference radiation. Equation \eqref{eqn:rbe_alpha} can be generalized as
\begin{equation}
\label{eqn:rbe_alpha_gen}
\text{RBE}_\alpha = k_1 + \frac{k_2}{R} \times y_{D}
\end{equation}
where $k_1$ and $k_2$ are phenomenological parameters.
Since $z_{D}$ is proportional to the dose-averaged lineal energy $y_d$ (equation \ref{eqn:link_zy}),  Eq.\ \ref{eqn:rbe_alpha_gen} is analogous to the linear models based on the dose average LET (LET$_D$) used for protons \cite{Carabe2012a,Wedenberg2014,Jones2015,McNamara2015}. Following the MK model premises, Eq.\ \ref{eqn:rbe_alpha_gen} could also be generally used for other ions to describe the linear growth of the RBE as a function of the LET$_D$ in the low-LET region (see Fig.\ \ref{rbea_poisson}). However the linear dependence on the LET fails to be adequate in the region of mid- and high-LET as found in experimental studies \cite{Friedrich2012b}. In these regions further corrections to the MK model are used to reproduce the experimental observations. Different corrective approaches for high-LET irradiation are described in the following sections.

\subsection{Non-Poisson correction}
\label{non-poisson}

In the approximation introduced in equation \eqref{EQN:SurvMKMPFin} it is assumed that the variance of the specific energy $z_n$ among cells is sufficiently small. In this assumption the number of lethal events follows the same Poisson distribution in each cells, with average $x_{n,I}$.      \\
However, in general, the specific energy received in the cell is a stochastic quantity, that varies from cell to cell, bringing also a deviation from the Poisson distribution when considering the whole population of irradiated cells. We remark that this deviation is present even if the radiation is perfectly mono-energetic. In this case the variance of the specific energy $z_n$ arises from the fluctuation of the number of particles that are hitting the cells. The fluctuations are particularly relevant when the LET of the particle is relatively high since, given a macroscopic dose $D$, the average number of high-LET particles interacting with the cell is lower than the number of low-LET particles. 
To account for the non-Poisson distribution of the lethal events a correction to the MK model has been introduced by Hawkins in 2003 \cite{hawkins2003} bringing a deviation from the linear behaviour of the RBE vs.\ LET, described in equation \eqref{eqn:rbe_alpha}, in the high LET region.

The effect of the non-Poisson distribution of lethal lesions is considered by explicitly evaluating the fraction of hit and non-hit cell nuclei. Considering a very low dose high-LET irradiation, $D \ll 1$, the probability for a cell to interact with more  {than} one particle is negligible. In this case the population of cells can be subdivided in a fraction $\Phi$ of cells that suffer a single particle interaction and a fraction $1-\Phi$ of cell with zero interactions.  We denote with $x_{I,n}(z_{n,D})$ the average number of type I lethal lesions in the fraction $\Phi$ of cells whose sensitive nucleus has been hit by a single particle imparting exactly a specific energy $z_{n,D}$ in the nucleus
Then, recalling equations \eqref{EQN:SurvMKMPFin}-- \eqref{EQN:SurvMKMPFin2_ab}, we obtain
\begin{equation}
x_{I,n}(z_{n,D}) =-\log S(z_{n,D}) = (\alpha_0 + z_{D}\beta_0)z_{n,D} + \beta_0 z_{n,D}^2\,.
\end{equation}
It is possible to explicitly write the global surviving fraction of cells (including both hit and non-hit nuclei) as:
\begin{equation}\label{eqn:fraction1}
    S(D) = (1-\Phi) + \Phi e^{-x_{I,n}(z_{n,D})}\,.
\end{equation}
This corresponds to consider a probability density function $f_n(z_n;D) = (1-\Phi)\delta(z_n) + \Phi \delta(z_n-z_{n,D})$ in equation \eqref{EQN:SurvMKMP}. Since the number of lethal lesions per cell averaged over the whole cell population (including both hit and non-hit nuclei) exposed to the macroscopic dose $D$ can be directly evaluated as
\begin{equation}
    \langle x_{I,n}(z_n)\rangle_c = \Phi x_{I,n}(z_{n,D})\,,
\end{equation}
equation \ref{eqn:fraction1} can be rewritten as
\begin{equation}
\begin{split}
S(D) &= 1 + \frac{\langle x_{I,n}(z_n) \rangle_c}{x_{I,n}(z_{n,D})} (e^{-x_{I,n}(z_{n,D})} - 1) \\
  &= 1 + \left[ \frac{e^{-(\alpha_0 + z_{D}\beta_0)z_{n,D} - \beta_0 z_{n,D}^2} - 1}{(\alpha_0 + z_{D}\beta_0)z_{n,D} + \beta_0 z_{n,D}^2} \right]\left((\alpha_0 + \beta_0 z_D)D + \beta_0 D^2\right)
\end{split}
\end{equation}
Taking the log of $S$, expanding around $D=0$ and dropping terms in $D^2$ or higher powers, the linear term of $\log S(D)$ can be written as
\begin{equation}\label{eqn:survMKMNP}
\begin{split}
\left. - \log S(D)\right|_{D \to 0} &\approx (\alpha_0 + z_D\beta_0) \times \left( \frac{1-e^{-(\alpha_0+z_D\beta_0)z_{n,D}-\beta_0z_{n,D}}}{(\alpha_0 + z_D\beta_0)z_{n,D} + \beta_0z_{n,D}^2} \right) \times D\\
&\approx \alpha_\mathrm{P} \times \left( \frac{1-e^{-\alpha_P z_{n,D}}}{\alpha_P z_{n,D} } \right) \times D\\
&=\alpha_\mathrm{NP} \times D
\end{split}
\end{equation}
where $\alpha_\mathrm{P}$ is the Poisson $\alpha$ coefficient defined in equation \eqref{EQN:SurvMKMPFin2_ab}, while the subscript NP denote the Non-Poisson corrected $\alpha$ coefficient. Following also the original formulation of Hawkins \cite{hawkins2003}, in equation \eqref{eqn:survMKMNP} the quadratic term $z_{n,D}^2 \ll z_{n,D}$ was also neglected.

Notice that in the above equations a subtle approximation is assumed in order to match the request of having a Poisson distribution in the lethal events, with only a single well defined value of $z_{n}=z_{n,D}$ when the particle hits the cell. Generally, this is not the case and the specific energy can also vary as a function of the impact parameter of the particle with respect to cell nucleus. Still, $x_{I,n}(z_{n,D})$ is used as an estimation of the average number of lethal lesions in those cells that have suffered a single event after exposure to a dose $D$. This assumption can be reasonable when low energy particles with high LET (see also section \ref{SSEC:amtrack}) are considered.

The Non-Poisson correction to the RBE in the limit of zero dose ($\mathrm{RBE}_{\alpha}$) is given by
\begin{equation}\label{eqn:np-rbe}
\mathrm{RBE}_{\alpha,\mathrm{NP}} =\frac{\alpha_\mathrm{NP}}{\alpha_{X}} =\left( \frac{1-e^{-\alpha_\mathrm{P} z_{n,D}}}{\alpha_\mathrm{P} z_{n,D} } \right) \mathrm{RBE}_{\alpha,\mathrm{P}}
\end{equation}
with $\mathrm{RBE}_{\alpha,\mathrm{P}}$ given by equation \eqref{eqn:rbe_alpha}. No corrections are applied to the $\mathrm{RBE}_\beta$ that is still assumed constant ($\mathrm{RBE}_\beta \sim 1$) and independent on the quality of the radiation.

\begin{figure}[h!]
\centering
\includegraphics[width= 0.6\textwidth]{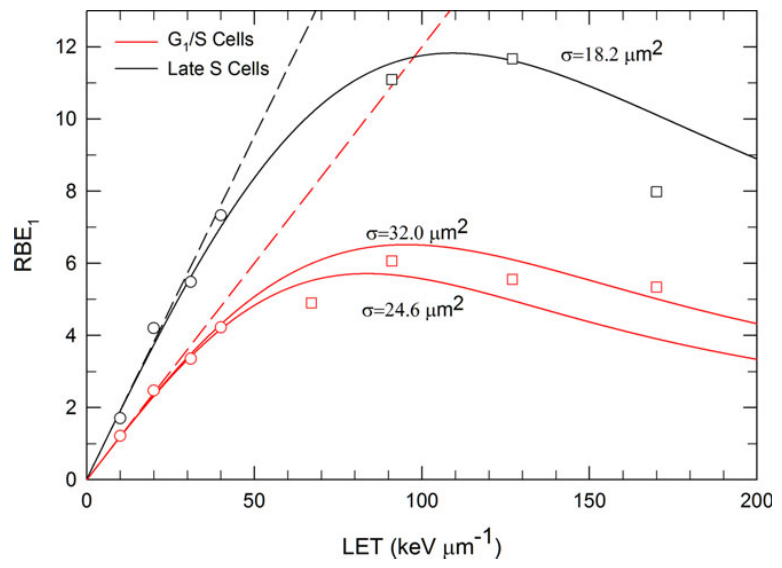}  	
\caption{
Comparison of MKM estimates of particle $\mathrm{RBE}_\alpha$ to experimental values for V79 cells. Red curves show $\mathrm{RBE}_\alpha$ for cells synchronized at G1-S transition for cross sections of $\sigma= 32.0$ and $24.6 \mu m^2$ (reference radiation: 250 kVp X-rays, $\alpha_R=0.234 \mathrm{Gy^{-1}}$, $\beta_R=0.042 \mathrm{Gy^{-2}}$).
Black curve shows $\mathrm{RBE}_\alpha$ for cells synchronized in late S phase for cross sections of $\sigma= 18.2 \mu m^2$ (reference radiation: 250 kVp X-rays, $\alpha_R=0.064 \mathrm{Gy^{-1}}$, $\beta_R=0.0165 \mathrm{Gy^{-2}}$).
Dashed lines represent $\mathrm{RBE}_\alpha=0.02+0.19 \times \text { LET. }$ in the Poission regime.
Experimental data are from \cite{Bird1980,Bird1983}, plot taken from \cite{Stewart2018}.
}
\label{fig:np_exp1}
\end{figure}

\begin{figure}[h!]
\centering
\includegraphics[width= 0.8\textwidth]{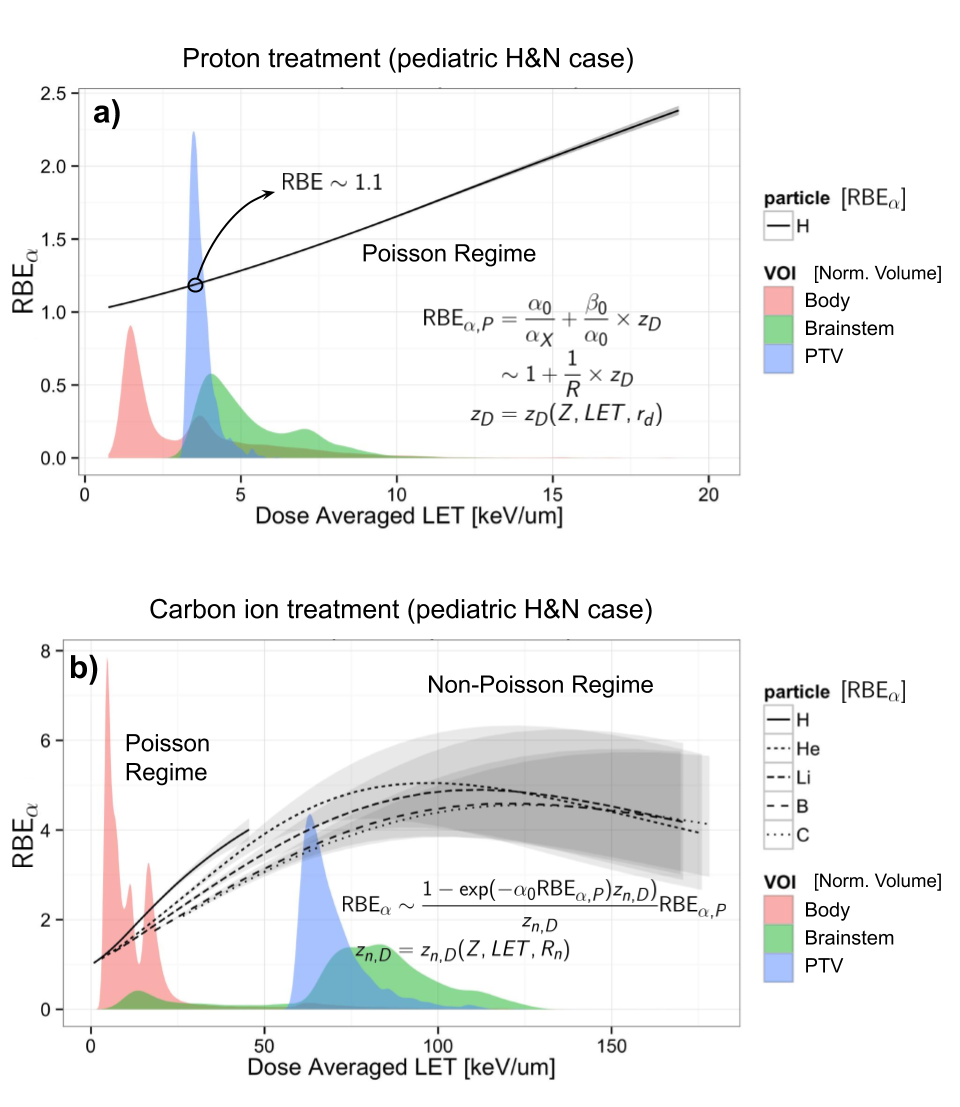}  	
\caption{Evaluation of the RBE$\alpha$ \textit{vs.} LET$_D$ evaluated via equation \eqref{eqn:np-rbe} in combination with an amorphous track model (see section \ref{SSEC:amtrack}) for proton (a) and for carbon and other ions (b). Using the same x axis of the plot is reported for comparison the LET$_D$ volumetric distribution (filled areas) found in a patient (a pediatric brain tumor case) irradiated with a primary beam of protons (a) and carbon ions (b). The LET$_D$ distribution is normalized and evaluated in 3 volumes: total body (red), brainsteam (green) and the planning target volume, PTV, (blue). Annotated in the plot are the low-LET range for the Poisson regime, applicable mainly for the proton treatment, and the Non-Poisson regime, in the case of high-LET carbon ion treatment (for both the primary carbon ions and the fragments). The gray bands represent the 95\% confidence band obtained with a bootstrap procedure to the fit of the input parameters $(\alpha_0, \beta_0, R_n, r_d)$ of the model to the \textit{in-vitro} experimental data taken for cells with $R \simeq 10$ Gy from the PIDE database \cite{Friedrich2012b}. The evaluations have been carried out with a research TPS \cite{russo2015}.
}
\label{fig:non-poisson}
\end{figure}

 The correction causes the $\mathrm{RBE_\alpha} $ to be less than indicated by the extrapolation of the linear relationship (equation \ref{eqn:rbe_alpha_gen}) to higher LET, and to pass through a maximum in the range of LET of 50 to 150 keV/$\mu$m. This behaviour is compatible with several experimental studies from the literature \cite{Friedrich2012b} and it shows also a sensitivity of the maximum of the RBE to the response of the cell at low-LET, related to the parameter $R = \alpha_X/\beta_X$ \cite{Weyrather2003}. An exemplification of the RBE behaviour and the prediction of the model is reported in Fig.\ \ref{fig:np_exp1}. In Fig.\ \ref{fig:non-poisson} some qualitative implications of the non-Poisson regime in high-LET ion beam therapy are depicted.

\subsection{The saturation correction}
\label{SSEC:satcor}

\begin{figure}[h!]
\centering
\includegraphics[width= 0.55\textwidth]{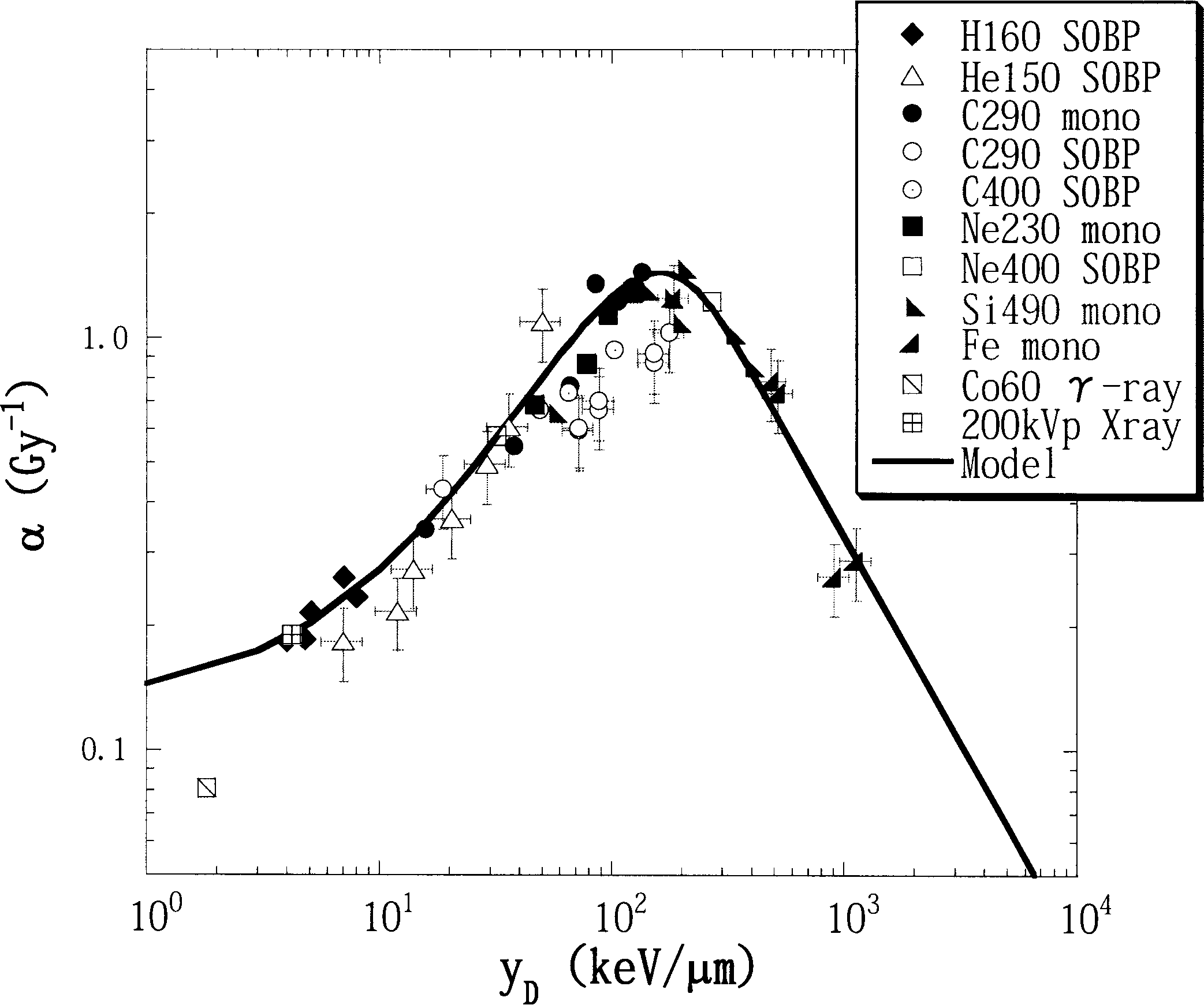}  	
\caption{Experimental $\alpha$ values, fitted by the linear-quadratic model from the survival curves of HSG cells value with $\beta_0 = 0.05\  \mathrm{Gy}^{-2}$, as a function of the dose mean lineal energy, $y_d$. The $y_d$ were measured by the TEPC with a simulated diameter of 1.0 $\mu $m. The solid line indicates the curve calculated using equation \eqref{Kase_alpha} with the following model parameters: $r_{d}=0.42\ \mu \mathrm{m}$, $R_{n}=4.1\ \mu \mathrm{m}$, $\alpha_{0}=0.13\ \mathrm{Gy}^{-1}$, and $\beta_0 = 0.05\ \mathrm{Gy}^{-2}$. Plot taken from \cite{kase2006}.
}
\label{alpha_comp}
\end{figure}

Kase \etal\ \cite{kase2006} introduced a correction factor in the MK model to account the decrease in RBE due to the overkill effect observed in high-LET radiations (see for example Fig.\ \ref{alpha_comp}). The correction factor was applied to dose-averaged saturation-corrected specific energy per event, $z_1^*$, for mixed radiation field with wide-ranging spectra.

In terms of lineal energy, the corrected value of $y_D$ (and hence $z_D$) was obtained by applying a correction for each lineal energy component of the lineal energy spectrum. The correction of the components was obtained by using an empirical \textit{saturation parameter} $y_0$ based on the saturation correction method introduced by \cite{Powers1968} and then used in the TDRA \cite{kellerer1978}
\begin{equation}
    \label{kase_y*}
    y_D^{*} =  \frac{y_{0}^{2} \int\left[1-\exp \left(-y^{2} / y_{0}^{2}\right)\right] f(y)\  \text{d}y}{\int y f(y)\ \text{d}y}
\end{equation}
The saturation parameter indicates the lineal energy above which the correction due to the overkill effects became important.

The correction to cell survival is then obtained by evaluating the corrected dose-averaged saturation specific energy per event $z^*_{D}$ in the domain, which can obtained from the corrected dose-averaged lineal energy \eqref{kase_y*} using the relationships reported in Eq. \eqref{EQN:Mom2zD} and \eqref{eqn:zD}:
\begin{equation}\label{eqn:y*_z*}
z^*_{D}=\frac{\bar{l}_d}{m_d} y^*_{D}=\frac{y^*_{D}}{\rho \pi r_{d}^{2}}
\end{equation}
where $ \rho$, $r_{d}$, $\bar{l}_d$ and $m_d$ are the density, radius, mean cord length, and mass of the domain, respectively. The equation for the cell survival (equation \eqref{EQN:SurvMKMPFin2}) is then modified as follows:
\begin{equation}\label{Kase_L_n}
-\ln(S) = \left(\alpha_{0}+\beta_0 z_{D}^{*}\right) D + \beta_0 D^2
\end{equation}
Considering the linear term in the macroscopic dose $D$, the corrected $\alpha^*$ coefficient is hence:
\begin{equation}\label{Kase_alpha}
\alpha^* =\left(\alpha_{0}+\beta_0 z_{D}^{*}\right)
\end{equation}
No correction is considered for the $\beta$ coefficient and it is still assumed to be independent on the energy spectrum.

An example of the prediction MK model modified with the saturation correction compared with experimental data is reported figure \ref{alpha_comp}, where the $\alpha$ vs.\ $y_D$ for HSG cells irradiated with carbon ions is shown. It is interesting to note, by comparing equations \eqref{eqn:survMKMNP} and \eqref{Kase_alpha}, that the saturation correction can be considered an alternative way to describe the non-Poisson correction defined in section \ref{non-poisson}, since both factors modulate the behaviour of $\mathrm{RBE_\alpha}$ in similar ways (see also figure \ref{fig:np_exp1}). In particular it was shown in \cite{kase2006} that in the case of monoenergetic spectra, the equations \eqref{eqn:survMKMNP} and \eqref{Kase_alpha} are functionally equivalent for $y < 500$ keV/$\mu$m. Thus, by matching these equations in the limit of low LET ($y_D \to 0$), and defining $z_{n,D} = y_{D} / \rho\pi R_n$, with $R_n$ the radius of the nucleus, analogously to equation \eqref{eqn:y*_z*} for the domain, is possible to link the saturation correction parameter $y_0$ with the other parameters of the model
\begin{equation}\label{eqn:y0}
y_0 = \frac{\rho \pi r_d R^2_n}{\sqrt{\beta_0(r^2_d + R^2_n)}}
\end{equation}
A typical used value of the saturation parameter was $y_0 = 150$ keV/$\mu$m \cite{kase2006,inaniwa2010}.

Other quantities that one needs to determine for the RBE evaluations are the lineal energy spectra, obtainable with a microdosimeter detector such as TEPC \cite{lindborg2017}, and the values of $ \alpha_0 $, $r_d$ and $R_N$ from which the correction to the $y^*_D$ is calculated. The $\alpha_0$ and $r_d$ coefficients can be extrapolated experimentally from the initial slope of the survival curves (equation \ref{Kase_alpha}) for low-LET irradiation (in the limit of $y_D \to 0$ and $D\to 0$)
\begin{equation}
\begin{aligned}
&r_{d}=\sqrt{\frac{\beta\left(y_{D}-(y_{D})_X\right)}{\rho \pi\left(\alpha-\alpha_{X}\right)}}\\
&\lim_{y_D\to 0} \alpha\equiv \alpha_{0}=\alpha_{x}-\left(\frac{\alpha-\alpha_{x}}{y_{D}-(y_{D})_X}\right) (y_{D})_X
\end{aligned}
\end{equation}
where $ \rho = 1.0\ \mathrm{g}/\mathrm{cm}^3 $, $ \alpha_{X} $ is the LQ parameter of the X-Ray, and $(y_{D})_X$ is the dose mean lineal energy for X-ray irradiation.

\begin{figure}[h!]
\centering
\includegraphics[width= 0.8\textwidth]{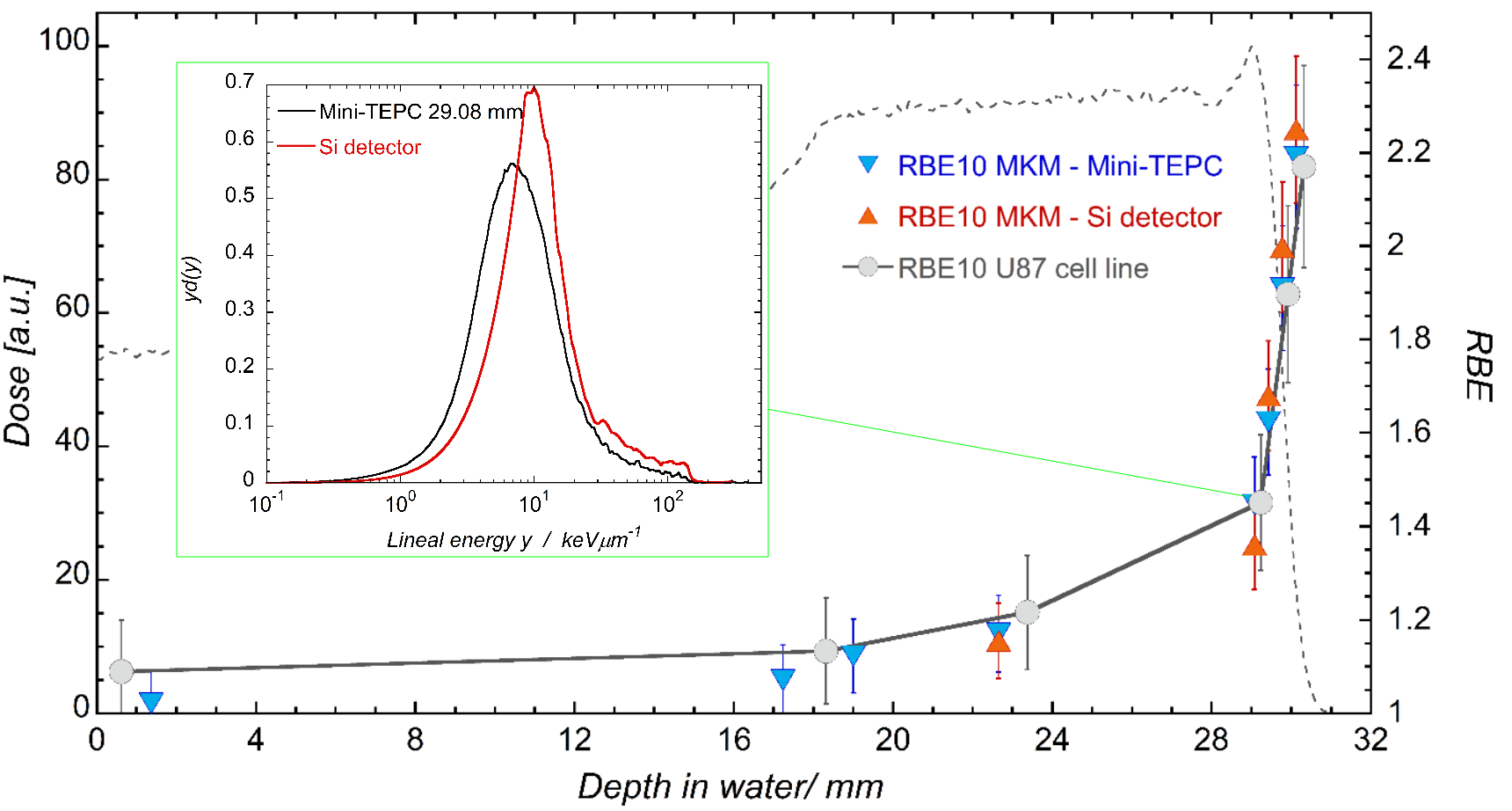}  	
\caption{Clinical SOBP of CATANA. Dose profile measured with the Markus chamber in black, total LET-dose from Geant4 MC simulation in blue. Crosses indicate positions at which both detectors measured. The box reports the normalized spectra obtained with the mini-TEPC (black) and the silicon telescope (red) at a depth of 29.08 mm. Plots taken from \cite{bianchi2020}}
\label{FIG:rbEtepc}
\end{figure}

The saturation-corrected formulation of the MK model is one of the most widely used approaches to estimate the RBE from microdosimetric measurements. Many studies have been published where the computed RBE is compared with the RBE measured along single Bragg Peaks or more complex mixed field irradiations \cite{rosenfeld2016,guardiola2015,kase2011,bianchi2020}. In figure \ref{FIG:rbEtepc} the RBE vs. depth for a proton spread-out Bragg Peak is reported as an example of these assessments \cite{bianchi2020}.

\subsection{Track structure model incorporation in the MK model}

\label{SSEC:amtrack}
\begin{figure}[h!]
\centering
\includegraphics[width=\textwidth]{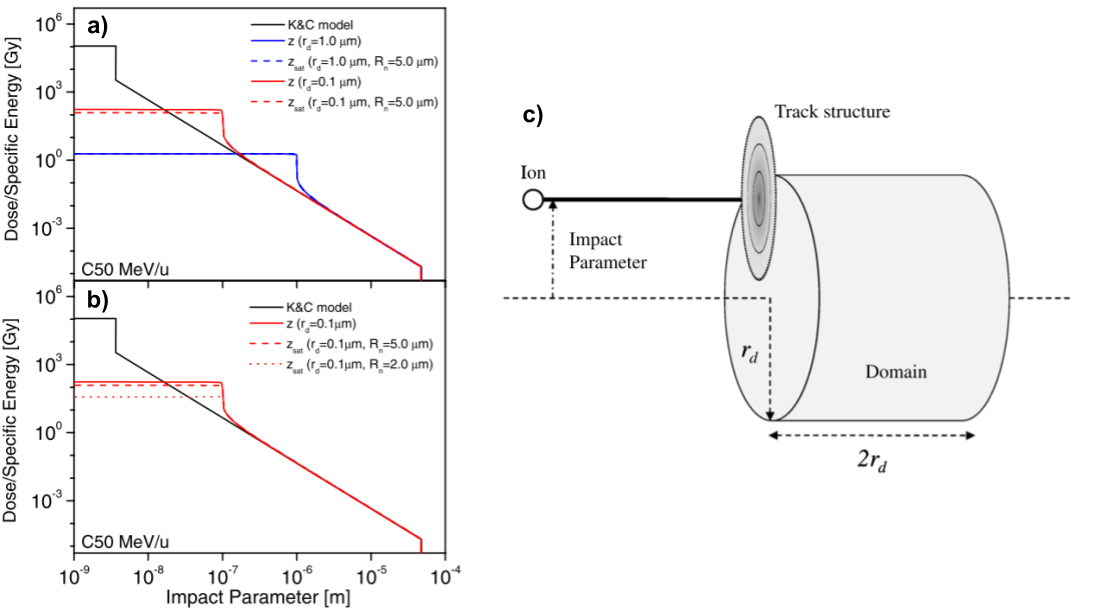}  	\caption{\textbf{(a)} and \textbf{(b)} Track structures for a carbon-ion beam with an energy of 50 MeV/u calculated with the Kiefer–Chatterjee model and the corresponding specific energy $z$ and saturation-corrected specific energy $z_\mathrm{sat}$ as functions of the impact parameter for domain and nucleus with different sizes. \textbf{(c)} Schematic of an incident ion with respect to a cylindrical sensitive volume. Plots taken from \cite{inaniwa2010}.}
\label{Kase_rad}
\end{figure}

In 2008, Kase  \textit{et al.}\ \cite{kase2007} introduced the usage of amorphous track structure models as an alternative numerical approach to evaluate theoretically the dose-averaged specific energy deposition per event in the nucleus $z_{n,D}$ and in the domain $z_{D}$ for the MK model calculations.  This approach has the advantage of bypassing the necessity to acquire experimental lineal energy spectra to evaluate the RBE; this approach is particularly useful when experimental spectra are not available, such as in TPS calculations. Another interesting aspect  is possibility to evince the dependence of LET-RBE curves on the ion type.
\\
The amorphous track model adopted for the MK calculation is based on a combination of the Kiefer model for the penumbra region \cite{kiefer1986} and the Chatterjee model for the core radius \cite{chatterjee1976}, introduced for explaining the responses of the diamond detector to heavy-ion beams \cite{sakama2005}. Here the core radius $R_c$ ($\mu$m), the penumbra radius $R_p$ ($\mu$m) and the dose $z_\mathrm{KC}$ as function of track radius $r$ ($\mu$m) are evaluated as follows
\begin{eqnarray}
R_c &=& 0.0116 \times \beta_\mathrm{ion}\\
R_p &=& 0.0616 \times (E_s)^{1.7}\\
z_\mathrm{KC}(r \le R_c) &=& \frac{1}{R_c^2} \left( \frac{\mathrm{LET}_{\infty}}{r\rho} - 2 \pi K_p \ln (R_p/R_c) \right)\\
z_\mathrm{KC}(r > R_c) &=& 1.25 \times 10^{-4}(Z^*/\beta_\mathrm{ion})^2 r^{-2}
\end{eqnarray}
where $E_s$ is the specific energy in MeV/u, $Z^*$ is the effective charge given by the Barkas expression, $\beta_\mathrm{ion}$ is the velocity relative to the speed of light, $\mathrm{LET}_\infty$ is the unrestricted LET and $\rho$ is the density of water. In order to evaluate the dose-averages of the specific energies, the domain and nucleus are assumed to have a cylindrical symmetry with the direction of the incident ion parallel to the cylinder axis. Using this geometry is possible to write explicitly the dose average specific energy for a single track (equation \ref{EQN:Mom2zD}) for both domain and nucleus as 
\begin{equation}
\label{eqn:z1D}
{z_{D} = \int_0^{b_\mathrm{max}}z_\mathrm{KC}(b)^2 b\,db \bigg/ \int_0^{b_\mathrm{max}} z_\mathrm{KC}(b) b\,db}
\end{equation}
\begin{equation}
\label{eqn:z1Dn}
{z_{n,D} = \int_0^{b^{(n)}_\mathrm{max}} z_\mathrm{KC}(b)^2 b\,db \bigg/ \int_0^{b^{(n)}_\mathrm{max}} z_\mathrm{KC}(b) b\,db}
\end{equation}
where $b_\mathrm{max}$ and $b^{(n)}_\mathrm{max}$ are the maximum impact parameters to have a non negligible energy deposition in the domain and in the nucleus respectively. These parameters, or equivalently the radius of the domain $r_d$ and of the nucleus $R_n$, represents two parameters of the model. Examples of these evaluations are shown in figure \ref{Kase_rad}. $R_n$ in principle it can be related to direct observations while $r_d$ does not represent a measurable quantity, since it cannot be uniquely identified with any structure in the cell or cell nucleus. $r_d$ can be used as a free parameter to be fixed by fitting the model to the experimental survival and RBE data.

Using the non-Poisson formulation for the biological effect as defined in \eqref{eqn:survMKMNP} is possible to obtain the corrected non-Poisson $\alpha$ coefficient as:
\begin{equation}
\label{kase_alpha_np}
\alpha_\mathrm{NP} \simeq \left( 1 - \exp(-(\alpha_0 + \beta_0 z_{D})z_{n,D}) \right) \left( \frac{1}{z_{n,D}} \right)
\end{equation}
According to \cite{hawkins2003}, one can approximate $z_{n,D} = \propto \mbox{LET}_\infty /A$ , where $\mbox{LET}_\infty$ is the unrestricted linear energy transfer in keV$\mu$m${}^{-1}$ for the incident particle and $A$ is the area of the cell nucleus in $\mu$m${}^2$, assuming $\rho = 1$ g cm${}^{-3}$ for the density of water. By assuming $A = \pi R_n^2$, equation \eqref{kase_alpha_np} becomes
\begin{equation}\label{kase_alpha_np2}
\alpha_\mathrm{NP} \simeq \left( 1 - \exp\left(-(\alpha_0 + \beta_0 z_{D})\frac{\mbox{LET}_\infty}{\rho \pi R^2_n}\right) \right) \left( \frac{\rho \pi R^2_n}{\mbox{LET}_\infty} \right)
\end{equation}
No correction is considered for the $\beta$ coefficient and it is still assumed to be independent on the energy and particle type.

\begin{figure}[h!]
	\centering
	\includegraphics[width= 0.9\textwidth]
	{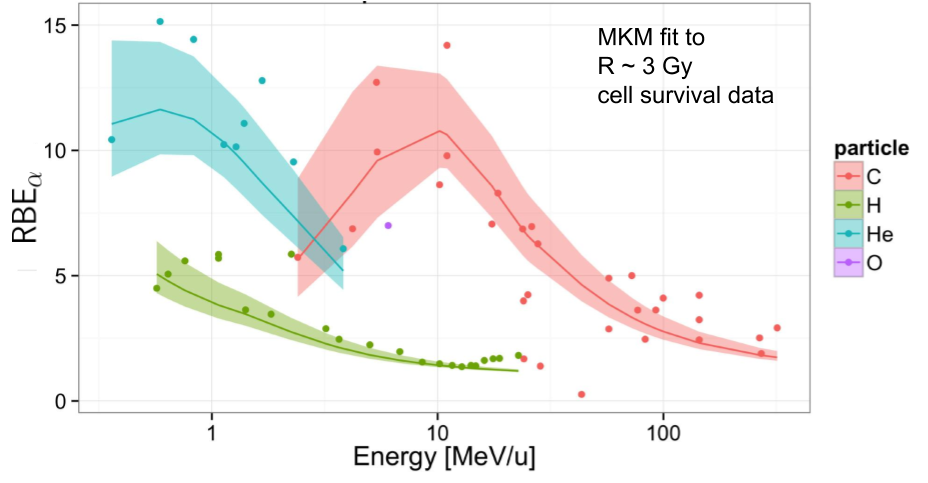}
	\caption{Global fitting of the MK model to experimental $\mathrm{RBE}_\alpha$ data for different ion irradiation subsetted from the PIDE database \cite{Friedrich2012b} with $R = \alpha_X/\beta_X \in [2.8, 3.2]$ Gy. The plotted points indicate the experimental results  and the lines represent the MKM results calculated with the Kiefer–Chatterjee track structure model. The bands represent the 90\% confidence band obtained from the MK model parameters $\{\alpha_0, \beta_0, R_n, r_d\}$ probability distribution, evaluated through a bootstrap procedure.}
	\label{Fig_kase08_surv}
\end{figure}

As seen in section \ref{non-poisson} in the case of high LET irradiation the $z_{D}$ value is comparably very large with respect to the $(z_{D})_X$ evaluated for photon; consequently, from Eq.\ \ref{EQN:mkmAB} $\alpha_0$ and $\beta_0$ can be approximated with the experimental $\alpha_X$ and $\beta_X$.  An example of the model evaluations compared to the experimental data is reported in figure \ref{Fig_kase08_surv}.

Interestingly, the explicit usage of a track model shows how some aspects of the MKM are conceptually similar to that of the LEM \cite{LEMscholz1997,LEMscholz1992,LEMscholz1996,LEMelsasser2007,LEMelsasser2008,LEMelsasser2010,friedrich2012}. In both MK model and LEM the principal target is the cell nucleus for any radiation quality, the nucleus is divided into small independent sub-volumes (infinitesimal volumes in the case of LEM and domains in the case of MK model) and a cell survival curve for x-rays is adopted as the local dose-effect curve of each sub-volume. Finally the summation of the local effect in all sub-volumes over the whole nucleus determines the cell survival probability.

The inclusion of the amorphous track model allow to evaluate directly $z_{D}$ and $z_{n,D}$ without the necessity to obtain these values by extrapolating them from microdosimetric measurements via scaling relationships such as equation \eqref{eqn:zD}. This aspect can be particularly advantageous for some applications where these spectra are generally not easily or partially available such as in the simulation and optimization of treatments in TPS applications, where the biological effect should be evaluated in the whole irradiated 3-D patient volume.

At present the MK model is implemented in the proton and carbon ion TPS used clinically at the National Institute of Radiological Sciences (NIRS) in Japan to evaluate the RBE and the RBE-weighted dose optimized for the individual patients. The computation method, developed by Inaniwa \textit{et al.} \cite{inaniwa2010,Inaniwa2015b} takes advantage of the incorporation of the amorphous track model in combination with the saturation corrected approach developed by Kase \textit{et al.} \cite{kase2006} described in section \ref{SSEC:satcor}, for evaluations in case of mixed field irradiation. In the TPS implementation a set of pre calculated look-up tables of the saturation corrected specific energies for monoenergetic beams are created using a generalization of equations \eqref{eqn:z1D}-\eqref{eqn:z1Dn} where the saturation effect is explicitly included for the dose averaged specific energy for the domain
\begin{equation}
\label{eqn:z1D*}
{z^*_{D} = \int_0^{b_\mathrm{max}}z_\mathrm{sat}(b)z_\mathrm{KC}(b) b\,db \bigg/ \int_0^{b_\mathrm{max}} z_\mathrm{KC}(b) b\,db}
\end{equation}
and equivalently for the nucleus, where $z_\mathrm{sat}$ is the saturation-corrected specific energy
\begin{equation}
z_\mathrm{sat}(b) = \frac{z_0^2}{z_\mathrm{KC}(b)}\left( 1- \exp\left(-\frac{z_\mathrm{KC}(b)^2}{z^2_0}\right) \right)
\end{equation}
with saturation coefficient $z_0 = y_0(\bar{l}_d/m_d)$ (see equation \ref{eqn:y0}) evaluated in the cylindrical geometry (see also panel (c) of figure \ref{Kase_rad}). The effect of the mixed-field of the treatment at a position is hence evaluated through a dose-weighted linear combination
\begin{equation}
    (z^*_{D,\mathrm{mix}})_i = \frac{\sum_{j=1}^{N_b} D_{ij} (z^*_D)_{ij}}{\sum_{j=1}^{N_b} D_ij}
\end{equation}
where $N_b$ is the number of beams of the treatments, $D_{i,j}$ is the dose released at position $i$ by the beam $j$, and
\begin{equation}
   (z^*_D)_{ij} =  \frac{\sum_{k=1}^{N_{ij}} e_k(z^*_D)_k}{\sum_{k=1}^{N_{ij}} e_k}
\end{equation}
is the saturation-corrected dose-mean specific energy of the domain of cells at position $i$ delivered by the $j$th beam, obtained through the mono-energetic evaluations, $(z^*_D)_k$, described in equation \eqref{eqn:z1D*}. To obtain the energy imparted $e_k$ and the number $N_{ij}$ of the deposition events, Monte Carlo simulations are used, taking advantage of available codes such, as example, those derived from the Geant4 libraries \cite{Agostinelli2003,Aso2007,zhu2019microdosimetric} and Fluka \cite{magro2017}.

\subsection{The dependence of the biological effect on the dose-rate time structure}\label{SSEC:timestructure}

One of the interesting features of the MK model, in contrast to other radiobiological models used in hadrontherapy such as the LEM, is the possibility to account inherently for arbitrary time dependent dose-rates, such as protracted irradiations and fractionations. This feature derives from the explicit description of the time depending response of the cell to the irradiation trough the kinetic equations (Eq.\  \ref{EQN:SolXI}).

Different approaches to investigate and to model the dose-rate time effects have been carried out using the MK model as a theoretical base. Some examples of these approaches can be found in \cite{hawkins1996,inaniwa2013,hawkins2013,hawkins2014,inaniwa2015,manganaro2017,matsuya2018}. In these studies, the kinetic equations \eqref{MKM_kinEQ} are slightly generalized to account an arbitrary time dependent specific energy deposition rate $\dot{z}^{(c,d)}$ in the domains:
\begin{eqnarray}\label{MKMt_kinEQ}
\left\{\begin{array}{l}
\dot{x}_{I}^{(c,d)} = \lambda \dot{z}^{(c,d)} + a x_{I I}^{(c,d)} + b\left(x_{I I}^{(c,d)}\right)^{2} \,,\\
\dot{x}_{I I}^{(c,d)} = \kappa\dot{z}^{(c,d)} -(a+r) x_{I I}^{(c,d)} - 2 b\left(x_{I I}^{(c,d)}\right)^{2} \simeq \kappa\dot{z}^{(c,d)} -(a+r) x_{I I}^{(c,d)}\,.
\end{array}\right.
\end{eqnarray}
where in the second equation, as described in section \ref{mkm}, the second order process describing the pairwise combination between type II lesions has been removed since it is considered negligible if compared to the first-order process.

In \cite{inaniwa2013} and \cite{inaniwa2015}, the effects of dose-delivery time structure on the RBE in a mixed radiation fields of therapeutic carbon ion beams is investigated using by the modified microdosimetric kinetic model introduced by Kase et al.\ \cite{kase2006,kase2007,inaniwa2010}. The study evaluate the biological effect of the irradiation in two different dose-rate conditions: a split-dose irradiation and a protracted continuous irradiation.

In the case of a \textit{split-dose irradiation}, a population of cells is considered exposed to a macroscopic dose $D_1$ at time $t = 0$ and a dose $D_2$ at time $ t=\Delta T $, where a domain absorbs $ z_1 $ and $ z_2 $ from the two separate irradiation respectively.
Evaluating $x_I$ in the limit $t \to \infty$ by integrating the equations \eqref{MKM_kinEQ} and using the saturation correction specific energy as given in equation \eqref{Kase_L_n}, we obtain for the cell survival: 
\begin{equation}
\begin{aligned}
\label{eqn:split1}
\ln S(D, \Delta T) =& -(\alpha_{0} + \beta (z_{D}^{*})_1 )D_1 - \beta D_1^2
-(\alpha_{0} + \beta (z_{D}^{*})_2 )D_2 - \beta D_1^2 \\
&- 2\beta D_1 D_2 e^{-(a+c)\Delta T}
\frac{1 - e^{-2(a+c)(t_r-\Delta T)}}{1 - e^{-2(a+c)t_r}}
\end{aligned}
\end{equation}
where $(z_{1 D}^{*})_1$ and $(z_{1 D}^{*})_2$ are the saturation corrected dose means of the first and second irradiation and the total dose $D = D_1 + D_2$. The time parameters $t_r$ indicates the the time after which all  sub-lethal lesions that are not still repaired are fixed in lethal lesions, according to assumption (9) introduced in section \ref{mkm}. If the quality of the radiation does not change between the two irradiations then $(z_{D}^{*})_1 = (z_{D}^{*})_2 = z^*_{D}$ and Eq.\ \ref{eqn:split1} can be simplified as
\begin{equation}
\label{eqn:split2}
\begin{aligned}
\ln S(D, \Delta T)=&-\alpha_{0}\left(D_{1}+D_{2}\right)-z_{1 D}^{*} \beta\left(D_{1}+D_{2}\right)-\beta\left(D_{1}+D_{2}\right)^{2} \\
&+2 \beta D_{1} D_{2}\left[1-e^{-(a+c) \tau} \frac{\left(1-e^{-2(a+c)\left(t_{r}-\tau\right)}\right)}{\left(1-e^{-2(a+c) t_{r}}\right)}\right]
\end{aligned}
\end{equation}
The values of $(a + c)$ and $t_r$ can be determined by using the following approximations (see Eqs.\ 35, 40 in \cite{inaniwa2013}):
\begin{equation}
(a+c) \cong \frac{1}{2 \beta D_{1} D_{2}}\left[\frac{1}{S} \frac{d S}{d \tau}\right]_{\tau=0}
\end{equation}
\begin{equation}
t_{r}=-\frac{\ln (\alpha / \kappa)}{(a+c)}
\end{equation} 
\\
In the case of a \textit{continuous protracted irradiation}, a population of cells receive a constant macroscopic dose rate of $\dot{D}$ starting at time $t=0$ and ending at $t=T$. In order to carry out the evaluation, the irradiation is assumed microscopically equivalent to a number $N$ of instantaneous irradiations with random doses to a domain delivered every infinitesimal interval. The time interval between these irradiations is $\delta t = T/(N-1)$, with $\delta t \ll 1/(a+c) $, and each domain absorbs $z_{1}\,, z_{2}\,, \dots,z_{N} $.
The the number of lethal lesions per domain $x_I$ is therefore obtained by integrating and summing the solution of Eq.\ \ref{MKM_kinEQ} for each time segment. The final cell survival probability is then obtained by introducing the corresponding saturation corrected dose means in a way analogous to the split dose evaluation. In the case in which the quality of the radiation does not change with time, the final log survival is given as:
\begin{equation}\label{eqn:continuous1}
	\ln S=-\left(\alpha_0+\beta_0 z_{d,D}^{*}\right) D-\beta^{\prime} D^{2}
\end{equation}
where 
\begin{equation}
\label{eqn:continuous2}
\beta^{\prime} \equiv \beta_0\left[1-\frac{2}{N^{2}} \sum_{n=1}^{N-1}\left\{(N-n)\left(1-e^{-(a+r)_{\vec{N}-1} T} \frac{\left(1-e^{-2(a+r)\left(t_{r}-\frac{n}{N-1} T\right)}\right)}{\left(1-e^{-2(a+r) t_{r}}\right)}\right)\right\}\right]
\end{equation}
The notation used in Eqs.\ \ref{eqn:continuous1} and \ref{eqn:continuous2} highlights the importance of the quadratic term $\beta$, which modulates the impact of the dose-rate time structure, according to the LQ interpretation of the biological effects \cite{LQ_deehan1988,Fowler1989,LQ_yaes1991,LQ_mcmahon2018}.

%

\begin{figure}[h!]
	\centering
	\includegraphics[width= 0.8\textwidth]
	{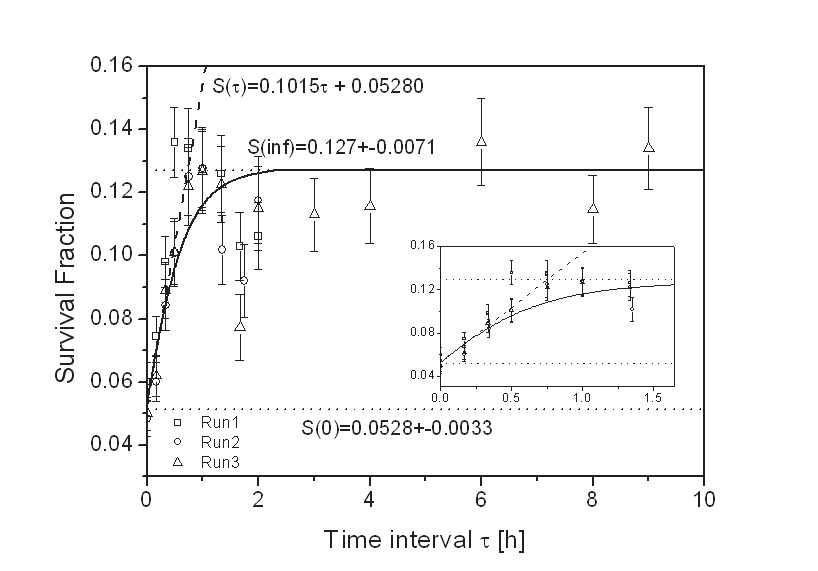}  	
	\caption{Survival fraction of HSG tumor cells after exposure to two equal doses of carbon beams with $D_1= D_2= 2.5$ Gy separated by time interval $\tau$ ($\Delta T$ in the text) from 0 to 9 h. Three different series of experiments (runs) are shown. The  estimated tangent at $\tau = 0$ h is reported in dashed line, while the solid curve is the predicted survival by Eq.\ \eqref{eqn:split2} with $\beta=0.0703\ \mathrm{Gy}^{-2}$ , $\beta=0.237\ \mathrm{Gy}^{-2}$, $(a+c)=2.187\ \mathrm{h}^{-1} $ and $ t_{r}=2.284$. Plot taken from \cite{inaniwa2013}.}
	\label{Fig:InaSplit}
\end{figure}

It is worth remarking that these MK-based temporal formulations of the cell survival derived from the kinetic equations \eqref{MKMt_kinEQ} do not accounts for re-population and cell cycle redistribution. Figure \ref{Fig:InaSplit} reports the evaluation via equation \eqref{eqn:split2} of the survival fraction of HSG cell line for various time intervals compared to experimental data. An initial rise in cell survival due to repair is visible until time interval $\Delta T = 0.75$ h followed by a decrease in survival due to cell cycle redistribution and a rise due to re-population ($1.4 < \Delta T <4$ h ) and by the saturation region for $\Delta T > 4$ h. The predicted cell survivals reasonably agrees with data in the first and last regions, while it does not accounts for re-population and redistribution. The temporal formulations described in this section have been also incorporated in the TPS used at NIRS \cite{inaniwa2010} (see also section \ref{SSEC:amtrack}), and successfully used to estimate the impact of the beam interruption in single-fractionated treatments with carbon ions for patients with prostate tumor \cite{inaniwa2015}.

\subsection{Stochastic approaches and variable $\beta$}\label{sec:stochastic}

As discussed in the previous sections, the MK model accounts for the stochastic aspects of the induction of damage in the cell by exploiting probability theory to develop simple formulas for the LQ coefficients of the cell survival [equations \eqref{EQN:SurvMKMPFin2}, \eqref{EQN:SurvMKMPFin2_ab}, \eqref{eqn:survMKMNP}, \eqref{Kase_L_n}, and \eqref{eqn:continuous1}]. These formulations of the model are obtained introducing approximations \cite{hawkins2003} or ad-hoc corrections \cite{kase2006} that shows some discrepancies with experiments for high-LET irradiation, in particular in the determination of the $\beta$ coefficient, since the measured $\beta$ tends to decrease at very high LET \cite{furusawa2000,Czub2008,Tsuruoka2005,Friedrich2012b}, while the $\beta$ derived from the MK model is considered constant.

The disagreements in the $\beta$ coefficient are ultimately acknowledged to be induced by the partial accounting of the stochastic nature of the specific energies in the MK model calculations that plays an important role for high-LET irradiation \cite{sato2012}. Following these considerations, attempts to improve the model, introducing more refined approaches to account a variable $\beta$, have been made \cite{sato2012,manganaro2017,chen2017, inaniwa2018}. In the rest of this section some of these developments, based on improved stochastic modelings of the specific energy depositions, are described in details.

\subsubsection{Monte Carlo based evaluations}
\label{Sec:Manganaro}
A method to account in a natural and straightforward way the inherent stochastic nature of the irradiation is to implement a Monte Carlo algorithm in the MK model, as recently shown by Manganaro \textit{et al.} \cite{manganaro2017,manganaro2018} in their formulation of the model named MCt-MKM (Monte Carlo temporal microdosimetric kinetic model). The implemented model accounts also the  stochastic temporal correlations characteristic of the irradiation process and the cellular repair kinetics by solving explicitly in the MC evaluations the kinetic equations \eqref{MKMt_kinEQ} where the time dependent specific energy rate $\dot{z}^{(c,d)}$ appears explicitly \cite{inaniwa2013,Hawkins2017}. 

In the MC approach the irradiation of a complete population of $N_c$ cells is simulated, where $N_c$ is supposed to be large enough in order to achieve statistical convergence. The irradiation is modelled as an ordered temporal sequence of particles (primaries and secondaries) that interact with the cell nucleus at random spatial coordinates and random times $t_0^{(c)} \le  t_1^{(c)} \le t_2^{(c)} \le \ldots$, compatible with the chosen irradiation setting and the definition of the time-dependent macroscopic dose rate $\dot{D}(t)$. The irradiation is hence modelled as a sequence of spikes in $\dot{z}^{(c,d)}(t)$, each one of them corresponding to the passage of a particle through or nearby the nucleus, delivering a sequence of random specific energies in the domains
\begin{equation}\label{eqn:zseq}
z_0^{(c,d)}(t_0), z_1^{(c,d)}(t_1), z_2^{(c,d)}(t_2), \ldots
\end{equation}
depending on the particle spectra and track impact positions with respect the cell. A depiction of the temporal evolution of the lesions ($x_I$ and $x_II$) due to the effects of the irradiation as described in equation \eqref{eqn:zseq} is reported in panel (b) of figure \ref{fig:lesions}. For a macroscopic dose $D$ and a specific component $e$ of the energy and particle type spectra, the total number of particles $N_e$ interacting with the nucleus follow a Poisson distribution with mean
\begin{equation}
\bar{N}_{e}=\frac{\rho \sigma D w_{e}}{\mathrm{LET}_F}
\label{man_hits}
\end{equation}
where $\mathrm{LET}_F$ is the frequency-mean LET of the radiation, $\sigma$ is the cross section of the nucleus, $\rho$ is the density of the tissue, and $w_e$ is the weight of the component $e$.

In principle, the tracks can directly sampled from the full measured microdosimetric spectra (\textit{i.e.} not only the first and second moment) from which the experimental $w_e$ can be obtained. From a practical point of view, the specific energies deposited randomly in each domain of each nucleus are evaluated by coupling general purpose MC tools (such as Geant4 \cite{Agostinelli2003} or Fluka \cite{Ferrari2005}), used to evaluate the spatio-temporal energy and particle type spectra in a specific irradiation configurations, with the amorphous track model, as described in \cite{kase2007, inaniwa2010}, with domains arranged according to a close packing hexagonal structure inside the nucleus.

Once the sequence of specific energies in the domains \eqref{eqn:zseq} are obtained, the kinetic equations \eqref{MKMt_kinEQ} can be formally solved for $t \to \infty$ to obtain the average lethal lesions for each cell-domain $(c,d)$:
\begin{equation}
-\log s^{(c,d)} = \left. x_{I}^{(c,d)} \right|_{t\to\infty}= \frac{\alpha_{0}}{N_{d}} \sum_{i=0}^{n_{c}-1} z_{i}^{(c,d)}+G_{c,d} \frac{\beta_{0}}{N_{d}}\left(\sum_{i=0}^{n_{c}-1} z_{i}^{(c,d)}\right)^{2}
\label{man_typeIG}
\end{equation}
where $n_c$ is a Poisson random variable indicating the number of particles that interacted with the cell $c$, $\tau = 1/(a+r)$ is the time constant that defines the first order repair kinetics, and $G$ is the generalized Lea-Catcheside time factor \cite{Lea1942,Kellerer1972} defined at the nanodosimetric level of the domain:
\begin{equation}\label{eqn:micro_lea}
G^{(c,d)}=1-\frac{2}{\left(\sum_{i=0}^{n_{c}-1} z_{i}^{(c,d)}\right)^{2}} \sum_{i=0}^{n_{c}-2} \sum_{j=i+1}^{n_{c}-1}\left(1-e^{-\frac{1}{\tau}\left(t_{j}^{(c)}-t_{i}^{(c)}\right)}\right) z_{i}^{(c,d)} z_{j}^{(c,d)}
\end{equation}

The survival fraction $S$ is obtained by averaging over the entire cell population the survival probability evaluated for each single cell $S_n{(c)}$ (see for comparison equations \eqref{EQN:SurvMKMC} and \eqref{EQN:SurvMKMP})
\begin{equation}\label{eqn:mc_survival}
    S(D)= \langle S_n^{(c)};D\rangle_c = \left< \exp\left( - \sum_{d=1}^{N_d} x_I^{(c,d)}\right)\right>_c = N_d\langle -\langle x_I^{(c,d)}\rangle^{(c)}_d \rangle_c
\end{equation}

\begin{figure}[h!]
\centering
\includegraphics[width= 0.8\textwidth]
{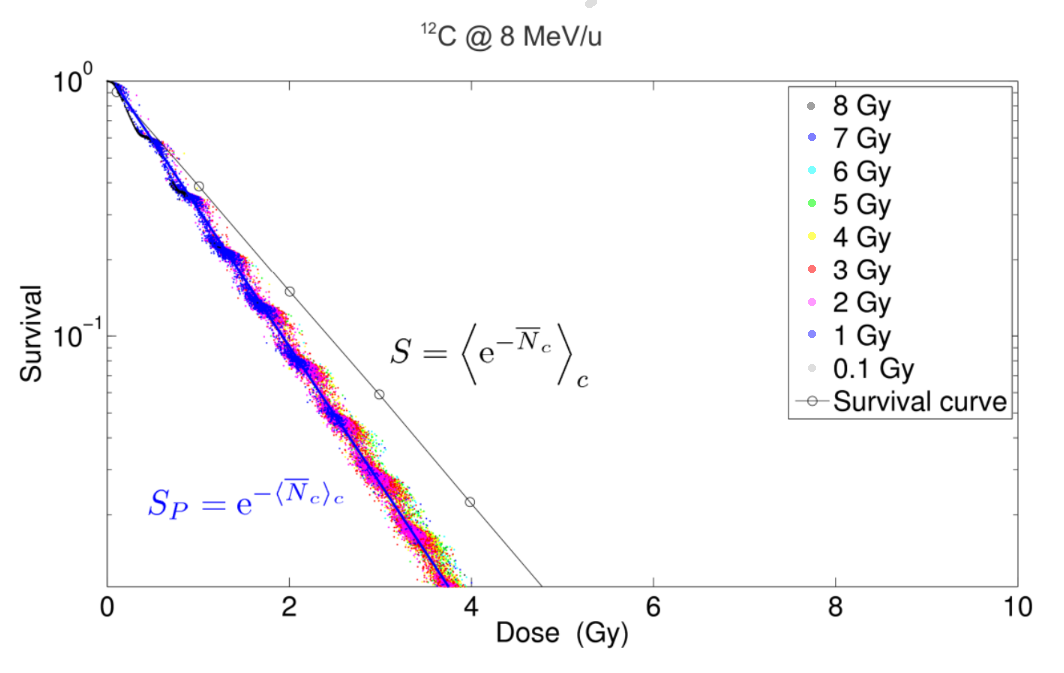}  	
\caption{Simulated survival curves obtained for acute irradiation, $(t_0 = t_1 = t_2 = \ldots)$, of monoenergetic carbon ion (8 MeV/u) with imposed macroscopic doses ranging from 0.1 to 8 Gy. The dots represent the values of cell survival $S^{(c)}_n (z_n^{(c)})$ and specific energy $z_n^{(c)}=  \/N_d\sum_{d=1}^{N_d} \sum_{i=0}^{(n_c-1)}z_i^{(c,d)}$ delivered in the cell $c$ (a small dot for each simulated cell); the variability of the delivered dose with respect to the imposed dose derives by the fluctuation determined by the MC simulation. The two curves were fitted using the LQ model (solid and dashed lines respectively) in order to get the LQ parameters. The blue curve is fitted directly to the $S^{(c)}_n(z_n)$ data, and correspond to the Poisson approximation described in equation \eqref{EQN:SurvMKMPFin}. The black line correspond to the fit to the population averages $\langle S^{(c)}_n; D \rangle_c$ defined in equation \eqref{eqn:mc_survival} for each imposed dose $D$ (open dots) and correspond to the non-Poisson formulation of equation \eqref{kase_alpha_np}. Plot taken from \cite{RussoPHD}.}
	\label{Fig_man_surv}
\end{figure}

Notice that the Monte Carlo approach does not compute directly the LQ coefficients $\alpha$ and $\beta$, in contrast the analytical approaches described in the previous sections. However is possible to derive the LQ coefficients by simulating a complete survival curve, \textit{i.e.} by evaluating \eqref{eqn:mc_survival} using different macroscopic doses $D_1 < D_2 < D_3 < \ldots$, and then fitting the curve with the LQ formula. An example of a complete simulated survival curve is reported in figure \ref{Fig_man_surv}. We remark also that, as done in section \ref{SSEC:timestructure}, the cell population generated by the solutions of the kinetic equations \eqref{MKMt_kinEQ} neglects re-population and cell cycle re-distributions.

One of the benefits of the MCt-MKM approach is that both  $\alpha$ and $\beta$ coefficients, obtained through the survival fitting, show the expected saturation behaviour for high-LET irradiation without adding any corrective factors, like the non-Poisson (section \ref{non-poisson}) or  (section \ref{SSEC:satcor}) saturation. The disadvantage of the approach, other than the inherent approximations specific to the used MC transport code and the adoption of an amorphous track model, is that it can be particularly computing intensive, although this is mitigated by exploiting the multi-core parallelism of modern CPUs \cite{manganaro2018}.

\begin{figure}[h!]
\centering
\includegraphics[width= 0.9\textwidth]
{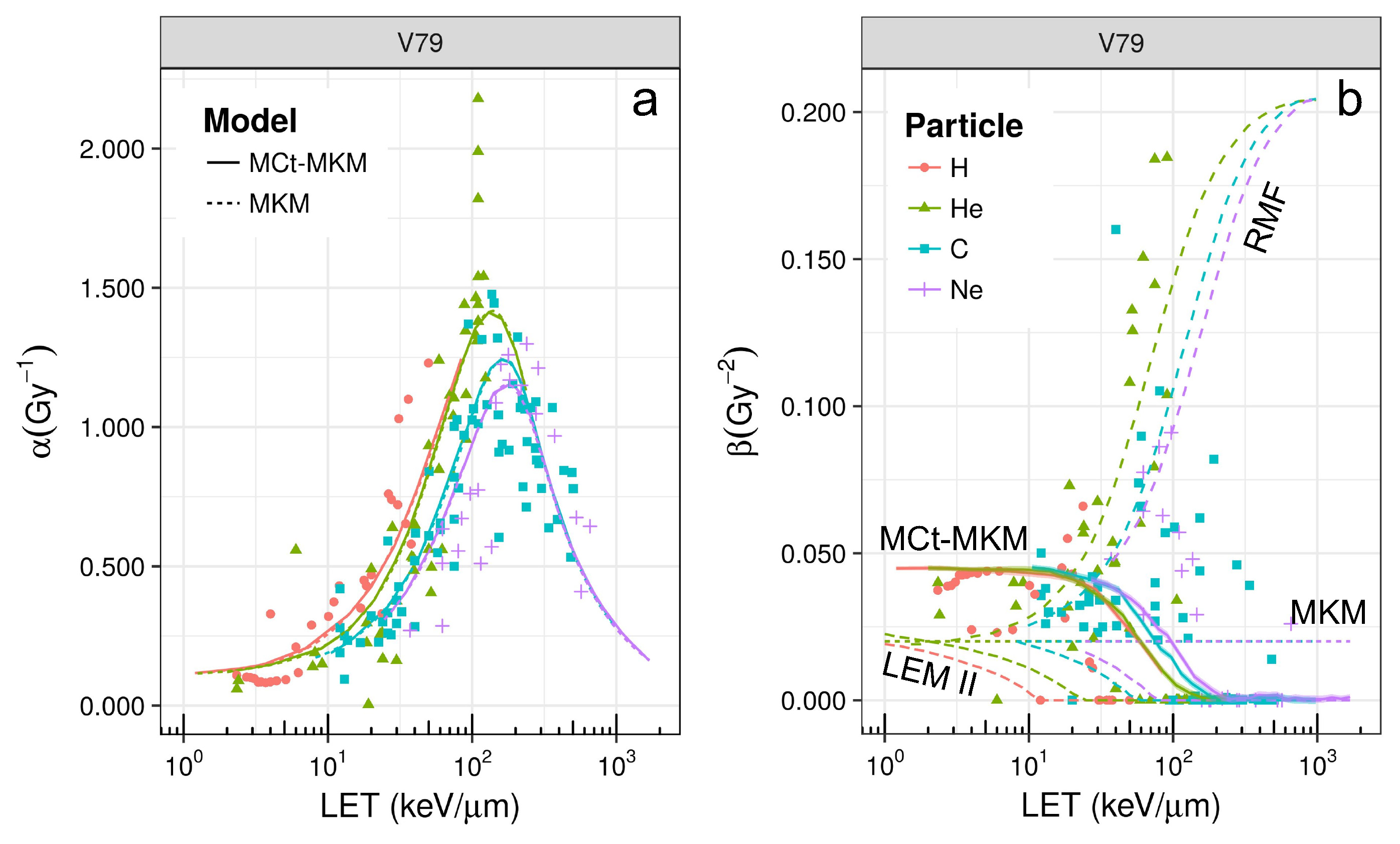}  	
\caption{Linear quadratic $ \alpha $ (panel a) and $ \beta $ (panel b) parameters as a function of LET for the irradiation of of V79 cells with different ions. Points represent experimental data taken from PIDE \cite{friedrich2013}, different colors/gray levels and shapes refer to H, He, C and Ne ions respectively (the colour/gray level and shape legend refers both to panel a and b). In panel a, solid and dashed lines represent respectively the extrapolation with the MCt-MKM and the original MKM models, while in panel b a comparison between different model is reported (namely MKM, MCt-MKM, LEM-II and RMF). In the case of the MCt-MKM, overlapped to the $ \alpha $ and $ \beta $ curves the MC statistical confidence bands (68\%) are reported. These bands are small due to the high statistics and they blends with the curves thickness. A saturation effect is observed for both  $\alpha$ and $\beta$ parameters. Plot taken from \cite{manganaro2017}.}
	\label{Fig_man_res}
\end{figure}

The MCt-MKM has been validated on \textit{in-vitro} experiments considering acute and \textit{split-dose} irradiation on HSG, T1 and V79 cell lines in aerobic conditions of H, He, C and Ne ion beams \cite{manganaro2017}. An example of the behaviour of the LQ $\alpha$ and $\beta$ coefficients is reported in figure \ref{Fig_man_res}, where also a comparison with the prediction of other models, a non-stochastic MKM evaluation, the LEM and the he Repair-Misrepair-Fixation (RMF) model (see section \ref{rmf}), is shown. The main difference with respect to the original MKM is
that the MCt-MKM predicts a non constant and vanishing $\beta$ with high LET values. This behaviour is ultimately due to the non-Poisson statistics inherently implemented in the model. 

The model was also implemented in a TPS \cite{russo2015} to evaluate the effect of the temporal protraction of treatments with different ion beams. The effect of the protraction, described microscopically by the equation \eqref{eqn:micro_lea}, was shown to be compatible with a macroscopic first order effect with temporal constant $\tau$ \cite{manganaro2017}. We remark that, in the framework of the LQ formalization, in the studies of high dose irradiation and the dose-rate effect, such as those reported in \cite{inaniwa2013,inaniwa2015,hawkins2013,hawkins2014,manganaro2017}, the specific way $\beta$ is modeled plays a fundamental role \cite{Fowler2004,carabe2010}. In particular, the behaviour of a vanishing $\beta$ for high LET is compatible with the experimental observation of a reduction of the sensitivity to the dose rate (including the fractionation) in healthy tissues for treatments with high-LET ions, and hence the potential advantage of hypofractionated treatments with these particles.

\subsubsection{The stochastic microdosimetic kinetic model}\label{SEC:sato2012}

The analytical computation method proposed by Sato and Furusawa in \cite{sato2012} introduces a correction to the original formulation of MK model, taking into account the stochastic nature of specific energy deposition in both the domain $z$ and the cell nucleus $z_n$, to improve the adherence of the model to the measured survival fractions for high-LET and high-dose irradiation. The new model is named double-stochastic micro-dosimetric kinetic (DSMK) model. In the same study \cite{sato2012}, a second  model, termed stochastic microdosimetric kinetic (SMK) model, is derived to represents the stochastic nature of domain specific energy $z$ by its approximated mean value and variance in order to reduce the computational time.


Based on radiobiological evidences that states that DNA damage saturates at high-LET regions \cite{friedland2006,hada2008}, the original assumption of the MKM, that the initial numbers of lethal and sub-lethal lesions produced in a domain to be proportional to the specific energy in the site, is changed in the DSMKM assuming that the initial numbers of lethal and sub-lethal lesions produced in a domain are proportional to the \textit{saturation-corrected specific energy}, $z^{*}$ in the domain, calculated as:
\begin{equation}\label{Sato_z'}
z^{*}=z_{0} \sqrt{1-\exp \left[-\left(z_{d} / z_{0}\right)^{2}\right]}
\end{equation}
where $z_0$ can be obtained from $z_0 = y_0(\bar{l}_d/m_d)$ with the saturation correction parameter $y_0$ defined in equation \eqref{eqn:y0} and with dose average $z_D^*$, the saturation correct dose averaged specific energy introduced in equation \eqref{eqn:y*_z*}.

By applying this new parameter, the natural logarithm of survival for a domain and the nucleus, equation \eqref{EQN:SurvDomain} and \eqref{EQN:SurvMKMC}, can be rewritten respectively as:
\begin{equation}
\begin{aligned}\label{Sato_surv}
\log \left(s\left(z\right)\right) &=-A z^{*}-B z^{* 2} \\
\log \left(S_n(z_n)\right) &= -\alpha_{0} \int_{0}^{\infty} z^{*} f(z;z_n)\,d z 
-\beta_{0} \int_{0}^{\infty} z^{* 2} f(z;z_n)\,dz
\end{aligned}
\end{equation}
with the natural log of the survival fraction of cell irradiated with dose $D$, $\log S(D) = \langle S_n \rangle_c$ given by substituting equation \eqref{Sato_surv} in equation \eqref{EQN:SurvMKMP}.

The evaluation of the multi-event probability density $f(z;z_n)$ is obtained numerically by applying a general $n-$fold convolution method such as the one presented in section \ref{SEC:MM} (equations \eqref{eqn:conv} and \eqref{EQN:MutiEvP}) to the single event probability density $f_{1}(z)$. The evaluation of $f_{1}(z)$ is performed exploiting a microdosimetric function implemented in the PHITS Monte Carlo code \cite{sato2018,iwamoto2017}. The sum in equation \eqref{EQN:MutiEvP} can be truncated for practical purposes, with 100 events being enough to evaluate the density probability function of cells with $z_n=100$. 

The same approach is used to calculate the multi-event probability density of the cell nucleus specific energy, for an absorbed macroscopic dose $D$, $f_n(z_n,D)$ from the single event function $f_{n,1}(z_n)$. However, since the nucleus radius is over the available range of the microdosimetric function implemented in PHITS, $f_{n,1}(z_n)$ is determined from the frequency distribution of the LET $L$, $F_L(L)$:
\begin{equation}
    f_{n,1}(z_n) = \int_0^{\infty} F_L(L)f_{n,1}(z_n, L)\,dL
\end{equation}
where $f_{n,1}(z_n, L)$ represents the probability density of $z_n$ from a particle with $\mathrm{LET}=L$.
Following the formalism carried out in \cite{Olko1990}, the expression for $f_{n,1}(z_n,L)$ is written as a Fermi function:
\begin{equation}
f_{n, 1}\left(z_n, L\right) =\frac{2 C}{(L \eta)^{2}} \frac{z_n}{\exp \left[\left(z_{n}-L \eta\right) / \gamma\right]+1}
\end{equation}
where $C$ is a normalization constant and $\eta$ is a units conversion coefficient. The parameter $\gamma$ tunes the slope of the Fermi function or, equivalently, the magnitude of the fluctuation of $z_n$ due to the straggling.

Once $f_{n,1}(z_n)$ is determined, the multi-event function $f_n(z_n;D)$ for the nucleus is obtained with the same $n$-fold convolution procedure used in the case of $f(z;z_n)$. In this case however, due to the higher average number of events that can happen in the nucleus compared to the domain, $\lambda_n(D) \gg \lambda(z_n)$, the computation is significantly more demanding and cannot be practically used for applications inherently complex such as TPS evaluations for ion beam treatments. To overcome this problem, in the case of high-dose irradiations, a pre-evaluated database of probability densities derived from mono energetic irradiations are used and combined in a central limit approximation.

To overcome the long computational time of DSMK model in a TPS workflow a further optimization is performed for the computation of the survival in \eqref{Sato_surv} that bypass the necessity to compute the $n$-fold convolution integral (equation \ref{EQN:MutiEvP}). In this formulation of the model (SMK) is assumed that saturation effect triggered by multiple hits of radiations to a domain is negligibly small so that the magnitude of the effect for the $n$-event energy deposition can be derived from the estimate with the single event density probabilities (see also equations \eqref{EQN:Mom1zF} and \eqref{EQN:Mom2zD}): 
\begin{equation}
\frac{\int_{0}^{\infty} z^* f(z,z)\,dz}{\int_{0}^{\infty} z f(z, z)\,dz} = \frac{\int_{0}^{\infty} z^* f_{1}(z)\,dz}{\int_{0}^{\infty} z f_{1}(z)\,dz} = \frac{z_F^{*}}{z_F}
\label{sato_zDfrac}
\end{equation}
and 
\begin{equation}
\frac{\int_{0}^{\infty} z^{* 2} f(z,z)\,dz}{\int_{0}^{\infty} z^2 f(z,z)\,dz} = \frac{\int_{0}^{\infty} z^{* 2} f_1(z)\,dz}{\int_{0}^{\infty} z^2 f_1(z)\,dz} = \frac{z_D^{*}}{z_D}
\label{sato_zFfrac}
\end{equation}
The natural log of the survival fraction of a cell can be calculated by substituting equations \eqref{sato_zDfrac} and \eqref{sato_zFfrac} in equation \eqref{Sato_surv} obtaining 
\begin{equation}
\log \left(S_n\left(z_n\right)\right) = -\alpha_{0} z_n \frac{\bar{z}_F^{*}}{z_F}-\beta_{0}\left(z_D z_n + z_n^{2}\right) \frac{z_D^{*}}{z_D}
\end{equation}

In order to reduce further the computational effort for TPS applications, assuming that in standard condition of charged-particle therapy the events inducing the saturation of complex DNA damages are rare, and hence $z^*_F/z_F \approx 1$, the following approximation of equation \eqref{EQN:SurvMKMP} was derived \cite{inaniwa2018},
\begin{equation}
    \log S(D) = \log\left\{ 1 + D \left[ -\beta_\mathrm{SMK} + \frac{1}{2}(\alpha_\mathrm{SMK} + 2 \beta_\mathrm{SMK}D)^2\right]z_{n,D} \right\} - \alpha_\mathrm{SMK}D - \beta_\mathrm{SMK}D^2
\end{equation}
with
\begin{align}
    \alpha_{\mathrm{SMK}} &:= (\alpha_0 + z^*_D \beta_0)\\
    \beta_{\mathrm{SMK}} &:=\beta_0(z^*_D/z_D)
\end{align}

\begin{figure}
    \centering
    \includegraphics[width= 0.6\textwidth]{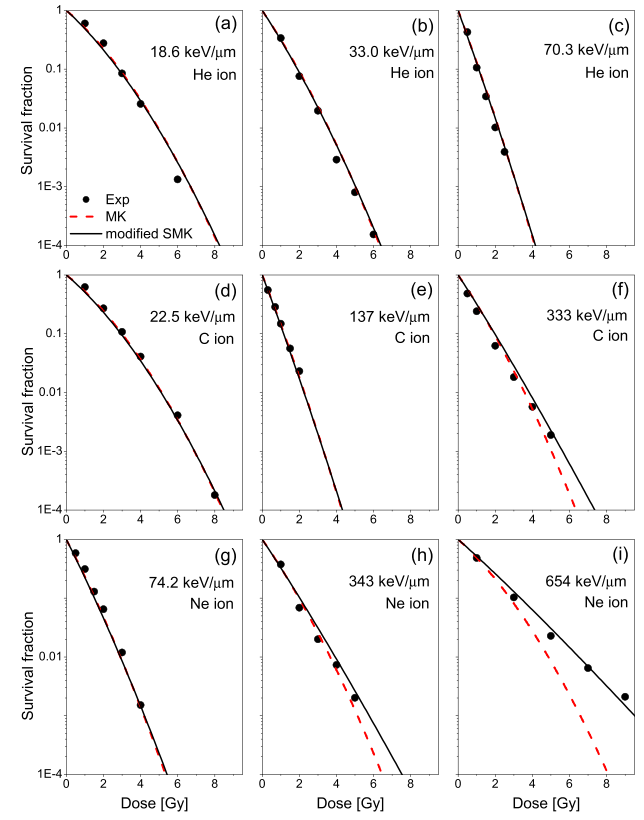}
    \caption{The measured survival fractions of HSG cell exposed to 3He- (a)–(c), 12C- (d)–(f) and 20Ne-ion beams (g)–(i) at different LETs reported by Furusawa textit{et al.} \cite{furusawa2000} (dots), compared with the estimations based on the modified SMK (solid curves) and the MK (dashed curves) models. Experimental data taken from \cite{furusawa2000}. Plot taken from \cite{inaniwa2018}}
    \label{fig:smk}
\end{figure}

Both DSMK and SMK models can reproduce the measured survival fractions, even for high-LET and high-dose irradiations, whereas the simple saturation-based MK model \cite{kase2006} predicts lower values for these irradiations due to the intrinsic ignorance of the stochastic nature of the cell nucleus specific energies (see figure \ref{fig:smk}). In particular, the DSMK model can account for the decrease in the $\beta$ parameter observed in high-dose irradiations over 10 Gy due to the saturation effect triggered by multiple hits of radiations to a domain.

\subsection{Extensions and further improvements of the MK model}\label{sec:extensions}

In recent years, a number of studies have been published reporting further refinements and extensions of the MK model. Among these are further improvement accounting of the non-Poissionian statistics \cite{Abolfath2017,inaniwa2018}, the inclusion of an explicit DNA modeling \cite{Matsuya2014,chen2017}, the effect of a heterogeneous cell population including the cell-cycle variance \cite{Hawkins2000,Hawkins2017,matsuya2018}, and the inclusion of non-target effects \cite{matsuya2018b}. Extension of the model have been also proposed to compute quantities beyond the RBE, such as the oxygen ehancement ratio (OER) \cite{Bopp2016, strigari2018}. In the following of this section, some details about a selection of these developments are described.

 
\subsubsection{Oxygen enhancement ratio (OER) modeling}

Several experimental evidences that cellular oxygenation condition strongly affects their response to ionizing radiation. In particular, a significantly lower cell death rate is observed after exposure to ionizing radiation in the presence of a reduced concentration of oxygen in the cells, \textit{i.e.} in hypoxic conditions. As clinically observed, solid tumours can contain oxygen-deficient regions, thus increasing their radioresistance and potentially leading to treatment failure \cite{Thomlinson1955,Hockel1993}. An understanding of why high-LET radiations are so effective at overcoming tumor hypoxia \cite{Ito2006,Nakano2006} is also particularly relevant for the individualization and optimization of hadron therapy. For this purpose, attempts to extend the MK model to describe the dependence of the the radiation effects on the oxygen concentration in cells and to model the oxygen enhancement ratio (OER) have been made
\cite{Bopp2016,strigari2018}.

It this interesting to note that these MK model-based approaches, although different, do not focus on OER modeling, a relative value, but directly on the prediction of hypoxic cell survival data, being the OER a derived quantity. In \cite{Bopp2016} the reduction of lethal ($x_I$) and potentially lethal ($x_{II}$) damage is linked in the low-LET region to the phenomenological photon OER, though the modulation of parameters $\lambda$ and $\kappa$ (equation \ref{eqn:MKM_kinEQ_initial}) and the effective sizes of domain and nucleus, with the Alper and Flanders functional formalism \cite{Alper1956} the introduce formally the Oxygen concentration dependence in the moedl. In \cite{strigari2018} the general approach proposed by Wenzl and Wilkens \cite{Wenzl2011} has been adapted to the amorphous track approach to the MK model \cite{kase2007} (the latter described in section \ref{SSEC:amtrack}). The inclusion of track model ultimately brings to the OER an explicit dependence on ion type while the Wenzl and Wilkens formalism brings and explicit dose and Oxygen concentration dependence int the $alpha$ and $beta$ parameters. These characteristics have been exploited, by integrating the model in a TPS, to evaluate the tumour control probability (TCP), to facilitate the identification of the optimal treatment conditions in term of ion choice and dose fractionation in presence of hypoxia.

\begin{figure}
    \centering
    \includegraphics[width= 0.9\textwidth]{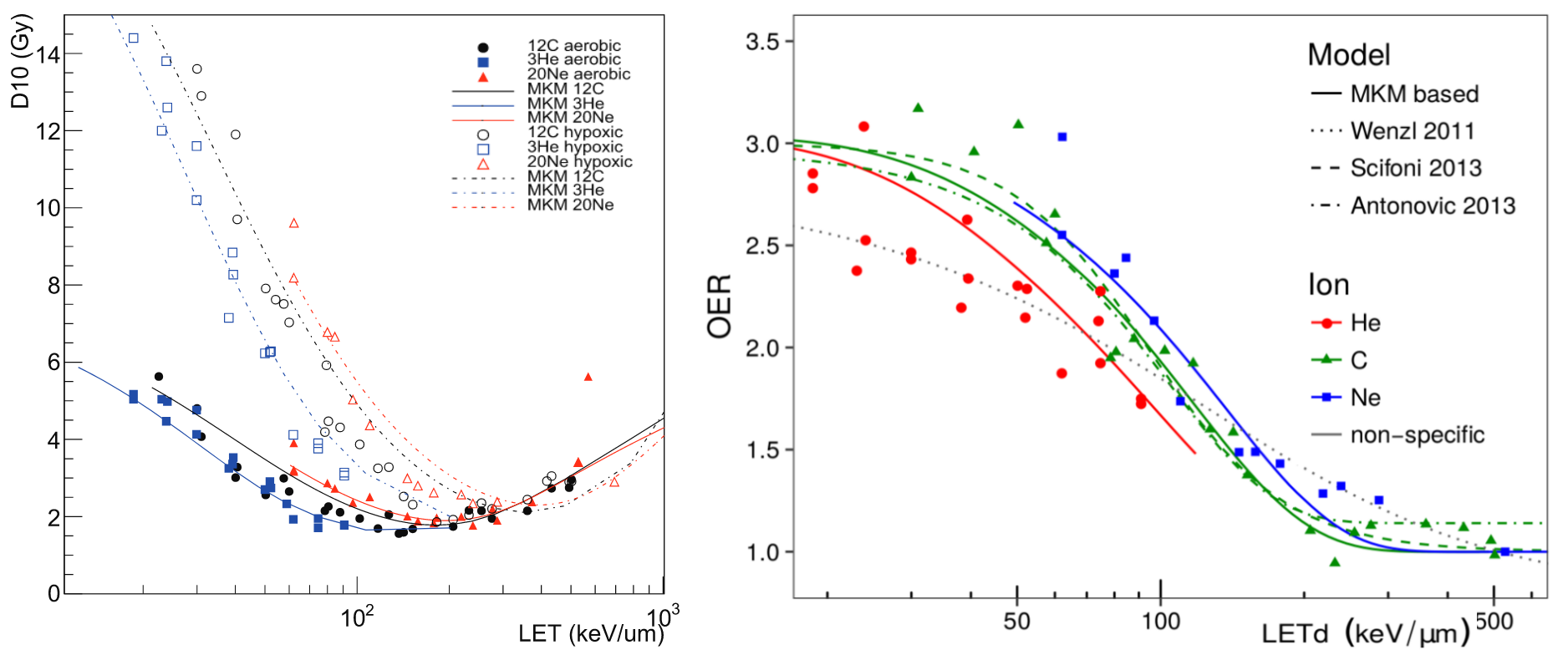}
    \caption{(a) $D_{10}\%$ values for oxic (closed symbols) and extremely hypoxic (open symbols) V79  cells  as  a  function  of  dose-averaged  LET \cite{furusawa2000}.  The  lines  represent the MK model calculations (solid lines) and the hypoxia-adapted MK model calculations with parameters optimised for the V79 cell line (dashed lines). Plot from \cite{Bopp2016}.
    (b) The OER as a function of the dose-averaged LET for the irradiation of HSG cells with different ions. The points represent the experimental data taken from Furusawa \textit{et al.} \cite{furusawa2000}. The continuous lines represent the OER simultaneously evaluated by means of the MKM-based model for He, C and Ne. In the same plot, the OER calculated by other models is reported for comparison (non-continuous lines). The model by Scifoni \textit{et al.} \cite{Scifoni2013a} is associated with carbon ions only, while the model by Wenzl and Wilkens \cite{Wenzl2011} is not associated with any specific ion. An evaluation with the model proposed in Antonovic \textit{et al.} \cite{Antonovic2013} is also shown. Plot taken from \cite{strigari2018}.
    }
    \label{fig:hipoxia}
\end{figure}

The MK-based OER models were verified against in-vitro data from HSG, V79 and CHO cells in aerobic and hypoxic conditions, irradiated with different ion beams \cite{furusawa2000}. Examples of the model prediction versus the experimental data are reported in figure \ref{fig:hipoxia}.

\subsubsection{Non-targeted effects}

In the majority of cell survival modeling approaches it is assumed that biological effects of radiation are exclusively due to direct DNA damage resulting from the ionization caused by the incident radiation. In recent years this assumption has been extensively challenged by considering a variety of indirect processes, also referred as bystander or non-targeted effects (NTE), that significantly impact on the cellular response to the radiation \cite{Mothersill2001}. NTEs have been interpreted as a result of intercellular communication with cell-killing signals between hit and non-hit cells \cite{HAMADA2007,Hamada2012} resulting in induced DNA damage in non-hit cells \cite{Hu2006}.

Attempts to derive kinetic equations to model the intercellular signalling which incorporates signal production and response kinetics have been made \cite{mcmahon2013,Kundrat2015,Mcmahon2016}. In recent studies, such as in Matsuya \textit{et al.} \cite{matsuya2018b}, an integration of these signalling kinetic equations in the MK model, has been proposed.

In this formulation, denoted integrated MK (IMK) model, the number of signalling activation events, $N_\mathrm{NT}$, in the domain is assumed to be a linear-quadratic function of the specific energy $z$. Thus, following the same procedure and the assumption of a Poisson statistics used to derive equations \eqref{EQN:SurvMKMPFin2}-\eqref{EQN:SurvMKMPFin2_ab}, the dose dependent fraction of receiving cells that are activated is written as:
\begin{equation}
    f_\mathrm{NT}(D) = 1 - e^{- \langle N_\mathrm{NT} \rangle_c}
    = 1 - e^{- (\alpha_\mathrm{NT} + z_D \beta_\mathrm{NT})D + \beta_\mathrm{NT} D^2}\,,
\end{equation}
where $\alpha_\mathrm{NT}$ and $\beta_\mathrm{NT}$ are the LQ coefficients of the signal activation process. The propagation of the cell-killing signal is modeled as a diffusion process with diffusion constant $\theta$ and a simple exponential decay with a rate constant $\lambda$. The signal concentration $\rho_s(\underline{r},t)$, where $\underline{r}$ is a spatial position and $t$ is time, can hence be obtained by solving the continuity equation:
\begin{equation}
    \frac{\partial \rho_s(\underline{r},t)}{\partial t} = \theta \nabla^2\rho_s(\underline{r},t) - \lambda \rho(\underline{r},t)\,.
\end{equation}
In non-hit cells the NTE sub-lethal lesions $[x_{II}]_\mathrm{NT}$ are assumed to be induced in proportion to the signal concentration $\rho_s(\underline{r},t)$ and then converted to lethal lesions $x_I$ with the same constant rate $a$ of equations \eqref{MKM_kinEQ} so that the number of sub-lethal lesion is written as:
\begin{equation}\label{eqn:nte}
    [\dot{x}_{II}(\underline{r},t)]_\mathrm{NT} = (1-f_\mathrm{NT}(D)) [x_{II}(\underline{r},t)]_\mathrm{NT} \kappa_\mathrm{NT} R_\mathrm{NT} \rho_s(\underline{r},t) - (a+r_\mathrm{NT})[x_{II}(\underline{r},t)]_\mathrm{NT}\,,
\end{equation}
where $R_\mathrm{NT}$ is the constant rate for cell-killing signals reacting with the nucleus of non-hit cells, $\kappa_\mathrm{NT}$ is the number of sub-lethal lesion per domain caused by the signals, and $r_\mathrm{NT}$ is a constant rate for repair in non-hit cells (in general $r_\mathrm{NT} \neq r$, \textit{i.e.} the repair rate in targeted cells).
In \cite{matsuya2018b} the following functional form for the cell surviving fraction by the NTE ($S_\mathrm{NT}$) has been proposed as an approximate solution of the previous equations
\begin{equation}\label{eqn:surv-imk}
    \log S_\mathrm{NT} = -\langle [x_{I,n}]_\mathrm{NTE} \rangle_c = - \delta \left(1 - e^{- (\alpha_\mathrm{NT} + z_D \beta_\mathrm{NT})D + \beta_\mathrm{NT} D^2} \right)e^{- (\alpha_\mathrm{NT} + z_D \beta_\mathrm{NT})D + \beta_\mathrm{NT} D^2}
\end{equation}
where $\delta$ is a function of the other parameters introduced in the former equations that characterize the intercellular signalling process.

\begin{figure}[h!]
    \centering
    \includegraphics[width= 0.85\textwidth]{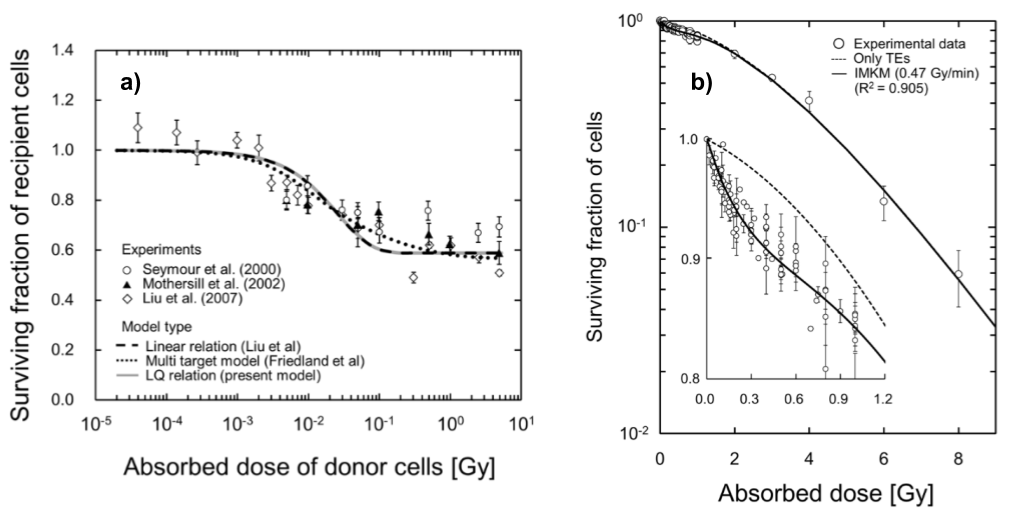}
    \caption{(a) Comparison between the IMK model (Eq. \ref{eqn:surv-imk}), continuous gray line in the plot, and experimental medium transfer bystander effect (MTBE) cell survival data. (b) Fitting of the IMK model to experimental cell survival data for V79-379A. Plot taken from \cite{matsuya2018b} (
    \href{http://creativecommons.org/licenses/by/4.0/}{http://creativecommons.org/licenses/by/4.0/} )m
    }
    \label{fig:imk}
\end{figure}

In order to practically compute the total cell survival $S$, which includes both targeted and NTEs, it is assumed that the probability of interaction between sub-lethal lesions $x_{II}$ and $[x_{II}]_\mathrm{NT}$ in the domain is negligible. This assumption factorizes the two system of equations \eqref{MKM_kinEQ} and \eqref{eqn:nte}, and hence considers the total cell survival as the product of $S = S_\mathrm{T} \times S_\mathrm{NT}$, where the survival for targeted cells, $S_\mathrm{T}$, is given by equation \eqref{EQN:SurvMKMPFin}. Figure \ref{fig:imk} shows an example of the fitting of the IMK model with experimental clonogenic data. It is interesting to note the possibility of the IMK model to account for deviations from the LQ formalism, reproducing the low-dose hypersensitivity behaviour of cell response and evincing its relation with DNA repair mechanisms.

\section{Other models}
\label{SEC:NonMKM}

This section presents alternative models to determine RBE based on microdosimeric approaches. 

\subsection{RBE weighting functions}

The microdosimetric RBE-weighting function approach has been proposed initially by Menzel, Pihet, Wambersie \etal\ \cite{pihet1990other,menzel1990,wambersie1994other,wambersie1990other} to compare
the beam quality of different neutron \cite{pihet1990other} and proton \cite{robertson1994other, coutrakon1997other} therapeutic installations using measured microdosimetric distributions of lineal energy. Based on previous studies on proton beams \cite{kliauga1978other,hall1978other}, this approach  uses measured microdosimetric distributions of lineal energy, $y$, combined with an experimental derived \textit{biological weighting function},  {for specific cell line and endpoints}, $r(y)$, to evaluate  the RBE.

Let $P(y)$  be the cellular response function for a population suffering the fraction of dose $d(y)dy$ corresponding to the lineal energy $y$. $d(y)$ is the dose probability density of $y$ and can be evaluated as $d(y) = \frac{y}{\bar{y}_{F}} f(y)$ \cite{icru1983report}. The linear $\alpha$ parameter, interpreted as the biological effect $E$ per unit dose, is expressed as:
\begin{equation}
\alpha= E/D = \int \frac{P(y)}{y}d(y)\,dy \equiv \int r(y) d(y)\,dy
\end{equation}
where $r(y)$ is defined as the response function. Therefore, the model is rigorously valid under the assumption if a low dose approximation where the cellular response function is linear.

$P(y)$ or directly $r(y)$ are experimental derived. A formulation for $r(y)$ is given in the following \cite{paganetti1997other,morstin1989other}
\begin{equation}
r(y)=\sigma_{E}(1-\exp \left(-a_{1}  y-a_{2}  y^{2}-a_{3} y^{3}\right)/ y\,,
\end{equation}
where the $\sigma_{E}$, $a_1$, $a_2$ and $a_3$ are  parameters specific of the radiobiological end points and are independent on the quality of the radiation. These parameters are determined experimentally by fitting a set of different measurements of $\alpha_i$ or $\mathrm{RBE}_{\alpha,i} = \alpha_i/\alpha_X$ using different irradiation modalities with different radiation qualities $i = 1,2,3, \ldots, N$.

\cite{paganetti1997other,morstin1989other}.
The set of relations that have to be fitted is hence

\begin{equation}
\mathrm{RBE}_{\alpha,i}=\int r(y)  d_{i}(y)\ d y ;\ i=1, \ldots N
\label{EQ:otherRBE}
\end{equation}
The solution of the system of Eq.\ \ref{EQ:otherRBE} can be obtained with different methods, such as non-parametric multi-objective optimization methods \cite{olko1989unfolding} or iterative procedures \cite{loncol1994} through which an initial guess function $ r(y) $ is iteratively updated to best fit the Eq.\ \ref{EQ:otherRBE}.

\subsection{The Repair-Misrepair-Fixation (RMF) model}\label{rmf}

The Repair-Misrepair-Fixation (RMF) model combines the RMR
and LPL models adding the consideration of intra- and intertrack\footnote{
	\textit{Intratrack} binary misrepair occurs when an energy deposition along the track forms two or more DSBs that interact in pairwise mode to form an exchange. \textit{Intertrack} instead, is a binary misrepair arises from the pairwise interaction of break ends associated with DSBs that were
	formed by two separate radiation tracks through a cell.}
binary misrepair to  predict the biological effect of LET \cite{Carlson2008,Frese2012,Stewart2018}. The RMF model considers the entire cell nucleus as the volume for pairwise DSB interactions. In the RMF model a coupled system of nonlinear ordinary differential equations is used to model the time-dependent kinetics of DSB induction, rejoining, and pairwise DSB interaction to form lethal (and nonlethal) chromosome damage. The model treats initial DSB formation as a compound Poisson process and postulates a first-order repair term that gives rise to exponential rejoining kinetics for most DSB ($> 98\%$) and a second-order (quadratic) term to account for the small subset of the initial DSB ($<2\%$) that undergo pairwise DSB interactions to form an exchange.

A LQ approximation of the solution of the RMF system of differential equations can be expressed as follows \cite{Carlson2008,Frese2012}:
\begin{align}
\label{eqn:rmf1}
\mathrm{RBE}_\alpha =& \frac{\Sigma}{\Sigma_X} \left(1 + \frac{2}{R\Sigma_X}(\Sigma z_{n,F} - \Sigma_X (z_{n,F})_X) \right) \\
\mathrm{RBE}_\beta =&  \frac{\Sigma}{\Sigma_X}\,,
\end{align}
where $\Sigma$ is the initial number of DBS per Gray per giga base pair (Gy$^{-1}$Gbp$^{-1}$) and $z_{F,n}$ is the frequency-mean specific energy evaluated in the nucleus (see Eq.\ \ref{EQN:Mom1zF}), for the particle radiation. The suffix $X$ indicates the same quantities for the reference photon radiation. $\mathrm{RBE}_\alpha$, $\mathrm{RBE}_\beta$ and $R$ are defined in Eq.\ \ref{eqn:rbe}, while the ratio $\Sigma/\Sigma_X = \mathrm{RBE}_\mathrm{DSB}$ can be interpreted as the RBE for DSB induction.

From Eq.\ \ref{eqn:rmf1} it follows that the radiation response of a cell exposed to a radiation is uniquely determined by one microdosimetric parameter ($z_{F,n}$) and two biological parameters ($\mathrm{RBE}_\mathrm{DSB}$) and $R$. $z_F$ can be derived form microdosimetric measurements or computed via Monte Carlo simulations. Implicit in the determination of $z_F$ is the knowledge of the size of the nucleus. For a spherical water target of diameter $d$, the frequency-mean specific energy can be approximated by Eq.\ \ref{eqn:sphere_zy}. A complication arises from the fact that, in general, $\mathrm{RBE}_\mathrm{DSB}$ also strongly depends on particle type and kinetic energy (and thus LET or lineal energy) although it is considered to be the same among all eukaryotes.  Consequently, $R$ is basically left as the main parameter of the RMF needed to discriminate the radiation response among different cell lines (compare also the MK model formulation, equation \eqref{eqn:rbe_alpha_gen}).

From a practical point of view, $\mathrm{RBE}_\mathrm{DSB}$ is obtained and stored in a look-up table as a function of particle type and kinetic energy by means of Monte Carlo computations. The MCDS Monte Carlo code \cite{MCDS1,MCDS2}, which is able to simulate also the dependence on the oxygen concentration, is typically used in these computations, so that the RMF has been also used to predict the OER \cite{stewart2011} along with the clonogenic data \cite{Frese2012} and DSB inductions estimates \cite{Carlson2008} for ion irradiations.

In panel (b) of figure \ref{Fig_man_res}  the RMF prediction of the $\mathrm{RBE}_\beta$  compared with experimental data and the evaluations of other models, is reported. It is worth noting that the RMF model predicts an increasing $\beta$ for increasing LET (see equation \eqref{eqn:rmf1}). This is in contrast with other models, such as the LEM, which predict a decreasing $\beta$, or the MK model, which, depending on the specific formulation, predicts both a constant or a decreasing $\beta$.

The RMF has been also implemented in a TPS to evaluate the 3D RBE distribution in irradiated patients \cite{kamp2015}. It is interesting to note that one of the appealing aspects of the RMF for for TPS studies is that the specific response of the tissue, both healty or tumoral, is explicitly determined by a single parameter, $R$. This is a simplification, but allow to study the effect of the specificity of the tissue response in a direct way, also allowing for a distribution of $R$ values and hence easily accounting for the variability and the uncertain associated to this clinical parameter.

\section{Summary}

In clinical treatment planning, the RBE has to be calculated by radiobiological mathematical models, which, in spite of all validation efforts, still involve significant sources of uncertainty.

The aim of this review was to present the theoretical aspects of a selection of radiobiological models that emphasize the link of in-vitro and in-vivo radiobiological outcomes, such as the RBE, to microdosimetric  {experimental} data. We approached these models through a conceptual sketch of their assumptions, highlighting the continuity and leaps of their mathematical formulations. For each model we addressed the limit of applicability and eventual improvements, and the link of their input parameters to experimental observables.

A particular emphasis to the microdosimetric kinetic (MK) -based models has been given. Starting from its first seminal formulation by Hawkins in 1994 \cite{hawkins1994} the MK model has represented an effective approach to link the microdosimetric quantities, which describe the quality of the radiation, to the radiobiological effects and, at present, it is one of the  {most} widely used model to evaluate the RBE in both research and clinical applications. The MK modeling approach for RBE evaluations has gained a particular interest in recent years, with the appearance of different studies aimed to improve the accuracy of the model and to extend its range of applicability in different biological contexts, such as the OER prediction and non-target effects.

Although sharing similar theoretical bases, the MK-based models make different assumptions and approximations in their implementation. Based on these differences, the models considered in this review (including also the RMF model) make, in particular, different predictions in the dependence of $\beta$ on particle LET, and the RBE for cell survival in the overkill regime, for particles with an $\mathrm{LET} \gtrsim 150$ keV/$\mu$m.

Two main aspects of the considered models, where recent efforts have brought interesting insight, and where further future studies could bring potential improvements, could be identified. One aspect is the ascertainment of a more accurate link of the theoretical descriptions to specific cellular mechanisms of DNA damage induction and its evolution, exploiting also information from nanodosimetric data. Another aspect is to improve the theoretical statistical description of the involved processes, be them either the stochastic nature of the energy deposition or the stochastic nature of the cell response to the irradiation.

Future comparisons of model predictions with experimental data are hence needed to fully discriminate among competing mechanisms to be incorporated for the improvement of these model to evaluate the RBE.

\section*{ACKNOWLEDGMENTS}
This work has been partially funded by MoVeIT and NEPTUNE INFN CSN5 projects. \\
We  {thank}  {Dr.} Matsuya \etal for Fig.\ref{fig:imk}

\cleardoublepage
\nocite{*}
\bibliographystyle{apalike}
\bibliography{MicroModRew}

\begin{thebibliography}{}

\bibitem[Abolfath et~al., 2017]{Abolfath2017}
Abolfath, R., Peeler, C.~R., Newpower, M., Bronk, L., Grosshans, D., and Mohan,
  R. (2017).
\newblock {A model for relative biological effectiveness of therapeutic proton
  beams based on a global fit of cell survival data}.
\newblock {\em Scientific Reports}, 7(1):1--12.

\bibitem[Agostinelli et~al., 2003]{Agostinelli2003}
Agostinelli, S., Allison, J., Amako, K., Apostolakis, J., Araujo, H., Arce, P.,
  Asai, M., Axen, D., Banerjee, S., Barrand, G., Behner, F., Bellagamba, L.,
  Boudreau, J., Broglia, L., Brunengo, A., Burkhardt, H., Chauvie, S., Chuma,
  J., Chytracek, R., Cooperman, G., Cosmo, G., Degtyarenko, P., Dell'Acqua, A.,
  Depaola, G., Dietrich, D., Enami, R., Feliciello, A., Ferguson, C.,
  Fesefeldt, H., Folger, G., Foppiano, F., Forti, A., Garelli, S., Giani, S.,
  Giannitrapani, R., Gibin, D., {Gomez Cadenas}, J.~J., Gonzalez, I., {Gracia
  Abril}, G., Greeniaus, G., Greiner, W., Grichine, V., Grossheim, A.,
  Guatelli, S., Gumplinger, P., Hamatsu, R., Hashimoto, K., Hasui, H.,
  Heikkinen, A., Howard, A., Ivanchenko, V., Johnson, A., Jones, F.~W.,
  Kallenbach, J., Kanaya, N., Kawabata, M., Kawabata, Y., Kawaguti, M., Kelner,
  S., Kent, P., Kimura, A., Kodama, T., Kokoulin, R., Kossov, M., Kurashige,
  H., Lamanna, E., Lampen, T., Lara, V., Lefebure, V., Lei, F., Liendl, M.,
  Lockman, W., Longo, F., Magni, S., Maire, M., Medernach, E., Minamimoto, K.,
  {Mora de Freitas}, P., Morita, Y., Murakami, K., Nagamatu, M., Nartallo, R.,
  Nieminen, P., Nishimura, T., Ohtsubo, K., Okamura, M., O'Neale, S., Oohata,
  Y., Paech, K., Perl, J., Pfeiffer, A., Pia, M.~G., Ranjard, F., Rybin, A.,
  Sadilov, S., di~Salvo, E., Santin, G., Sasaki, T., Savvas, N., Sawada, Y.,
  Scherer, S., Sei, S., Sirotenko, V., Smith, D., Starkov, N., Stoecker, H.,
  Sulkimo, J., Takahata, M., Tanaka, S., Tcherniaev, E., {Safai Tehrani}, E.,
  Tropeano, M., Truscott, P., Uno, H., Urban, L., Urban, P., Verderi, M.,
  Walkden, A., Wander, W., Weber, H., Wellisch, J.~P., Wenaus, T., Williams,
  D.~C., Wright, D., Yamada, T., Yoshida, H., and Zschiesche, D. (2003).
\newblock {GEANT4 - A simulation toolkit}.
\newblock {\em Nuclear Instruments and Methods in Physics Research, Section A:
  Accelerators, Spectrometers, Detectors and Associated Equipment}.

\bibitem[Al-Affan and Watt, 1987]{al1987other}
Al-Affan, I. and Watt, D. (1987).
\newblock Determination of quality factors by microdosimetry.
\newblock {\em Nuclear Instruments and Methods in Physics Research Section A:
  Accelerators, Spectrometers, Detectors and Associated Equipment},
  255(1-2):338--340.

\bibitem[Albright, 1989]{albright1989}
Albright, N. (1989).
\newblock A markov formulation of the repair-misrepair model of cell survival.
\newblock {\em Radiation research}, 118(1):1--20.

\bibitem[Albright and Tobias, 1985]{albright1985}
Albright, N. and Tobias, C. (1985).
\newblock Extension of the time-independent repair-misrepair model of cell
  survival to high let and multicomponent radiation.
\newblock In {\em Proceedings of the Berkeley conference in honor of Jerzy
  Neyman and Jack Kiefer}, pages 397--424.

\bibitem[Alper and Howard-Flanders, 1956]{Alper1956}
Alper, T. and Howard-Flanders, P. (1956).
\newblock {Role of oxygen in modifying the radiosensitivity of E. coli B.}
\newblock {\em Nature}, 178(4540):978--979.

\bibitem[Anderson et~al., 2010]{anderson2010}
Anderson, J.~A., Harper, J.~V., Cucinotta, F.~A., and O'Neill, P. (2010).
\newblock Participation of dna-pkcs in dsb repair after exposure to high-and
  low-let radiation.
\newblock {\em Radiation research}, 174(2):195--205.

\bibitem[Antonovic et~al., 2013]{Antonovic2013}
Antonovic, L., Brahme, A., Furusawa, Y., and Toma-Dasu, I. (2013).
\newblock {Radiobiological description of the LET dependence of the cell
  survival of oxic and anoxic cells irradiated by carbon ions}.
\newblock {\em Journal of Radiation Research}, 54(1):18--26.

\bibitem[Asaithamby et~al., 2011]{Asaithamby2011}
Asaithamby, A., Hu, B., and Chen, D.~J. (2011).
\newblock {Unrepaired clustered DNA lesions induce chromosome breakage in human
  cells}.
\newblock {\em Proceedings of the National Academy of Sciences of the United
  States of America}, 108(20):8293--8298.

\bibitem[Asaithamby et~al., 2008]{Asaithamby2008}
Asaithamby, A., Uematsu, N., Chatterjee, A., Story, M.~D., Burma, S., and Chen,
  D.~J. (2008).
\newblock {Repair of HZE-Particle-Induced DNA Double-Strand Breaks in Normal
  Human Fibroblasts}.
\newblock {\em Radiation Research}, 169(4):437--446.

\bibitem[Aso et~al., 2007]{Aso2007}
Aso, T., Kimura, A., Kameoka, S., Murakami, K., Sasaki, T., and Yamashita, T.
  (2007).
\newblock {GEANT4 based simulation framework for particle therapy system}.
\newblock In {\em IEEE Nuclear Science Symposium Conference Record}.

\bibitem[Bianchi et~al., 2020]{bianchi2020}
Bianchi, A., Selva, A., Colautti, P., Bortot, D., Mazzucconi, D., Pola, A.,
  Agosteo, S., Petringa, G., Cirrone, G.~P., Reniers, B., et~al. (2020).
\newblock Microdosimetry with a sealed mini-tepc and a silicon telescope at a
  clinical proton sobp of catana.
\newblock {\em Radiation Physics and Chemistry}, 171:108730.

\bibitem[Bichsel, 1990]{barkas1990}
Bichsel, H. (1990).
\newblock Barkas effect and effective charge in the theory of stopping power.
\newblock {\em Physical Review A}, 41(7):3642.

\bibitem[Bird, 1980]{Bird1980}
Bird, R.~P. (1980).
\newblock {Cysteamine as a Protective Agent with High-LET Radiations}.
\newblock {\em Radiation Research}.

\bibitem[Bird et~al., 1983]{Bird1983}
Bird, R.~P., Zaider, M., Rossi, H.~H., Hall, E.~J., Marino, S.~A., and Rohrig,
  N. (1983).
\newblock {The Sequential Irradiation of Mammalian Cells with X Rays and
  Charged Particles of High LET}.
\newblock {\em Radiation Research}.

\bibitem[Booz et~al., 1983]{icru1983report}
Booz, J., Braby, L., Coyne, J., Kliauga, P., Lindborg, L., Menzel, H., and
  Parmentier, N. (1983).
\newblock Report 36.
\newblock {\em Journal of the International Commission on Radiation Units and
  Measurements}, (1):NP--NP.

\bibitem[Bopp et~al., 2016]{Bopp2016}
Bopp, C., Hirayama, R., Inaniwa, T., Kitagawa, A., Matsufuji, N., and Noda, K.
  (2016).
\newblock Adaptation of the microdosimetric kinetic model to hypoxia.
\newblock {\em Physics in Medicine \& Biology}, 61(21):7586.

\bibitem[Braby, 2015]{braby2015EXP}
Braby, L. (2015).
\newblock Experimental microdosimetry: history, applications and recent
  technical advances.
\newblock {\em Radiation protection dosimetry}, 166(1-4):3--9.

\bibitem[Bradley et~al., 2001]{bradley2001}
Bradley, P., Rosenfeld, A., and Zaider, M. (2001).
\newblock Solid state microdosimetry.
\newblock {\em Nuclear Instruments and Methods in Physics Research Section B:
  Beam Interactions with Materials and Atoms}, 184(1-2):135--157.

\bibitem[Byun et~al., 2009]{byun2009GEM}
Byun, S.~H., Spirou, G.~M., Hanu, A., Prestwich, W.~V., and Waker, A.~J.
  (2009).
\newblock Simulation and first test of a microdosimetric detector based on a
  thick gas electron multiplier.
\newblock {\em IEEE Transactions on Nuclear Science}, 56(3):1108--1113.

\bibitem[Carabe et~al., 2012]{Carabe2012a}
Carabe, A., Moteabbed, M., Depauw, N., Schuemann, J., and Paganetti, H. (2012).
\newblock {Range uncertainty in proton therapy due to variable biological
  effectiveness.}
\newblock {\em Physics in Medicine and Biology}, 57(5):1159--72.

\bibitem[Carabe-Fernandez et~al., 2010]{carabe2010}
Carabe-Fernandez, A., Dale, R., Hopewell, J., Jones, B., and Paganetti, H.
  (2010).
\newblock Fractionation effects in particle radiotherapy: implications for
  hypo-fractionation regimes.
\newblock {\em Physics in Medicine \& Biology}, 55(19):5685.

\bibitem[Carabe-Fernandez et~al., 2007]{Carabe-Fernandez2007}
Carabe-Fernandez, A., Dale, R.~G., and Jones, B. (2007).
\newblock {The incorporation of the concept of minimum RBE (RbEmin) into the
  linear-quadratic model and the potential for improved radiobiological
  analysis of high-LET treatments.}
\newblock {\em International journal of radiation biology}, 83:27--39.

\bibitem[Carabe-Fernandez et~al., 2011]{Carabe-Fernandez2011}
Carabe-Fernandez, A., Dale, R.~G., and Paganetti, H. (2011).
\newblock {Repair kinetic considerations in particle beam radiotherapy.}
\newblock {\em The British journal of radiology}, 84(1002):546--55.

\bibitem[Carlson et~al., 2008]{Carlson2008}
Carlson, D.~J., Stewart, R.~D., Semenenko, V.~A., and Sandison, G.~A. (2008).
\newblock Combined use of monte carlo dna damage simulations and deterministic
  repair models to examine putative mechanisms of cell killing.
\newblock {\em Radiation research}, 169(4):447--459.

\bibitem[Chatterjee and Schaefer, 1976]{chatterjee1976}
Chatterjee, A. and Schaefer, H. (1976).
\newblock Microdosimetric structure of heavy ion tracks in tissue.
\newblock {\em Radiation and environmental biophysics}, 13(3):215--227.

\bibitem[Chen et~al., 2017]{chen2017}
Chen, Y., Li, J., Li, C., Qiu, R., and Wu, Z. (2017).
\newblock A modified microdosimetric kinetic model for relative biological
  effectiveness calculation.
\newblock {\em Physics in Medicine \& Biology}, 63(1):015008.

\bibitem[Conte et~al., 2019]{conte2019data}
Conte, V., Bianchi, A., Selva, A., Petringa, G., Cirrone, G., Parisi, A.,
  Vanhavere, F., and Colautti, P. (2019).
\newblock Microdosimetry at the catana 62 mev proton beam with a sealed
  miniaturized tepc.
\newblock {\em Physica Medica}, 64:114--122.

\bibitem[Conte et~al., 2013]{conte2013}
Conte, V., Moro, D., Grosswendt, B., and Colautti, P. (2013).
\newblock Lineal energy calibration of mini tissue-equivalent gas-proportional
  counters (tepc).
\newblock In {\em AIP Conference Proceedings}, volume 1530, pages 171--178.
  American Institute of Physics.

\bibitem[Coutrakon et~al., 1997]{coutrakon1997other}
Coutrakon, G., Cortese, J., Ghebremedhin, A., Hubbard, J., Johanning, J., Koss,
  P., Maudsley, G., Slater, C., Zuccarelli, C., and Robertson, J. (1997).
\newblock Microdosimetry spectra of the loma linda proton beam and relative
  biological effectiveness comparisons.
\newblock {\em Medical physics}, 24(9):1499--1506.

\bibitem[Cunha et~al., 2017]{cunha2017nanox}
Cunha, M., Monini, C., Testa, E., and Beuve, M. (2017).
\newblock Nanox, a new model to predict cell survival in the context of
  particle therapy.
\newblock {\em Physics in Medicine \& Biology}, 62(4):1248.

\bibitem[Curtis, 1986]{curtis1986}
Curtis, S.~B. (1986).
\newblock Lethal and potentially lethal lesions induced by radiation---a
  unified repair model.
\newblock {\em Radiation research}, 106(2):252--270.

\bibitem[Curtis, 1988]{curtis1988}
Curtis, S.~B. (1988).
\newblock The lethal and potentially lethal model—a review and recent
  development.
\newblock In {\em Quantitative mathematical models in radiation biology}, pages
  137--146. Springer.

\bibitem[Czub et~al., 2008]{Czub2008}
Czub, J., Bana{\'{s}}, D., B{\l}aszczyk, A., Braziewicz, J., Buraczewska, I.,
  Choi{\'{n}}ski, J., G{\'{o}}rak, U., Jask{\'{o}}{\l}a, M., Korman, A.,
  Lankoff, A., Lisowska, H., {\L}ukaszek, A., Szefli{\'{n}}ski, Z., and
  W{\'{o}}jcik, A. (2008).
\newblock {Biological effectiveness of 12C and 20Ne ions with very high LET}.
\newblock {\em International Journal of Radiation Biology}.

\bibitem[Dale et~al., 1999]{Dale1999b}
Dale, R.~G., Fowler, J.~F., and Jones, B. (1999).
\newblock {A new incomplete-repair model based on a 'reciprocal-time' pattern
  of sublethal damage repair}.
\newblock {\em Acta Oncologica}, 38(7):919--929.

\bibitem[Dale and Jones, 1999]{Dale1999}
Dale, R.~G. and Jones, B. (1999).
\newblock {The assessment of RBE effects using the concept of biologically
  effective dose.}
\newblock {\em International journal of radiation oncology, biology, physics},
  43(3):639--45.

\bibitem[Deehan and O'Donoghue, 1988]{LQ_deehan1988}
Deehan, C. and O'Donoghue, J. (1988).
\newblock Biological equivalence between fractionated radiotherapy treatments
  using the linear-quadratic model.
\newblock {\em The British journal of radiology}, 61(732):1187--1188.

\bibitem[Dikomey and Franzke, 1986]{Dikomey1986}
Dikomey, E. and Franzke, J. (1986).
\newblock {DNA repair kinetics after exposure to X-irradiation and to internal
  $\beta$-rays in CHO cells}.
\newblock {\em Radiation and Environmental Biophysics}, 25(3):189--194.

\bibitem[Durante and Loeffler, 2010]{durante2010}
Durante, M. and Loeffler, J.~S. (2010).
\newblock Charged particles in radiation oncology.
\newblock {\em Nature reviews Clinical oncology}, 7(1):37.

\bibitem[Durante et~al., 2017]{durante2017}
Durante, M., Orecchia, R., and Loeffler, J.~S. (2017).
\newblock Charged-particle therapy in cancer: clinical uses and future
  perspectives.
\newblock {\em Nature Reviews Clinical Oncology}, 14(8):483.

\bibitem[Els{\"a}sser et~al., 2008]{LEMelsasser2008}
Els{\"a}sser, T., Kr{\"a}mer, M., and Scholz, M. (2008).
\newblock Accuracy of the local effect model for the prediction of biologic
  effects of carbon ion beams in vitro and in vivo.
\newblock {\em International Journal of Radiation Oncology* Biology* Physics},
  71(3):866--872.

\bibitem[Els{\"a}sser and Scholz, 2007]{LEMelsasser2007}
Els{\"a}sser, T. and Scholz, M. (2007).
\newblock Cluster effects within the local effect model.
\newblock {\em Radiation research}, 167(3):319--329.

\bibitem[Els{\"a}sser et~al., 2010]{LEMelsasser2010}
Els{\"a}sser, T., Weyrather, W.~K., Friedrich, T., Durante, M., Iancu, G.,
  Kr{\"a}mer, M., Kragl, G., Brons, S., Winter, M., Weber, K.-J., et~al.
  (2010).
\newblock Quantification of the relative biological effectiveness for ion beam
  radiotherapy: direct experimental comparison of proton and carbon ion beams
  and a novel approach for treatment planning.
\newblock {\em International Journal of Radiation Oncology* Biology* Physics},
  78(4):1177--1183.

\bibitem[{Ferrari, A. Sala, P.R. Fasso, A. Ranft}, 2005]{Ferrari2005}
{Ferrari, A. Sala, P.R. Fasso, A. Ranft}, J. (2005).
\newblock {FLUKA: A multi-particle transport code (Program version 2005)}.
\newblock {\em Cern-2005-010}.

\bibitem[Fowler, 1989]{Fowler1989}
Fowler, J.~F. (1989).
\newblock {The linear-quadratic formula and progress in fractionated
  radiotherapy}.

\bibitem[Fowler, 1999]{Fowler1999}
Fowler, J.~F. (1999).
\newblock {Is repair of DNA strand break damage from ionizing radiation
  second-order rather than first-order? A simpler explanation of apparently
  multiexponential repair.}
\newblock {\em Radiation research}, 152(2):124--36.

\bibitem[Fowler et~al., 2004]{Fowler2004}
Fowler, J.~F., Welsh, J.~S., and Howard, S.~P. (2004).
\newblock {Loss of biological effect in prolonged fraction delivery.}
\newblock {\em International Journal of Radiation Oncology, Biology, Physics},
  59(1):242--249.

\bibitem[Frese et~al., 2012]{Frese2012}
Frese, M.~C., Victor, K.~Y., Stewart, R.~D., and Carlson, D.~J. (2012).
\newblock A mechanism-based approach to predict the relative biological
  effectiveness of protons and carbon ions in radiation therapy.
\newblock {\em International Journal of Radiation Oncology* Biology* Physics},
  83(1):442--450.

\bibitem[Friedland et~al., 2006]{friedland2006}
Friedland, W., Jacob, P., Paretzke, H.~G., Ottolenghi, A., Ballarini, F., and
  Liotta, M. (2006).
\newblock Simulation of light ion induced dna damage patterns.
\newblock {\em Radiation protection dosimetry}, 122(1-4):116--120.

\bibitem[Friedrich et~al., 2013a]{LEMfriedrich2013}
Friedrich, T., Durante, M., and Scholz, M. (2013a).
\newblock The local effect model-principles and applications.
\newblock {\em The Health Risks of Extraterrestrial Environments}.

\bibitem[Friedrich et~al., 2012a]{friedrich2012}
Friedrich, T., Scholz, U., Els{\"a}sser, T., Durante, M., and Scholz, M.
  (2012a).
\newblock Calculation of the biological effects of ion beams based on the
  microscopic spatial damage distribution pattern.
\newblock {\em International journal of radiation biology}, 88(1-2):103--107.

\bibitem[Friedrich et~al., 2012b]{Friedrich2012b}
Friedrich, T., Scholz, U., Els{\"{a}}sser, T., Durante, M., and Scholz, M.
  (2012b).
\newblock {Systematic analysis of RBE and related quantities using a database
  of cell survival experiments with ion beam irradiation.}
\newblock {\em Journal of radiation research}.

\bibitem[Friedrich et~al., 2013b]{friedrich2013}
Friedrich, T., Scholz, U., Els{\"a}Sser, T., Durante, M., and Scholz, M.
  (2013b).
\newblock Systematic analysis of rbe and related quantities using a database of
  cell survival experiments with ion beam irradiation.
\newblock {\em Journal of radiation research}, 54(3):494--514.

\bibitem[Furusawa et~al., 2000]{furusawa2000}
Furusawa, Y., Fukutsu, K., Aoki, M., Itsukaichi, H., Eguchi-Kasai, K., Ohara,
  H., Yatagai, F., Kanai, T., and Ando, K. (2000).
\newblock Inactivation of aerobic and hypoxic cells from three different cell
  lines by accelerated 3he-, 12c-and 20ne-ion beams.
\newblock {\em Radiation research}, 154(5):485--496.

\bibitem[Grimes, 2020]{grimes2020OER}
Grimes, D.~R. (2020).
\newblock Estimation of the oxygen enhancement ratio for charged particle
  radiation.
\newblock {\em Physics in Medicine \& Biology}.

\bibitem[Guardiola et~al., 2015]{guardiola2015}
Guardiola, C., Fleta, C., Rodr{\'\i}guez, J., Lozano, M., and G{\'o}mez, F.
  (2015).
\newblock Preliminary microdosimetric measurements with ultra-thin 3d silicon
  detectors of a 62 mev proton beam.
\newblock {\em Journal of Instrumentation}, 10(01):P01008.

\bibitem[Gueulette et~al., 2007]{gueulette2007}
Gueulette, J., Wambersie, A., Octave-Prignot, M., De~Coster, B., and
  Gr{\'e}goire, V. (2007).
\newblock Radiobiological characterisation of clinical beams: Importance for
  the quality assurance (qa) programme in ion beam therapy.
\newblock {\em IAEA-TECDOC-1560}, page~29.

\bibitem[Hada and Georgakilas, 2008]{hada2008}
Hada, M. and Georgakilas, A.~G. (2008).
\newblock Formation of clustered dna damage after high-let irradiation: a
  review.
\newblock {\em Journal of radiation research}, pages 0804090035--0804090035.

\bibitem[Hall et~al., 1978]{hall1978other}
Hall, E.~J., Kellerer, A.~M., Rossi, H.~H., and Lam, Y.-M.~P. (1978).
\newblock The relative biological effectiveness of 160 mev protons—ii
  biological data and their interpretation in terms of microdosimetry.
\newblock {\em International Journal of Radiation Oncology* Biology* Physics},
  4(11-12):1009--1013.

\bibitem[Hamada et~al., 2012]{Hamada2012}
Hamada, N., Maeda, M., Otsuka, K., and Tomita, M. (2012).
\newblock {Signaling Pathways Underpinning the Manifestations of Ionizing
  Radiation-Induced Bystander Effects}.
\newblock {\em Current Molecular Pharmacology}.

\bibitem[HAMADA et~al., 2007]{HAMADA2007}
HAMADA, N., MATSUMOTO, H., HARA, T., and KOBAYASHI, Y. (2007).
\newblock {Intercellular and Intracellular Signaling Pathways Mediating
  Ionizing Radiation-Induced Bystander Effects}.
\newblock {\em Journal of Radiation Research}.

\bibitem[Hawkins, 1996]{hawkins1996}
Hawkins, R. (1996).
\newblock A microdosimetric-kinetic model of cell death from exposure to
  ionizing radiation of any let, with experimental and clinical applications.
\newblock {\em International journal of radiation biology}, 69(6):739--755.

\bibitem[Hawkins, 1994]{hawkins1994}
Hawkins, R.~B. (1994).
\newblock A statistical theory of cell killing by radiation of varying linear
  energy transfer.
\newblock {\em Radiation research}, 140(3):366--374.

\bibitem[Hawkins, 1998]{hawkins1998}
Hawkins, R.~B. (1998).
\newblock A microdosimetric-kinetic theory of the dependence of the rbe for
  cell death on let.
\newblock {\em Medical physics}, 25(7):1157--1170.

\bibitem[Hawkins, 2000]{Hawkins2000}
Hawkins, R.~B. (2000).
\newblock {Survival of a Mixture of Cells of Variable Linear-Quadratic
  Sensitivity to Radiation}.
\newblock {\em Radiation Research}, 153(6):840--843.

\bibitem[Hawkins, 2003]{hawkins2003}
Hawkins, R.~B. (2003).
\newblock A microdosimetric-kinetic model for the effect of non-poisson
  distribution of lethal lesions on the variation of rbe with let.
\newblock {\em Radiation research}, 160(1):61--69.

\bibitem[Hawkins, 2017]{Hawkins2017}
Hawkins, R.~B. (2017).
\newblock {Effect of heterogeneous radio sensitivity on the survival, alpha
  beta ratio and biologic effective dose calculation of irradiated mammalian
  cell populations}.
\newblock {\em Clinical and Translational Radiation Oncology}, 4:32--38.

\bibitem[Hawkins and Inaniwa, 2013]{hawkins2013}
Hawkins, R.~B. and Inaniwa, T. (2013).
\newblock A microdosimetric-kinetic model for cell killing by protracted
  continuous irradiation including dependence on let i: repair in cultured
  mammalian cells.
\newblock {\em Radiation research}, 180(6):584--594.

\bibitem[Hawkins and Inaniwa, 2014]{hawkins2014}
Hawkins, R.~B. and Inaniwa, T. (2014).
\newblock {A Microdosimetric-Kinetic Model for Cell Killing by Protracted
  Continuous Irradiation II: Brachytherapy and Biologic Effective Dose}.
\newblock {\em Radiation Research}, 182(1):72--82.

\bibitem[Hockel et~al., 1993]{Hockel1993}
Hockel, M., Knoop, C., Schlenger, K., Vorndran, B., Baussmann, E., Mitze, M.,
  Knapstein, P.~G., and Vaupel, P. (1993).
\newblock {Intratumoral pO2 predicts survival in advanced cancer of the uterine
  cervix}.
\newblock {\em Radiotherapy {\&} Oncology}, 26(1):45--50.

\bibitem[Hu et~al., 2006]{Hu2006}
Hu, B., Wu, L., Han, W., Zhang, L., Chen, S., Xu, A., Hei, T.~K., and Yu, Z.
  (2006).
\newblock {The time and spatial effects of bystander response in mammalian
  cells induced by low dose radiation}.
\newblock {\em Carcinogenesis}.

\bibitem[Inaniwa et~al., 2010]{inaniwa2010}
Inaniwa, T., Furukawa, T., Kase, Y., Matsufuji, N., Toshito, T., Matsumoto, Y.,
  Furusawa, Y., and Noda, K. (2010).
\newblock Treatment planning for a scanned carbon beam with a modified
  microdosimetric kinetic model.
\newblock {\em Physics in Medicine \& Biology}, 55(22):6721.

\bibitem[Inaniwa et~al., 2008]{inaniwa2008}
Inaniwa, T., Furukawa, T., Sato, S., Tomitani, T., Kobayashi, M., Minohara, S.,
  Noda, K., and Kanai, T. (2008).
\newblock Development of treatment planning for scanning irradiation at himac.
\newblock {\em Nuclear Instruments and Methods in Physics Research Section B:
  Beam Interactions with Materials and Atoms}, 266(10):2194--2198.

\bibitem[Inaniwa et~al., 2007]{inaniwa2007}
Inaniwa, T., Furukawa, T., Tomitani, T., Sato, S., Noda, K., and Kanai, T.
  (2007).
\newblock Optimization for fast-scanning irradiation in particle therapy.
\newblock {\em Medical physics}, 34(8):3302--3311.

\bibitem[Inaniwa and Kanematsu, 2018]{inaniwa2018}
Inaniwa, T. and Kanematsu, N. (2018).
\newblock Adaptation of stochastic microdosimetric kinetic model for
  charged-particle therapy treatment planning.
\newblock {\em Physics in Medicine \& Biology}, 63(9):095011.

\bibitem[Inaniwa et~al., 2014a]{inaniwa2014a}
Inaniwa, T., Kanematsu, N., Hara, Y., and Furukawa, T. (2014a).
\newblock Nuclear-interaction correction of integrated depth dose in carbon-ion
  radiotherapy treatment planning.
\newblock {\em Physics in Medicine \& Biology}, 60(1):421.

\bibitem[Inaniwa et~al., 2014b]{inaniwa2014}
Inaniwa, T., Kanematsu, N., Hara, Y., Furukawa, T., Fukahori, M., Nakao, M.,
  and Shirai, T. (2014b).
\newblock Implementation of a triple gaussian beam model with subdivision and
  redefinition against density heterogeneities in treatment planning for
  scanned carbon-ion radiotherapy.
\newblock {\em Physics in Medicine \& Biology}, 59(18):5361.

\bibitem[Inaniwa et~al., 2015a]{Inaniwa2015b}
Inaniwa, T., Kanematsu, N., Matsufuji, N., Kanai, T., Shirai, T., Noda, K.,
  Tsuji, H., Kamada, T., and Tsujii, H. (2015a).
\newblock {Reformulation of a clinical-dose system for carbon-ion radiotherapy
  treatment planning at the National Institute of Radiological Sciences,
  Japan}.
\newblock {\em Physics in Medicine and Biology}, 60(8):3271--3286.

\bibitem[Inaniwa et~al., 2015b]{inaniwa2015}
Inaniwa, T., Kanematsu, N., Suzuki, M., and Hawkins, R. (2015b).
\newblock Effects of beam interruption time on tumor control probability in
  single-fractionated carbon-ion radiotherapy for non-small cell lung cancer.
\newblock {\em Physics in Medicine \& Biology}, 60(10):4105.

\bibitem[Inaniwa et~al., 2013]{inaniwa2013}
Inaniwa, T., Suzuki, M., Furukawa, T., Kase, Y., Kanematsu, N., Shirai, T., and
  Hawkins, R.~B. (2013).
\newblock Effects of dose-delivery time structure on biological effectiveness
  for therapeutic carbon-ion beams evaluated with microdosimetric kinetic
  model.
\newblock {\em Radiation research}, 180(1):44--59.

\bibitem[Ito et~al., 2006]{Ito2006}
Ito, A., Nakano, H., Kusano, Y., Hirayama, R., Furusawa, Y., Murayama, C.,
  Mori, T., Katsumura, Y., and Shinohara, K. (2006).
\newblock {Contribution of Indirect Action to Radiation-Induced Mammalian Cell
  Inactivation: Dependence on Photon Energy and Heavy-Ion LET}.
\newblock {\em Radiation Research}.

\bibitem[Iwamoto et~al., 2017]{iwamoto2017}
Iwamoto, Y., Sato, T., Hashimoto, S., Ogawa, T., Furuta, T., Abe, S.-i., Kai,
  T., Matsuda, N., Hosoyamada, R., and Niita, K. (2017).
\newblock Benchmark study of the recent version of the phits code.
\newblock {\em Journal of Nuclear Science and Technology}, 54(5):617--635.

\bibitem[Jones, 2015]{Jones2015}
Jones, B. (2015).
\newblock {Towards Achieving the Full Clinical Potential of Proton Therapy by
  Inclusion of LET and RBE Models}.
\newblock {\em Cancers}, 7(1):460--480.

\bibitem[Kamp et~al., 2015a]{kamp2015}
Kamp, F., Cabal, G., Mairani, A., Parodi, K., Wilkens, J.~J., and Carlson,
  D.~J. (2015a).
\newblock Fast biological modeling for voxel-based heavy ion treatment planning
  using the mechanistic repair-misrepair-fixation model and nuclear fragment
  spectra.
\newblock {\em International Journal of Radiation Oncology* Biology* Physics},
  93(3):557--568.

\bibitem[Kamp et~al., 2015b]{RMFtps}
Kamp, F., Cabal, G., Mairani, A., Parodi, K., Wilkens, J.~J., and Carlson,
  D.~J. (2015b).
\newblock Fast biological modeling for voxel-based heavy ion treatment planning
  using the mechanistic repair-misrepair-fixation model and nuclear fragment
  spectra.
\newblock {\em International Journal of Radiation Oncology* Biology* Physics},
  93(3):557--568.

\bibitem[Kanai et~al., 1999]{kanai1999}
Kanai, T., Endo, M., Minohara, S., Miyahara, N., Koyama-Ito, H., Tomura, H.,
  Matsufuji, N., Futami, Y., Fukumura, A., Hiraoka, T., et~al. (1999).
\newblock Biophysical characteristics of himac clinical irradiation system for
  heavy-ion radiation therapy.
\newblock {\em International Journal of Radiation Oncology* Biology* Physics},
  44(1):201--210.

\bibitem[Kanai et~al., 1997]{Kanai1997}
Kanai, T., Furusawa, Y., Fukutsu, K., Itsukaichi, H., Eguchi-Kasai, K., and
  Ohara, H. (1997).
\newblock Irradiation of mixed beam and design of spread-out bragg peak for
  heavy-ion radiotherapy.
\newblock {\em Radiation research}, 147(1):78--85.

\bibitem[Karger and Peschke, 2017]{karger2017}
Karger, C.~P. and Peschke, P. (2017).
\newblock Rbe and related modeling in carbon-ion therapy.
\newblock {\em Physics in Medicine \& Biology}, 63(1):01TR02.

\bibitem[Kase, 2012]{kase2012dosimetry}
Kase, K. (2012).
\newblock {\em The dosimetry of ionizing radiation}.
\newblock Elsevier.

\bibitem[Kase et~al., 2007]{kase2007}
Kase, Y., Kanai, T., Matsufuji, N., Furusawa, Y., Els{\"a}sser, T., and Scholz,
  M. (2007).
\newblock Biophysical calculation of cell survival probabilities using
  amorphous track structure models for heavy-ion irradiation.
\newblock {\em Physics in Medicine \& Biology}, 53(1):37.

\bibitem[Kase et~al., 2006]{kase2006}
Kase, Y., Kanai, T., Matsumoto, Y., Furusawa, Y., Okamoto, H., Asaba, T.,
  Sakama, M., and Shinoda, H. (2006).
\newblock Microdosimetric measurements and estimation of human cell survival
  for heavy-ion beams.
\newblock {\em Radiation research}, 166(4):629--638.

\bibitem[Kase et~al., 2011]{kase2011}
Kase, Y., Kanai, T., Sakama, M., Tameshige, Y., Himukai, T., Nose, H., and
  Matsufuji, N. (2011).
\newblock Microdosimetric approach to nirs-defined biological dose measurement
  for carbon-ion treatment beam.
\newblock {\em Journal of radiation research}, 52(1):59--68.

\bibitem[Kellerer and Rossi, 1972]{Kellerer1972}
Kellerer, A.~M. and Rossi, H.~H. (1972).
\newblock {The theory of dual radiation action}.
\newblock {\em Current topics in radiation research}, 8:85--158.

\bibitem[Kellerer and Rossi, 1978]{kellerer1978}
Kellerer, A.~M. and Rossi, H.~H. (1978).
\newblock A generalized formulation of dual radiation action.
\newblock {\em Radiation research}, 75(3):471--488.

\bibitem[Kiefer and Straaten, 1986]{kiefer1986}
Kiefer, J. and Straaten, H. (1986).
\newblock A model of ion track structure based on classical collision dynamics
  (radiobiology application).
\newblock {\em Physics in Medicine \& Biology}, 31(11):1201.

\bibitem[Kliauga et~al., 1978]{kliauga1978other}
Kliauga, P.~J., Colvett, R.~D., Lam, Y.-M.~P., and Rossi, H.~H. (1978).
\newblock The relative biological effectiveness of 160 mev protons i.
  microdosimetry.
\newblock {\em International Journal of Radiation Oncology• Biology•
  Physics}, 4(11):1001--1008.

\bibitem[Kr{\"a}mer and Scholz, 2006]{kramer2006}
Kr{\"a}mer, M. and Scholz, M. (2006).
\newblock Rapid calculation of biological effects in ion radiotherapy.
\newblock {\em Physics in Medicine \& Biology}, 51(8):1959.

\bibitem[Kuang et~al., 2016]{kuang2016}
Kuang, Y., Nagy, J.~D., and Eikenberry, S.~E. (2016).
\newblock {\em Introduction to mathematical oncology}, volume~59.
\newblock CRC Press.

\bibitem[Kundr{\'{a}}t and Friedland, 2015]{Kundrat2015}
Kundr{\'{a}}t, P. and Friedland, W. (2015).
\newblock {Mechanistic modelling of radiation-induced bystander effects.}
\newblock {\em Radiation protection dosimetry}, 166(1-4):148--51.

\bibitem[Lea and Catcheside, 1942]{Lea1942}
Lea, D.~E. and Catcheside, D.~G. (1942).
\newblock {The mechanism of the induction by radiation of chromosome
  aberrations in Tradescantia}.
\newblock {\em Journal of Genetics}.

\bibitem[Leonard, 2007]{leonard2007}
Leonard, B.~E. (2007).
\newblock Adaptive response and human benefit: Part i. a microdosimetry
  dose-dependent model.
\newblock {\em International Journal of Radiation Biology}, 83(2):115--131.

\bibitem[Lindborg et~al., 1999]{lindborg1999}
Lindborg, L., Kyll{\"o}nen, J.-E., Beck, P., Bottollier-Depois, J.-F., Gerdung,
  S., Grillmaier, R., and Schrewe, U. (1999).
\newblock The use of tepc for reference dosimetry.
\newblock {\em Radiation protection dosimetry}, 86(4):285--288.

\bibitem[Lindborg and Waker, 2017]{lindborg2017}
Lindborg, L. and Waker, A. (2017).
\newblock {\em Microdosimetry: experimental methods and applications}.
\newblock CRC Press.

\bibitem[Loncol et~al., 1994]{loncol1994}
Loncol, T., Cosgrove, V., Denis, J., Gueulette, J., Mazal, A., Menzel, H.,
  Pihet, P., and Sabattier, R. (1994).
\newblock Radiobiological effectiveness of radiation beams with broad let
  spectra: microdosimetric analysis using biological weighting functions.
\newblock {\em Radiation Protection Dosimetry}, 52(1-4):347--352.

\bibitem[Magrin, 2018]{Magrin2018}
Magrin, G. (2018).
\newblock {A method to convert spectra from slab microdosimeters in therapeutic
  ion-beams to the spectra referring to microdosimeters of different shapes and
  material}.
\newblock {\em Physics in Medicine {\&} Biology}, 63(21):215021.

\bibitem[Magro et~al., 2017]{magro2017}
Magro, G., Dahle, T., Molinelli, S., Ciocca, M., Fossati, P., Ferrari, A.,
  Inaniwa, T., Matsufuji, N., Ytre-Hauge, K., and Mairani, A. (2017).
\newblock The fluka monte carlo code coupled with the nirs approach for
  clinical dose calculations in carbon ion therapy.
\newblock {\em Physics in Medicine \& Biology}, 62(9):3814.

\bibitem[Mairani et~al., 2017]{mairani2017}
Mairani, A., Magro, G., Tessonnier, T., B{\"o}hlen, T., Molinelli, S., Ferrari,
  A., Parodi, K., Debus, J., and Haberer, T. (2017).
\newblock Optimizing the modified microdosimetric kinetic model input
  parameters for proton and 4he ion beam therapy application.
\newblock {\em Physics in Medicine \& Biology}, 62(11):N244.

\bibitem[Manganaro, 2018]{manganaroPHD}
Manganaro, L. (2018).
\newblock {\em Dose delivery time structure effects in particle therapy:
  development of a time-resolved microdosimetric-kinetic model and
  implementation of spatiotemporal treatment plan optimization}.
\newblock PhD thesis, University of Turin, Italy.

\bibitem[Manganaro et~al., 2018]{manganaro2018}
Manganaro, L., Russo, G., Bourhaleb, F., Fausti, F., Giordanengo, S., Monaco,
  V., Sacchi, R., Vignati, A., Cirio, R., and Attili, A. (2018).
\newblock ‘survival’: a simulation toolkit introducing a modular approach
  for radiobiological evaluations in ion beam therapy.
\newblock {\em Physics in Medicine \& Biology}, 63(8):08NT01.

\bibitem[Manganaro et~al., 2017]{manganaro2017}
Manganaro, L., Russo, G., Cirio, R., Dalmasso, F., Giordanengo, S., Monaco, V.,
  Muraro, S., Sacchi, R., Vignati, A., and Attili, A. (2017).
\newblock A monte carlo approach to the microdosimetric kinetic model to
  account for dose rate time structure effects in ion beam therapy with
  application in treatment planning simulations.
\newblock {\em Medical physics}, 44(4):1577--1589.

\bibitem[Mariotti et~al., 2013]{Mariotti2013}
Mariotti, L.~G., Pirovano, G., Savage, K.~I., Ghita, M., Ottolenghi, A., Prise,
  K.~M., and Schettino, G. (2013).
\newblock {Use of the $\gamma$-H2AX assay to investigate DNA repair dynamics
  following multiple radiation exposures}.
\newblock {\em PLoS ONE}, 8(11):1--12.

\bibitem[Matsuya et~al., 2018a]{matsuya2018}
Matsuya, Y., McMahon, S.~J., Tsutsumi, K., Sasaki, K., Okuyama, G., Yoshii, Y.,
  Mori, R., Oikawa, J., Prise, K.~M., and Date, H. (2018a).
\newblock Investigation of dose-rate effects and cell-cycle distribution under
  protracted exposure to ionizing radiation for various dose-rates.
\newblock {\em Scientific reports}, 8(1):1--14.

\bibitem[Matsuya et~al., 2014]{Matsuya2014}
Matsuya, Y., Ohtsubo, Y., Tsutsumi, K., Sasaki, K., Yamazaki, R., and Date, H.
  (2014).
\newblock {Quantitative estimation of DNA damage by photon irradiation based on
  the microdosimetric-kinetic model}.
\newblock {\em Journal of Radiation Research}, 55(3):484--493.

\bibitem[Matsuya et~al., 2018b]{matsuya2018b}
Matsuya, Y., Sasaki, K., Yoshii, Y., Okuyama, G., and Date, H. (2018b).
\newblock Integrated modelling of cell responses after irradiation for
  dna-targeted effects and non-targeted effects.
\newblock {\em Scientific reports}, 8(1):1--14.

\bibitem[McMahon, 2018]{LQ_mcmahon2018}
McMahon, S.~J. (2018).
\newblock The linear quadratic model: usage, interpretation and challenges.
\newblock {\em Physics in Medicine \& Biology}, 64(1):01TR01.

\bibitem[McMahon et~al., 2013]{mcmahon2013}
McMahon, S.~J., Butterworth, K.~T., Trainor, C., McGarry, C.~K., O’Sullivan,
  J.~M., Schettino, G., Hounsell, A.~R., and Prise, K.~M. (2013).
\newblock A kinetic-based model of radiation-induced intercellular signalling.
\newblock {\em PloS one}, 8(1).

\bibitem[Mcmahon et~al., 2016]{Mcmahon2016}
Mcmahon, S.~J., Schuemann, J., Paganetti, H., and Prise, K.~M. (2016).
\newblock {Mechanistic Modelling of DNA Repair and Cellular Survival Following
  Radiation-Induced DNA Damage}.
\newblock {\em Scientific Reports}.

\bibitem[McNair, 1981]{mcnair1981icru}
McNair, A. (1981).
\newblock {ICRU} report 33-radiation quantities and units pub: International
  commission on radiation units and measurements, washington dc usa issued 15
  april 1980, pp. 25.
\newblock {\em Journal of Labelled Compounds and Radiopharmaceuticals},
  18(9):1398--1398.

\bibitem[McNamara et~al., 2015]{McNamara2015}
McNamara, A.~L., Schuemann, J., and Paganetti, H. (2015).
\newblock {A phenomenological relative biological effectiveness (RBE) model for
  proton therapy based on all published in vitro cell survival data}.
\newblock {\em Physics in Medicine and Biology}, 60(21):8399--8416.

\bibitem[Menzel et~al., 1990]{menzel1990}
Menzel, H., Pihet, P., and Wambersie, A. (1990).
\newblock Microdosimetric specification of radiation quality in neutron
  radiation therapy.
\newblock {\em International journal of radiation biology}, 57(4):865--883.

\bibitem[Morstin et~al., 1989]{morstin1989other}
Morstin, K., Bond, V., and Baum, J. (1989).
\newblock Probabilistic approach to obtain hit-size effectiveness functions
  which relate microdosimetry and radiobiology.
\newblock {\em Radiation research}, 120(3):383--402.

\bibitem[Mothersill and Seymour, 2001]{Mothersill2001}
Mothersill, C. and Seymour, C. (2001).
\newblock {Radiation-Induced Bystander Effects: Past History and Future
  Directions}.
\newblock {\em Radiation Research}.

\bibitem[Nakano et~al., 2006]{Nakano2006}
Nakano, T., Suzuki, Y., Ohno, T., Kato, S., Suzuki, M., Morita, S., Sato, S.,
  Oka, K., and Tsujii, H. (2006).
\newblock {Carbon beam therapy overcomes the radiation resistance of uterine
  cervical cancer originating from hypoxia}.
\newblock {\em Clinical Cancer Research}, 12(7 I):2185--2190.

\bibitem[Newpower et~al., 2019]{newpower2019}
Newpower, M., Patel, D., Bronk, L., Guan, F., Chaudhary, P., McMahon, S.~J.,
  Prise, K.~M., Schettino, G., Grosshans, D.~R., and Mohan, R. (2019).
\newblock Using the proton energy spectrum and microdosimetry to model proton
  relative biological effectiveness.
\newblock {\em International Journal of Radiation Oncology* Biology* Physics},
  104(2):316--324.

\bibitem[Nikjoo et~al., 2002]{nikjoo2002}
Nikjoo, H., Khvostunov, I.~K., and Cucinotta, F.~A. (2002).
\newblock The response of tissue-equivalent proportional counters to heavy
  ions.
\newblock {\em Radiation research}, 157(4):435--445.

\bibitem[Olko, 1989]{olko1989unfolding}
Olko, P. (1989).
\newblock {\em Fluctuations of energy deposited in biological targets by
  ionising radiation}.
\newblock PhD thesis, PhD thesis, Institute of Medicine, KFA J{\"u}lich.

\bibitem[Olko and Booz, 1990]{Olko1990}
Olko, P. and Booz, J. (1990).
\newblock {Energy deposition by protons and alpha particles in spherical sites
  of nanometer to micrometer diameter}.
\newblock {\em Radiation and Environmental Biophysics}.

\bibitem[Orchard et~al., 2011]{orchard2011GEM}
Orchard, G., Chin, K., Prestwich, W., Waker, A., and Byun, S. (2011).
\newblock Development of a thick gas electron multiplier for microdosimetry.
\newblock {\em Nuclear Instruments and Methods in Physics Research Section A:
  Accelerators, Spectrometers, Detectors and Associated Equipment},
  638(1):122--126.

\bibitem[Paganetti et~al., 1997]{paganetti1997other}
Paganetti, H., Olko, P., Kobus, H., Becker, R., Schmitz, T., Waligorski, M.~P.,
  Filges, D., and M{\"u}ller-G{\"a}rtner, H.-W. (1997).
\newblock Calculation of relative biological effectiveness for proton beams
  using biological weighting functions.
\newblock {\em International Journal of Radiation Oncology* Biology* Physics},
  37(3):719--729.

\bibitem[Pihet et~al., 1990]{pihet1990other}
Pihet, P., Menzel, H., Schmidt, R., Beauduin, M., and Wambersie, A. (1990).
\newblock Biological weighting function for rbe specification of neutron
  therapy beams. intercomparison of 9 european centres.
\newblock {\em Radiation Protection Dosimetry}, 31(1-4):437--442.

\bibitem[Powers et~al., 1968]{Powers1968}
Powers, E.~L., Lyman, J.~T., and Tobias, C.~A. (1968).
\newblock {Some effects of accelerated charged particles on bacterial spores}.
\newblock {\em International Journal of Radiation Biology}, 14(4):313--330.

\bibitem[Prise, 1998]{prise1998review}
Prise, K. (1998).
\newblock A review of dsb induction data for varying quality radiations.
\newblock {\em International journal of radiation biology}, 74(2):173--184.

\bibitem[Robertson et~al., 1994]{robertson1994other}
Robertson, J., Eaddy, J., Archambeau, J., Coutrakon, G., Miller, D., Moyers,
  M., Siebers, J., Slater, J., and Dicello, J.~F. (1994).
\newblock Relative biological effectiveness and microdosimetry of a mixed
  energy field of protons up to 200 mev.
\newblock {\em Advances in Space Research}, 14(10):271--275.

\bibitem[Rosenfeld, 2016]{rosenfeld2016}
Rosenfeld, A.~B. (2016).
\newblock Novel detectors for silicon based microdosimetry, their concepts and
  applications.
\newblock {\em Nuclear Instruments and Methods in Physics Research Section A:
  Accelerators, Spectrometers, Detectors and Associated Equipment},
  809:156--170.

\bibitem[Rossi and Zaider, 1991]{rossi1991}
Rossi, H.~H. and Zaider, M. (1991).
\newblock Elements of microdosimetry.
\newblock {\em Medical physics}, 18(6):1085--1092.

\bibitem[Russo, 2011]{RussoPHD}
Russo, G. (2011).
\newblock {\em Development of a radiobiological database for carbon ion
  Treatment Planning Systems}.
\newblock PhD thesis, University of Turin, Italy.

\bibitem[Russo et~al., 2015]{russo2015}
Russo, G., Attili, A., Battistoni, G., Bertrand, D., Bourhaleb, F., Cappucci,
  F., Ciocca, M., Mairani, A., Milian, F., Molinelli, S., et~al. (2015).
\newblock A novel algorithm for the calculation of physical and biological
  irradiation quantities in scanned ion beam therapy: the beamlet superposition
  approach.
\newblock {\em Physics in Medicine \& Biology}, 61(1):183.

\bibitem[Sakama et~al., 2005]{sakama2005}
Sakama, M., Kanai, T., Kase, Y., Komori, M., Fukumura, A., and Kohno, T.
  (2005).
\newblock Responses of a diamond detector to high-let charged particles.
\newblock {\em Physics in Medicine \& Biology}, 50(10):2275.

\bibitem[Sato and Furusawa, 2012]{sato2012}
Sato, T. and Furusawa, Y. (2012).
\newblock Cell survival fraction estimation based on the probability densities
  of domain and cell nucleus specific energies using improved microdosimetric
  kinetic models.
\newblock {\em Radiation research}, 178(4):341--356.

\bibitem[Sato et~al., 2018]{sato2018}
Sato, T., Iwamoto, Y., Hashimoto, S., Ogawa, T., Furuta, T., Abe, S.-i., Kai,
  T., Tsai, P.-E., Matsuda, N., Iwase, H., et~al. (2018).
\newblock Features of particle and heavy ion transport code system (phits)
  version 3.02.
\newblock {\em Journal of Nuclear Science and Technology}, 55(6):684--690.

\bibitem[Schettino et~al., 2011]{schettino2011}
Schettino, G., Ghita, M., Richard, D.~J., and Prise, K.~M. (2011).
\newblock {Spatiotemporal investigations of DNA damage repair using
  microbeams.}
\newblock {\em Radiation Protection Dosimetry}, 143(2-4):340--343.

\bibitem[Scholz et~al., 1997]{LEMscholz1997}
Scholz, M., Kellerer, A., Kraft-Weyrather, W., and Kraft, G. (1997).
\newblock Computation of cell survival in heavy ion beams for therapy.
\newblock {\em Radiation and environmental biophysics}, 36(1):59--66.

\bibitem[Scholz and Kraft, 1992]{LEMscholz1992}
Scholz, M. and Kraft, G. (1992).
\newblock A parameter-free track structure model for heavy ion action cross
  sections.
\newblock In {\em Biophysical modelling of radiation effects}.

\bibitem[Scholz and Kraft, 1996a]{LEMscholz1996}
Scholz, M. and Kraft, G. (1996a).
\newblock Track structure and the calculation of biological effects of heavy
  charged particles.
\newblock {\em Advances in Space Research}, 18(1-2):5--14.

\bibitem[Scholz and Kraft, 1996b]{scholz199}
Scholz, M. and Kraft, G. (1996b).
\newblock Track structure and the calculation of biological effects of heavy
  charged particles.
\newblock {\em Advances in Space Research}, 18(1-2):5--14.

\bibitem[Schuhmacher and Dangendorf, 2002]{schuhmacher2002EXP}
Schuhmacher, H. and Dangendorf, V. (2002).
\newblock Experimental tools for track structure investigations: new approaches
  for dosimetry and microdosimetry.
\newblock {\em Radiation protection dosimetry}, 99(1-4):317--323.

\bibitem[Sch{\"u}rmann et~al., 2018]{schurmann2018}
Sch{\"u}rmann, R., Vogel, S., Ebel, K., and Bald, I. (2018).
\newblock The physico-chemical basis of dna radiosensitization: Implications
  for cancer radiation therapy.
\newblock {\em Chemistry--A European Journal}, 24(41):10271--10279.

\bibitem[Scifoni et~al., 2013]{Scifoni2013a}
Scifoni, E., Tinganelli, W., Weyrather, W.~K., Durante, M., Maier, a., and
  Kr{\"{a}}mer, M. (2013).
\newblock {Including oxygen enhancement ratio in ion beam treatment planning:
  model implementation and experimental verification.}
\newblock {\em Physics in medicine and biology}, 58:3871--95.

\bibitem[Semenenko and Stewart, 2004]{MCDS1}
Semenenko, V. and Stewart, R. (2004).
\newblock A fast monte carlo algorithm to simulate the spectrum of dna damages
  formed by ionizing radiation.
\newblock {\em Radiation research}, 161(4):451--457.

\bibitem[Stewart et~al., 2018]{Stewart2018}
Stewart, R.~D., Carlson, D.~J., Butkus, M.~P., Hawkins, R., Friedrich, T., and
  Scholz, M. (2018).
\newblock A comparison of mechanism-inspired models for particle relative
  biological effectiveness (rbe).
\newblock {\em Medical physics}, 45(11):e925--e952.

\bibitem[Stewart et~al., 2011]{stewart2011}
Stewart, R.~D., Yu, V.~K., Georgakilas, A.~G., Koumenis, C., Park, J.~H., and
  Carlson, D.~J. (2011).
\newblock Effects of radiation quality and oxygen on clustered dna lesions and
  cell death.
\newblock {\em Radiation research}, 176(5):587--602.

\bibitem[Strigari et~al., 2018]{strigari2018}
Strigari, L., Torriani, F., Manganaro, L., Inaniwa, T., Dalmasso, F., Cirio,
  R., and Attili, A. (2018).
\newblock Tumour control in ion beam radiotherapy with different ions in the
  presence of hypoxia: an oxygen enhancement ratio model based on the
  microdosimetric kinetic model.
\newblock {\em Physics in Medicine \& Biology}, 63(6):065012.

\bibitem[Takada et~al., 2018]{Takada2018}
Takada, K., Sato, T., Kumada, H., Koketsu, J., Takei, H., Sakurai, H., and
  Sakae, T. (2018).
\newblock {Validation of the physical and RBE-weighted dose estimator based on
  PHITS coupled with a microdosimetric kinetic model for proton therapy.}
\newblock {\em Journal of radiation research}, 59(1):91--99.

\bibitem[Thomlinson and Gray, 1955]{Thomlinson1955}
Thomlinson, R.~H. and Gray, L.~H. (1955).
\newblock {The Histological Structure of Some Human Lung Cancers and the
  Possible Implications for Radiotherapy}.
\newblock {\em British journal of cancer}, 9(4):539--549.

\bibitem[Tobias, 1980]{tobias1980}
Tobias, C.~A. (1980).
\newblock The repair-misrepair model of cell survival.

\bibitem[Tobias, 1985]{tobias1985}
Tobias, C.~A. (1985).
\newblock The repair-misrepair model in radiobiology: comparison to other
  models.
\newblock {\em Radiation Research}, 104(2s):S77--S95.

\bibitem[Tobias et~al., 1983]{tobias1983}
Tobias, C.~A., Albright, N.~W., and Yang, T.~C. (1983).
\newblock Roles of ionizing radiation in cell transformation.
\newblock Technical report, Lawrence Berkeley Lab.

\bibitem[Tsujii et~al., 2004]{tsujii2004}
Tsujii, H., Mizoe, J.-e., Kamada, T., Baba, M., Kato, S., Kato, H., Tsuji, H.,
  Yamada, S., Yasuda, S., Ohno, T., et~al. (2004).
\newblock Overview of clinical experiences on carbon ion radiotherapy at nirs.
\newblock {\em Radiotherapy and Oncology}, 73:S41--S49.

\bibitem[Tsuruoka et~al., 2005]{Tsuruoka2005}
Tsuruoka, C., Suzuki, M., Kanai, T., and Fujitaka, K. (2005).
\newblock {LET and Ion Species Dependence for Cell Killing in Normal Human Skin
  Fibroblasts}.
\newblock {\em Radiation Research}.

\bibitem[Van~Houten et~al., 2018]{van2018}
Van~Houten, B., Santa-Gonzalez, G.~A., and Camargo, M. (2018).
\newblock Dna repair after oxidative stress: current challenges.
\newblock {\em Current opinion in toxicology}, 7:9--16.

\bibitem[W{\"a}lzlein et~al., 2014]{walzlein2014}
W{\"a}lzlein, C., Kr{\"a}mer, M., Scifoni, E., and Durante, M. (2014).
\newblock Advancing the modeling in particle therapy: From track structure to
  treatment planning.
\newblock {\em Applied Radiation and Isotopes}, 83:171--176.

\bibitem[Wambersie, 1994]{wambersie1994other}
Wambersie, A. (1994).
\newblock Contribution of microdosimetry to the specification of neutron beam
  quality for the choice of the clinical rbe'in fast neutron therapy.
\newblock {\em Radiation Protection Dosimetry}, 52(1-4):453--460.

\bibitem[Wambersie et~al., 1990]{wambersie1990other}
Wambersie, A., Pihet, P., and Menzel, H. (1990).
\newblock The role of microdosimetry in radiotherapy.
\newblock {\em Radiation Protection Dosimetry}, 31(1-4):421--432.

\bibitem[Wang et~al., 2012]{MCDS2}
Wang, C.-C., Hsiao, Y., Lee, C.-C., Chao, T.-C., Wang, C.-C., and Tung, C.-J.
  (2012).
\newblock Monte carlo simulations of therapeutic proton beams for relative
  biological effectiveness of double-strand break.
\newblock {\em International journal of radiation biology}, 88(1-2):158--163.

\bibitem[Wedenberg and Toma-Dasu, 2014]{Wedenberg2014}
Wedenberg, M. and Toma-Dasu, I. (2014).
\newblock {Disregarding RBE variation in treatment plan comparison may lead to
  bias in favor of proton plans}.
\newblock {\em Medical Physics}.

\bibitem[Wenzl and Wilkens, 2011]{Wenzl2011}
Wenzl, T. and Wilkens, J.~J. (2011).
\newblock {Modelling of the oxygen enhancement ratio for ion beam radiation
  therapy.}
\newblock {\em Physics in medicine and biology}, 56(11):3251--68.

\bibitem[Weyrather and Debus, 2003]{Weyrather2003}
Weyrather, W. and Debus, J. (2003).
\newblock {Particle Beams for Cancer Therapy}.
\newblock {\em Clinical Oncology}, 15(1):S23--S28.

\bibitem[Wilson and Field, 1970]{wilson1970}
Wilson, K. and Field, S. (1970).
\newblock Measurement of let spectra using a spherical tissue-equivalent
  proportional counter.
\newblock {\em Physics in Medicine \& Biology}, 15(4):657.

\bibitem[Yaes et~al., 1991]{LQ_yaes1991}
Yaes, R.~J., Patel, P., and Maruyama, Y. (1991).
\newblock On using the linear-quadratic model in daily clinical practice.
\newblock {\em International Journal of Radiation Oncology• Biology•
  Physics}, 20(6):1353--1362.

\bibitem[Zaider and Brenner, 1985]{zaider1985other}
Zaider, M. and Brenner, D.~J. (1985).
\newblock On the microdosimetric definition of quality factors.
\newblock {\em Radiation research}, 103(3):302--316.

\bibitem[Zaider et~al., 1996]{zaider1996}
Zaider, M., Rossi, B. H.~H., and Zaider, M. (1996).
\newblock {\em Microdosimetry and its Applications}.
\newblock Springer.

\bibitem[Zaider and Rossi, 1980]{zaider1980}
Zaider, M. and Rossi, H. (1980).
\newblock The synergistic effects of different radiations.
\newblock {\em Radiation research}, 83(3):732--739.

\bibitem[Zhu et~al., 2019]{zhu2019microdosimetric}
Zhu, H., Chen, Y., Sung, W., McNamara, A.~L., Tran, L.~T., Burigo, L.~N.,
  Rosenfeld, A.~B., Li, J., Faddegon, B., Schuemann, J., et~al. (2019).
\newblock The microdosimetric extension in topas: development and comparison
  with published data.
\newblock {\em Physics in Medicine \& Biology}, 64(14):145004.

\end{thebibliography}

\end{document}